\documentclass[12pt,aps,tightenlines,notitlepage,superscriptaddress]{revtex4-1}

\usepackage{amsmath}
\usepackage{amssymb}
\usepackage{bm}
\usepackage[mathscr]{eucal}
\usepackage{theorem}
\usepackage{graphicx}
\usepackage{psfrag}
\usepackage{subfigure}
\usepackage{color}
\usepackage{wasysym}
\include{graphix}
\usepackage{framed}
\usepackage{enumitem}%
\usepackage{lipsum}
\usepackage{float}
\usepackage{stmaryrd}
\usepackage{MnSymbol}
\definecolor{orange}{rgb}{1,0.5,0}

\graphicspath{{figs_pdf/}}

\newcommand{\imag}{\mathrm{i}}
\newcommand{\mathe}{\mathrm{e}}
\newcommand{\mathd}{\mathrm{d}}

\newcommand{\vecv}{\bm{v}}
\newcommand{\vecx}{\bm{x}}

\newcommand{\mxQ}{{\bf{Q}}}
\newcommand{\mxOm}{{\bf{\Omega}}}
\newcommand{\mxD}{{\bf{D}}}
\newcommand{\mxT}{{\bf{T}}}

\newcommand{\myre}{\mathrm{Re}}
\newcommand{\myeps}{\epsilon}
\newcommand{\mytu}{\mathrm{Tu}}
\newcommand{\mybf}{\mathrm{Br}}
\newcommand{\myop}{\mathcal{S}}

\newcommand{\myq}{\rho}
\newcommand{\mylep}{\Lambda}
\newcommand{\crazylambda}{\nu}

\begin{document}

\bibliographystyle{apsrev}

\title{Advection of nematic liquid crystals by chaotic flow}

\author{Lennon \'O N\'araigh}
\email{onaraigh@maths.ucd.ie}

\affiliation{School of Mathematics and Statistics, University College Dublin, Belfield, Dublin 4}


\date{\today}

\begin{abstract}
Consideration is given to the effects of inhomogeneous shear flow (both regular and chaotic) on nematic liquid crystals in a planar geometry.  The Landau--de Gennes equation coupled to an externally-prescribed flow field is the basis for the study: this is solved numerically in a periodic spatial domain.  The focus is on a limiting case where the advection is passive, such that variations in the liquid-crystal properties do not feed back into the equation for the fluid velocity.   The main tool for analyzing the results (both with and without flow) is the identification of the fixed points of the dynamical equations without flow, which are relevant (to varying degrees) when flow is introduced.  The fixed points are classified as stable/unstable and further as either uniaxial or biaxial.
Accordingly, various models of passive shear flow are investigated, with the main focus being on the case where tumbling is absent from the model.  In this scenario, not only must advection of the $Q$-tensor be considered, but also its co-rotation relative to the local vorticity field.  Thus, only the biaxial fixed point survives as a solution of the $Q$-tensor dynamics under the imposition of a general flow field.    For certain highly specific flows, for which co-rotation effects can effectively be ignored along trajectories, both families of fixed points survive.  In this scenario, the system exhibits coarsening arrest, whereby the liquid-crystal domains are `frozen in' to the flow structures and the growth in their size is thus limited.  The outcome (biaxial final state or a mixture of both families of fixed points) can be understood in terms of the flow time- and length-scales that manifest themselves through the co-rotational derivative.
Some consideration is also given to the specific case where tumbling is present; here, the flow has a strong effect on the liquid-crystal morphology.  

\end{abstract}

\pacs{47.57.Lj, 47.52.+j}

\maketitle

\section{Introduction}
\label{sec:intro}

When a nematic liquid crystal is cooled below a critical temperature,
the isotropic state is energetically unfavorable and the
system spontaneously forms domains wherein the constituent rod-like molecules possess orientational order.  The structure of these domains can be modified in a variety of ways for practical applications~\cite{sluckin2004crystals}, for instance by electric or magnetic fields~\cite{chaikin2000principles}, or by coupling to hydrodynamics~\cite{qian1998}.  The focus of the present work is on the latter, using a mathematical modelling approach.
  Specifically, this work examines a particular configuration where the spatial variations in the liquid-crystal sample are effectively two-dimensional.  This enables one to apply well-established model flows from the theory of chaotic advection of a passive scalar~\cite{lattice_PH1,solomon2003lagrangian}, with a view to controlling the liquid-crystal morphology by stirring.  It is emphasized that although the considered problem geometry (as detailed below) is rather specific, it is physically realizable, and indeed, physically relevant (in particular, in studying the rheology of liquid cystals~\cite{oswald2005nematic}).  Therefore, a further aim of the present study is to open up the possibility to characterize a whole class of liquid-crystal systems with using a range of tools from the theory of chaotic advection~\cite{ottino1989kinematics} and lubrication theory~\cite{oron1997long}, similar to the programme of work already carried out for complex (albeit scalar-valued) order parameters in Reference~\cite{phdlennon}, thereby enhancing the general understanding of such liquid-crystal systems.  Before presenting the results of the present study, in the remainder of the introduction the work is placed in the context of the existing literature.
	
The chosen mathematical modelling framework for the study is the Landau-de Gennes model~\cite{de1993physics} coupled to a flow field, which uses a tensorial order paramter (the $Q$-tensor) to characterize the local orientational order of the sample.  Energy arguments (specifically, a gradient free energy approach) are used to derive an evolution equation for the $Q$-tensor.  This evolution equation for the $Q$-tensor dynamics can be coupled to an incompressible fluid velocity field by extension of these energy arguments~\cite{qian1998,sonnet2001dynamics,sonnet2004continuum}, which results in a full set of equations for the $Q$-tensor and velocity-field evolution.  The coupled set of equations correspond to a so-called active mixture~\cite{giomi2014defect}, whereby gradients in the $Q$-tensor feed back into the flow.   It is emphasized the focus of the present work is on a passive case, whereby the coupling is broken, such that the $Q$-tensor is advected by a velocity field that evolves independently of the $Q$-tensor.  Such a regime is justified by analogy with passively-stirred complex fluids with scalar order parameters~\cite{naraigh2007bubbles,chaos_Berthier}.  Moreover, this regime can also be derived as limiting case of the fully coupled equations for the active mixture; this is done below in Section~\ref{sec:math}.

Notwithstanding the focus of the present work on the passive case, it is worthwhile to discuss the literature on active mixtures.  The coupled equations for the active mixture 
 are amenable to numerical simulation, for which a popular solution method is a  combination of the lattice--Boltzmann method for the hydrodynamics and finite-differences for the $Q$-tensor equation to study domain growth in liquid crystals.  Reference~\cite{henrich2013rheology} 
examines the formation of the so-called `blue phase' in cholesteric liquid crystals, Reference~\cite{lintuvuori2010colloids} examines the hydrodynamic interactions between a liquid crystal and a colloidal suspension.
Notably, these works involve full three-dimensional simulations, which are computationally expensive, although necessary to characterize the system of coupled equations in its full generality.  In contrast, the focus of the present paper is on situations wherein physically sensible simulations in two dimensions are justifiable.
 
Also in contrast to these fully three-dimensional simulations, Reference~\cite{giomi2015geometry} again focuses on numerical simulations of the coupled equations in two dimensions where the flow is turbulent.  Here, the turbulent vortices are maintained by the  defect structure of the liquid crystal, which in turn advect the vortices, leading to enhanced turbulence production and chaotic mixing.  The existence of the inverse energy cascade in two-dimensional turbulence makes this process self-sustaining.    Reference~\cite{Thampi2015} compares active and passive mixtures in a two-dimensional channel geometry, for simple shear flows, for which the results on the passive case are relevant to the present study.  As such, the authors in Rererence~\cite{Thampi2015} observe that the passively-advected liquid crystal possesses no topological defects.  Indeed, in situations where the inhomogeneous term in the $Q$-tensor evolution equation is small (corresponding to low levels of `tumbling'), the liquid crystal orients with the flow to produce a uniform state.   Only when the channel walls are deformed so as to possess a non-uniform cross section do topological defects appear.   These results are of direct relevance to the present study wherein a mathematical model for a stirred two-dimensional liquid crystal is presented: for certain model flows, only a uniform state persists at late times, whereas for others, liquid-crystal defects are permitted.  Understanding the key differences between these two scenarios provides the motivation for this study, in particular, Section~\ref{sec:discussion} herein.

This article is organized as follows.  In Section~\ref{sec:math} the Landau--de Gennes model of the liquid-crystal dynamics is recalled, along with a formalism to couple the resulting $Q$-tensor equation to hydrodynamics, and a description of the model tensorial equations in the relevant planar geometry.  Passive advection of the $Q$-tensor dynamics appears as a limiting case and is identified via dimensional analysis; this limiting case provides the main focus for the present work.  The fixed points of the unadvected system are analysed in Section~\ref{sec:fixed}: this provides the main theoretical tool for the work; numerical simulations for the case without flow are also presented herein: this provides a benchmark against which to compare the scenarios with flow but also confirms the existence of coarsening in the system, whereby small-scale liquid-crystal domains grow algebraically in time.  The results for the cases with flow are presented in Section~\ref{sec:flow}, wherein several model flows are compared and contrasted.  A theoretical explanation for the results is presented in Section~\ref{sec:discussion}, along with concluding remarks.

\section{Mathematical Formulation}
\label{sec:math}

In this section the Landau--de Gennes model for a nematic liquid crystal is introduced.  A specialization to thin samples is made, such that the complexity of the problem reduces substantially. The homogeneous case is studied  in detail and the fixed points of the corresponding equations are identified.  Finally,  the relevant advection terms are introduced to complete the model description.

A standard model for nematic liquid crystals starts with the traceless symmetric $Q$-tensor $\mxQ$, $ij^{\mathrm{th}}$ entry $Q_{ij}$.  In a spectral decomposition, the $Q$-tensor can be written as
\begin{equation}
Q_{ij}=(2\lambda_1+\lambda_2) \left(n_i^{(1)} n_j^{(1)} -\tfrac{1}{3}\delta_{ij}\right)+
(2\lambda_2+\lambda_1) \left(n_i^{(2)} n_j^{(2)} -\tfrac{1}{3}\delta_{ij}\right),
\label{eq:spectral}
\end{equation}
where the $\lambda_1$ and $\lambda_2$ are eigenvalues of $\mxQ$ with corresponding eigenvectors $\bm{n}^{(1)}$ and $\bm{n}^{(2)}$ (the third eigenvalue is determined by the traceless condition, $\lambda_1+\lambda_2+\lambda_3=0$).  The system is characterized by its scalar free energy,
\begin{equation}
F=F_0+F_K,
\end{equation}
comprising bulk and elastic contributions.  Following the standard approach,  take the following form for the bulk contribution $F_0$ is taken:
\begin{equation}
F_0= \int_\Omega \mathd^3 x \chi(\mxQ),\qquad
\chi(\mxQ)=\tfrac{1}{2}\alpha_F \mathrm{tr}(\mxQ^2)-\beta_F \mathrm{tr}(\mxQ^3)+\gamma_F \left[\mathrm{tr}(\mxQ^2)\right]^2,
\label{eq:f0}
\end{equation}
where $\Omega$ is the container volume (surface contributions are neglected for the time being).
Here, $\alpha_F$ is taken to be temperature-dependent, with $\alpha_F=12a(T-T_*)$.  Here $a$ is a positive constant, $T$ is temperature, and $T_*$ is a reference temperature.  Also, $\beta_F$ and $\gamma_F$ are taken to be positive constants.  In this way, it is energetically favourable for the system to be in a nematic state with $\mxQ\neq 0$ below a certain critical temperature $T_{IN}$; the presence of the cubic term in Equation~\eqref{eq:f0} (corresponding to $\beta_F\neq 0$) means that $T_{IN}\neq T_*$ -- the precise value of $T_{IN}$ is computed below in Section~\ref{sec:fixed}.
Concerning the elastic contribution to the free energy,  consideration is given to a simple model wherein each elastic constant is the same and the cholesteric parameter is zero.  In this case, the elastic contribution to the free energy can be written simply (up to divergence terms) as
\begin{equation}
F_K=\int_\Omega \mathd^3 x W(\nabla \mxQ),\qquad
W(\nabla \mxQ)=\tfrac{1}{2}k\|\nabla \mxQ\|_2^2,
%
\end{equation}
where $k$ is a constant related to the Frank constant $K$ by $K=kS_0^2/2$, where $S_0$ is a typical value of the nematic order parameter $\myop=\sqrt{6\mathrm{tr}(\mxQ^2)}$.

In the absence of flow, the system then evolves through time so as to minimize the total free energy, subject to the constraints that the $Q$-tensor remain traceless and symmetric.  This results in the following gradient dynamics:
\begin{subequations}
\begin{equation}
\zeta_1\frac{\partial Q_{ij}}{\partial t}=-\left[\frac{\delta F}{\delta Q_{ij}}- \tfrac{1}{3}\mathrm{tr}\left(\frac{\delta F}{\delta Q_{ij}}\right)\delta_{ij}\right],
\end{equation}
where $\zeta_1^{-1}$ is the relaxation rate.
Explicitly (in index-free notation), this reads
\begin{equation}
\zeta_1\frac{\mxQ}{\partial t}=\left[k \nabla^2\mxQ-\left(\alpha_F \mxQ-3\beta_F\mxQ^2+4\gamma_F\mathrm{tr}(\mxQ^2)\mxQ-2\beta_F\mathrm{tr}(\mxQ^2)\mathbb{I}\right)\right].
\end{equation}%
\label{eq:dqdt}%
\end{subequations}%
In contrast, in this work, consideration is given to the case where the $Q$-tensor is transported and subjected to rotations via an incompressible velocity field $\vecv(\vecx,t)$.  In the general case, gradients in the $Q$-tensor feed back into the velocity field itself, via Navier--Stokes-type equations.  The general equations for the coupled motions of the $Q$-tensor and the fluid are as follows~\cite{qian1998,sonnet2001dynamics,sonnet2004continuum} (see also Appendix~\ref{app:derivation}):
\begin{subequations}
\begin{multline}
\zeta_1\left(\partial\frac{\partial\mxQ}{\partial t}+\vecv\cdot\nabla\mxQ-\mxOm\mxQ-\mxQ\mxOm\right)+\zeta_2\mxD=
\\
k\nabla^2\mxQ-\left(\alpha_F\mxQ-3\beta_F\mxQ^2+4\gamma_F\mathrm{tr}(\mxQ^2)\mxQ\right)+\tfrac{1}{3}\mathbb{I}
\left[\zeta_2 \mathrm{tr}(\mxD)-3\beta_F \mathrm{tr}(\mxQ^2)\right].
\label{eq:q_tensor}
\end{multline}
%
%
\begin{multline}
\rho_0\left(\frac{\partial\vecv}{\partial t}+\vecv\cdot\nabla\vecv\right)=\nabla\cdot\mxT,\\
\mxT=-p\mathbb{I}-k\nabla\mxQ\odot\nabla\mxQ+\zeta_2\stackrel{\circ}{\mxQ}+\zeta_3\mxD+\zeta_{31}(\mxD\mxQ+\mxQ\mxD)+\zeta_{32}(\mxD\cdot\mxQ)\mxQ.
\label{eq:ttensor}
\end{multline}
\label{eq:dynam2}%
\end{subequations}%
Equation~\eqref{eq:ttensor} for the hydrodynamics is supplemented with the incompressibility condition $\nabla\cdot\vecv=0$.  Here, $\mxD$ denotes the symmetric part of the rate-of-strain tensor, $D_{ij}=(1/2)(\partial_i u_j+\partial_j u_i)$ and $\mxOm$ denotes the antisymmetric part, such that $\Omega_{ij}=(1/2)(\partial_i u_j-\partial_j u_i)$.  Here also, $\rho_0$ denotes the constant fluid density and finally, $\zeta_1$, $\zeta_2$, $\zeta_3$, $\zeta_{31}$ and $\zeta_{32}$  are phenomenological viscosity coefficients which can be related to the Leslie viscosities~\cite{sonnet2001dynamics,sonnet2004continuum}.
Finally, the tensor products in Equation~\eqref{eq:dynam2} are defined via the relations
\begin{equation}
(\nabla\mxQ\odot\nabla\mxQ)_{ij}=\frac{\partial Q_{i'j'}}{\partial x_i}\frac{\partial Q_{i'j'}}{\partial x_j},
\label{eq:odotdef}
\end{equation}
and $\mxD\cdot\mxQ=D_{ij}Q_{ij}$; the context in which such terms arise is explained in detail in Appendix~\ref{app:derivation}.

The main subject of this work is  \textit{passive advection}, whereby the $Q$-tensor is advected by the flow, but does not feed back into the dynamical equation for the velocity field itself.  Therefore, in this section  a dimensional analysis is performed on Equation~\eqref{eq:dynam2} to derive the precise scaling relation for this flow regime to be valid.
Accordingly,  Equations~\eqref{eq:dynam2} are non-dimensionalized based on a characteristic timescale $t_0=\zeta_1/(8\gamma_F)$ and a characteristic lengthscale $L$, which results in the following dimensionless dynamical equations (dimensionless quantities denoted by a tilde, with $\mxQ$ already being inherently dimensionless).  For the $Q$-tensor, one obtains
\begin{subequations}
\begin{multline}
%
%
%
\frac{\partial\mxQ}{\partial \widetilde{t}}+\widetilde{\vecv}\cdot\widetilde{\nabla}\mxQ-\widetilde{\mxOm}\mxQ-\mxQ\widetilde{\mxOm}
+(\zeta_2/\zeta_1)\widetilde{\mxD}\\
=\myeps^2\widetilde{\nabla}^2\mxQ+g_1(1-\theta)\mxQ+3g_2\mxQ^2-\tfrac{1}{2}\mathrm{tr}(\mxQ^2)\mxQ
+\tfrac{1}{3}\mathbb{I}\left[ (\zeta_2/\zeta_1)\mathrm{tr}(\widetilde{\mxD})-3g_2\mathrm{tr}(\mxQ^2)\right],
\label{eq:qtensor}%
\end{multline}%
where 
\begin{equation}
\alpha_F/(8\gamma_F)=-g_1[1-(T/T_*)]\equiv -g_1(1-\theta),\qquad
\beta_F/(8\gamma_F)=g_2,\qquad
\myeps^2=k/(8L^2\gamma_F).
\end{equation}
Accordingly, the ratio 
\begin{equation}
\mytu=\zeta_2/\zeta_1
\end{equation} is identified as the \textit{tumbling parameter}~\cite{ramage2015}.   In a simple shear one can expect flow alignment for $\mytu>1$ wherein the liquid crystal aligns at a fixed angle with the direction of the local flow gradient~\cite{leslie1968}.  On the other hand, for $\mytu<1$ it can be expected that both transport and rotation (`tumbling') of the liquid-crystal sample should dominate.  For the hydrodynamic equation, the nondimensionalization yields
\begin{multline}
\widetilde{\dot\vecv}\equiv \frac{\partial\widetilde{\vecv}}{\partial \widetilde{t}}+\widetilde{\vecv}\cdot\widetilde{\nabla}\widetilde{\vecv}=
\\
-\widetilde{\nabla} \widetilde{p}+\frac{1}{\myre}\nabla\cdot\widetilde{\mxD}+\mybf\nabla\cdot\left[-\epsilon^2\widetilde{\nabla}\mxQ\odot\widetilde{\nabla}\mxQ+ \mytu \widetilde{\stackrel{\circ}{\mxQ}}+(\zeta_{31}/\zeta_1)(\widetilde{\mxD}\mxQ+\mxQ\widetilde{\mxD})+(\zeta_{32}/\zeta_1)(\widetilde{\mxD}\cdot\mxQ)\mxQ\right],
\end{multline}
where
\begin{equation}
\mybf=\frac{\zeta_1}{\rho_0 L(L/t_0)},\qquad \myre=\frac{\rho_0 L(L/t_0)}{\zeta_3}.
\end{equation}%
\label{eq:hydro_dimless}%
\end{subequations}%
Here, $\mybf$ is the backreaction strength, and represents the feedback of the $Q$-tensor and its gradients on the flow, while $\myre$ is the standard hydrodynamic Reynolds number.  Following standard practice,   the ornamentation over the dimesionless quantities is dropped, with the implicit assumption that all calculations are henceforth carried out using the dimensionless formulation.

Equations~\eqref{eq:hydro_dimless} now make manifest the precise regime of passive advection: it is $\mybf=0$.  This is the regime of interest for the rest of this work.  It is also assumed that the hydrodynamics are forced by a particular protocol, such that $\vecv(\vecx,t)$ corresponds to a desired model chaotic flow, used elsewhere in the literature to scalar transport~\cite{ottino1989kinematics} and also, advection of the Cahn--Hilliard scalar order parameter~\cite{chaos_Berthier,naraigh2007bubbles}, of which more details are presented below.
To make numerical progress (and to illuminate the physics in simple scenarios), this study is furthermore restricted  to two-dimensional geometries.  The problem geometry is selected such that the liquid-crystal sample is sandwiched between two narrowly-separated parallel plates.  For definiteness,  the direction normal to the plates is labelled as the $z$-direction.  Furthermore, boundary conditions are chosen such that the molecules on each of the bounding plates are anchored in the same fashion so that all $z$-variation is suppressed.  The director is parallel to the plates, i.e. has no component in the $z$-direction.  In this geometry, the $Q$-tensor possesses only four independent components~\cite{schopohl1987defect}, which are written here as
\begin{equation}
\mxQ=\left(\begin{array}{ccc}q & r & 0\\
                               r & s & 0\\
															 0 & 0 & -(q+s)\end{array}\right),
\label{eq:qtwod}
\end{equation}%
with $\myop^2=6 \mathrm{tr}(\mxQ^2)=12(q^2+r^2+s^2+qs)$.  The spectral decomposition~\eqref{eq:spectral} gives
\begin{multline}
\mxQ=\left(\begin{array}{ccc} \lambda_1 n_x^2+\lambda_2 n_y^2& (\lambda_1-\lambda_2)n_xn_y&0\\
(\lambda_1-\lambda_2)n_xn_y& \lambda_1 n_y^2+\lambda_2 n_x^2 & 0\\
0 & 0 & -(\lambda_1+\lambda_2)\end{array}\right),\qquad
\lambda_{1,2}=\tfrac{1}{2}\left(q+s\right)\pm \sqrt{\tfrac{1}{4}\left(q-s\right)^2+r^2},
\label{eq:qtwodspectral}
\end{multline}
where $\lambda_{1,2}$ are the eigenvalues of the $Q$-tensor that lie entirely in the $xy$-plane and $\bm{n}^{(1)}=(n_1,n_2,0)$  is the eigenvalue corresponding to $\lambda_1$.   The tensor $\mxQ$ is uni-axial if and only if it can be written as $Q_{ij}\propto m_im_j-(1/3)\delta_{ij}$, for some unit vector $\bm{m}=(m_1,m_2,m_3)$.  Based on Equation~\eqref{eq:qtwodspectral}, the tensor is uni-axial if and only if $2\lambda_1+\lambda_2=0$ or $2\lambda_2+\lambda_1=0$, hence
\begin{equation}
r^2=(2s+q)(2q+s)\qquad \text{(uni-axiality condition)}.
\label{eq:uni}
\end{equation}
For the present two-dimensional geometry, the uni-axialiaty condition leads to the following explicit form for the $Q$-tensor:
\begin{equation}
\mxQ=\tfrac{1}{2}\myop\left(\begin{array}{ccc}\cos^2\varphi-1/3 & \cos\varphi\sin\varphi&0\\
                                          \cos\varphi\sin\varphi&\sin^2\varphi-1/3&0\\                                            0&0&-1/3\end{array}\right),\qquad \text{(uni-axial case only)}
\label{eq:qunitwod}
\end{equation}
where $\varphi\in[0,2\pi)$ is a parameter.

\section{Fixed-point analysis and evolution in the absence of flow}
\label{sec:fixed}

Before studying the case wherein the $Q$-tensor is subject to passive advection, it is of interest to study the fixed points of the spatially homogeneous version of Equation~\eqref{eq:qtensor}, since this forms a key part of the analysis of the flow-driven case; the same applies to studying the spatially inhomogeneous version of Equation~\eqref{eq:qtensor} in the absence of flow.

Based on Equation~\eqref{eq:qtensor}, the pertinent $Q$-tensor dynamics in the absence of flow are
\begin{subequations}
\begin{eqnarray}
\frac{\partial q}{\partial t}&=&\epsilon^2 \nabla^2 q +g_1(1-\theta) q + 3g_2 (q^2+r^2)-(q^2+r^2+s^2+qs)q-2g_2(q^2+r^2+s^2+qs),\\
\frac{\partial r}{\partial t}&=&\epsilon^2  \nabla^2r +g_1(1-\theta) r + 3g_2 (q+s)r-(q^2+r^2+s^2+qs)r,\\
\frac{\partial s}{\partial t}&=&\epsilon^2 \nabla^2s +g_1(1-\theta) s + 3g_2 (r^2+s^2)-(q^2+r^2+s^2+qs)s-2g_2(q^2+r^2+s^2+qs);
\end{eqnarray}%
\label{eq:noflow2d}%
\end{subequations}%
and the fixed points corresponding to the constant-in-time spatially homogeneous case can be obtained by setting $\partial/\partial t\rightarrow 0$ and $\nabla^2\rightarrow 0$ in the above.  By direct computation, one obtains several families of fixed points in closed form, enumerated as follows (the trivial fixed point is also a possibility, with $r=s=q=0$).
\paragraph*{Case 1.} This involves $r=0$ and $q=as$, where (a) $a=1$, (b) $a=-2$, or (c) $a=-1/2$.  The value of $s$ is given in closed form as follows:
\[
s=-g_2\left( 1-\tfrac{3}{2}\frac{1}{1+a+a^2}\right)\pm \sqrt{g_2^2\left( 1-\tfrac{3}{2}\frac{1}{1+a+a^2}\right)^2+\frac{g_1(1-\theta)}{1+a+a^2}}.
\]
For $a=1$ this reduces to the simple form $s=A_{\pm}-g_2$, where
\begin{equation}
A_{\pm}=\tfrac{1}{2}g_2\pm \sqrt{\tfrac{1}{4}g_2^2+\tfrac{1}{3}g_1(1-\theta)}
\label{eq:adef}
\end{equation}

\paragraph*{Case 2.} With $r\neq 0$, $q$ is arbitrary,  $s=-q+A_{\pm}$, 
and
\[
r^2=g_1(1-\theta)+3g_2A_{\pm}-A_{\pm}^2+A_{\pm}q-q^2.
\]
%
These calculations give rise to a precise statement of the condition for the system to possess a nematic state: from Equation~\eqref{eq:adef} it is required that $A$  be real, hence
\[
\theta\leq \theta_\mathrm{c},\qquad \theta_{\mathrm{c}}=\tfrac{3}{4}\frac{g_2^2}{g_1}+1.
\]
%
%
A very brief classification of these fixed points is given here as follows:
\begin{itemize}
\item Based on Equation~\eqref{eq:uni}, it can then be seen that the fixed point in Case 1(a) is bi-axial while the rest are all uni-axial.
\item Based on Equation~\eqref{eq:qunitwod}, Cases 1(b)--(c) give either $\varphi=0$ or $\varphi=\pm\pi/2$.  For case (2) $\varphi$ takes arbitrary values, since this corresponds to an infinite family of solutions parametrized by $q$.
\item For all fixed points, the scalar order parameter takes the values $6|A_{\pm}|$.
\end{itemize}

Additionally,  the fixed points are classified according to their stability.  The time-dependent homogeneous version of Equation~\eqref{eq:noflow2d} is considered (i.e. with allowing for time-dependence but with $\nabla^2\rightarrow 0$), and a solution $(q,r,s)=(q_0,r_0,s_0)+(\delta q,\delta r,\delta s)$ is imposed.  Here, the subscript zero denotes the fixed-point solution.  Linearizing around the fixed point gives a system of equations 
\begin{equation}
\frac{\mathd}{\mathd t}\left(\begin{array}{c}\delta q\\\delta r\\\delta s\end{array}\right)=J(q_0,r_0,s_0)\left(\begin{array}{c}\delta q\\\delta r\\\delta s\end{array}\right),
\end{equation}
where $J$ is the Jacobian of the system.  Based on the eigenvalues $(\mu_1,\mu_2,\mu_3)= \mathrm{spec}(J)$, the fixed points are classified according to linear stability theory: the fixed point is stable if all eigenvalues of $J$ have negative real part; the fixed point is neutral if $J$ has at least one eigenvalue with zero real part (with the other eigenvalues having negative real part) while finally, the fixed point is said to be unstable if $J$ has at least one eigenvalue with a positive real part.
In this way,  a brief linear stability analysis is carried out for the families of fixed points enumerated in the above list.   Case 1(a) gives a combination of stable and unstable fixed points with $s=A_{-}-g_2$ giving stable states and $s=A_+-g_2$ giving unstable states.  In particular, the eigenvalues of the stability analysis are
\[
\mu_1=g_1(1-\theta)-6g_2 s-9s^2,\qquad \mu_{2,3}=g_1(1-\theta) +g_2s-3s^2,\qquad s=A_{\pm}-g_2,
\]
with
\begin{equation}
\mathrm{min}(\mu_1,\mu_2,\mu_3)=\mu_1|_{s=A_-}= -6A_{+} \sqrt{\tfrac{1}{4}g_2^2+\tfrac{1}{3}g_1(1-\theta)}\leq 0.
\label{eq:phidash0}
\end{equation}
Correspondingly, the order parameter can be obtained as $\myop=6|s+q|=12|A_{\pm}-g_2|$.  The negative branch gives the larger of the two values of $\myop$ and corresponds to the stable state.
%
%
Similar analysis for the cases 1(b)--(c) and case 2 indicates that the remaining fixed points are neutral or unstable, with $\myop=6|A_-|$ neutral and $\myop=6|A_+|$ unstable.  

\paragraph*{Case 3.}  This case is relevant both as a fixed point and an evolutionary state and arises when the scalar order parameter $\myop$ changes in time but the director is fixed.  This case corresponds to a uni-axial state.  Thus,
\[
q=\tfrac{1}{2}\myop(\cos^2\varphi-\tfrac{1}{3}),\qquad s=\tfrac{1}{2}\myop(\sin^2\varphi-\tfrac{1}{3}),
\]
hence $s/q=(3\cos^2\varphi-1)/(3\sin^2\varphi-1)\equiv \lambda$.   The quantity $\lambda$ is treated as a constant, which gives
\begin{multline}
\frac{ds}{dt}=g_1(1-\theta)s-3g_2(s^2+r^2)+\mathbb{T}s+2g_2\mathbb{T}\\
             =g_1(1-\theta)\lambda q-3g_2(\lambda^2q^2+r^2)+\mathbb{T}\lambda q+2g_2\mathbb{T}\\
						=\lambda \frac{dq}{dt}
						=g_1(1-\theta)\lambda q-3g_2(\lambda q^2+\lambda r^2)+\mathbb{T}\lambda q+2g_2\lambda\mathbb{T},
\end{multline}
where $\mathbb{T}=q^2+2r^2+s^2+(s+q)^2$.
Comparing these equations gives
\[
q^2\left(2\lambda^3+3\lambda^2-3\lambda-2\right)=\left(\lambda-1\right)r^2,
\]
For $\lambda\neq 1$ this gives
\[
r^2=\crazylambda^2 q^2,\qquad \crazylambda^2=2\lambda^2+5\lambda +2.
\]
which is a manifestly uni-axial state because $r^2-(2s+q)(2q+s)=q^2\crazylambda^2-q^2(2\lambda^2+5\lambda + 2)=0$.
These relations are now substituted back into the dynamic equation for $dq/dt$:
\begin{equation}
\frac{dq}{dt}=g_1(1-\theta)q+3g_2(1+\lambda)q^2-3(1+\lambda)^2q^3\equiv \Phi(q).
\end{equation}
This gives further uni-axial fixed points $q_0=A_{\pm}(1+\lambda)^{-1}$,
%
%
the stability of which can be determined by the growth rate $\mu=\Phi'(q_0)$, where
\begin{equation}
\Phi'(q_0)=\mp 6A_{\pm} \sqrt{\tfrac{1}{4}g_2^2+\tfrac{1}{3}g_1(1-\theta)}\leq 0.
\label{eq:phidash}
\end{equation}
with $\min(\Phi'(q_0))$ corresponding to the growth rate of the most-stable biaxial mode (Equation~\eqref{eq:phidash0}).  
%

It can be noted that the present uni-axial fixed point (Case 3) gives $q+s=A_{\pm}$.  Thus, the fixed points in Case 3 coincide the uni-axial fixed points from Cases 1--2 (specifically, $\lambda=-1/2,-2$ reproduces Cases 1(b)--(c) and $\lambda\in\mathbb{R}/\{-2,-1/2\}$ reproduces Case 2).
This begs the question: if Cases 1--2 give unstable fixed points, then why is Case 3 a stable fixed point?  The explanation 
is that the trajectory in Case 3  is constrained to be one dimensional and to lie along  stable eigendirections of the uni-axial fixed points in Case 1--2, and is therefore stable.  Finally,  Cases 1(a) and 3 are both linearly stable and have the same  negative growth rate, meaning both are energetically equally favourable.

It is of further interest to examine the influence of the diffusion on the evolution of the problem, before studying the effects of flow.  Therefore, the full system of equations~\eqref{eq:noflow2d} is solved using a pseudospectral numerical method~\cite{Zhu_numerics}  with corresponding periodic boundary conditions in both directions.  Specifically,  the numerical method involves computing all spatial derivatives in Fourier space and all products (e.g. $q^2$, $s^2$, $r^2$, $\mathbb{T}$, etc.) that make up the nonlinear terms in real space.  Concerning the temporal discretization, the  diffusion term is treated using the Crank--Nicholson method for maximum numerical stability while all remaining terms in the equations are treated using a third-order Adams--Bashforth scheme.

The simulations are performed on a $256\times 256$ grid
and the chosen timestep is $\Delta t=10^{-2}$.  Both the times step and the size of the spatial grid are chosen so as to ensure that the numerical simulations have converged. 
The constants $g_1$ and $g_2$ are fixed based on the liquid crystal 5CB and $\theta$ is viewed as a parameter.   Hence, using standard values for 5CB~\cite{sanati2003landau}, it follows that
\begin{equation}
g_1=\tfrac{3}{2}\frac{a T_*}{\gamma_F}\approx 0.8529,\qquad
g_2=\frac{\beta_F}{8\gamma_F}\approx 0.0684.
\label{eq:g1g2}
\end{equation}
It is revealing to study two sets of distinct initial conditions.  In the first case, the initial data are chosen to be uni-axial (i.e. based on Equation~\eqref{eq:qunitwod}) with a phase $\varphi\in[0,2\pi)$ and a value of $\myop\in[0,1]$ both chosen randomly from separate independent identical uniform distributions at each point in the domain.  In the second case, the initial data are chosen to be bi-axial (i.e. based on Equation~\eqref{eq:qtwodspectral}, with $(n_x,n_y)=(\cos\varphi,\sin\varphi)$ and again with $\varphi\in[0,2\pi)$ chosen randomly from independent identical uniform distributions at each point in the domain.  Also, $\lambda_1=\rho\cos\vartheta$ and $\lambda_2=\rho\sin\vartheta$, where $\rho$ is  chosen such that $6\mathrm{tr}(\mxQ^2)\leq 1$.  This is achieved by letting $\rho=\hat{\rho}\{6[1+(\cos\vartheta+\sin\vartheta)^2]\}^{-1/2}$,
where $0\leq \vartheta<2\pi$ and $0\leq \hat{\rho}\leq 1$ are further variables chosen from the uniform random distribution at each point in the domain. 

\begin{figure}
	\centering
		\subfigure[$\,\,t=100$]{\includegraphics[width=0.32\textwidth]{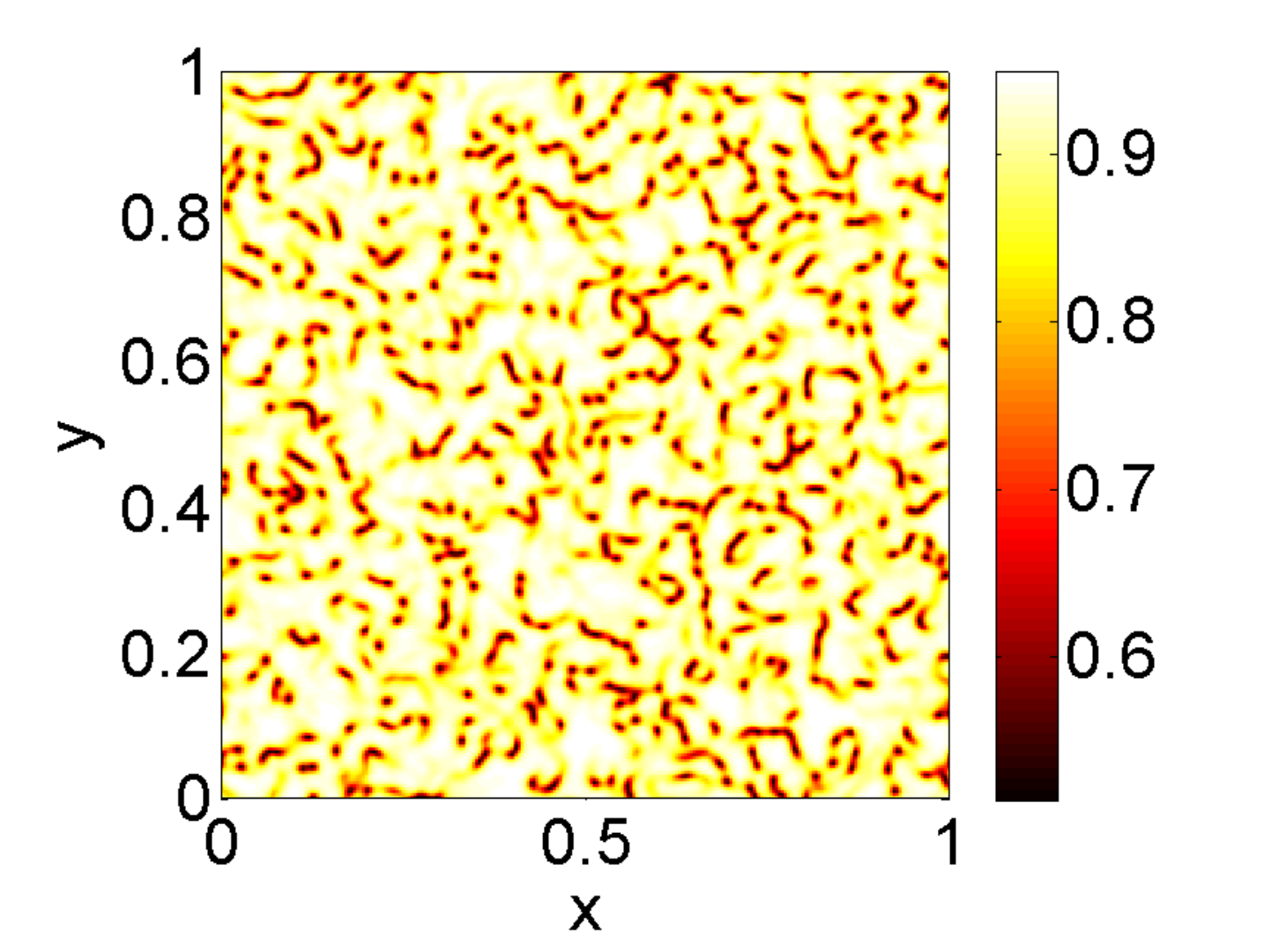}}
		\subfigure[$\,\,t=1000$]{\includegraphics[width=0.32\textwidth]{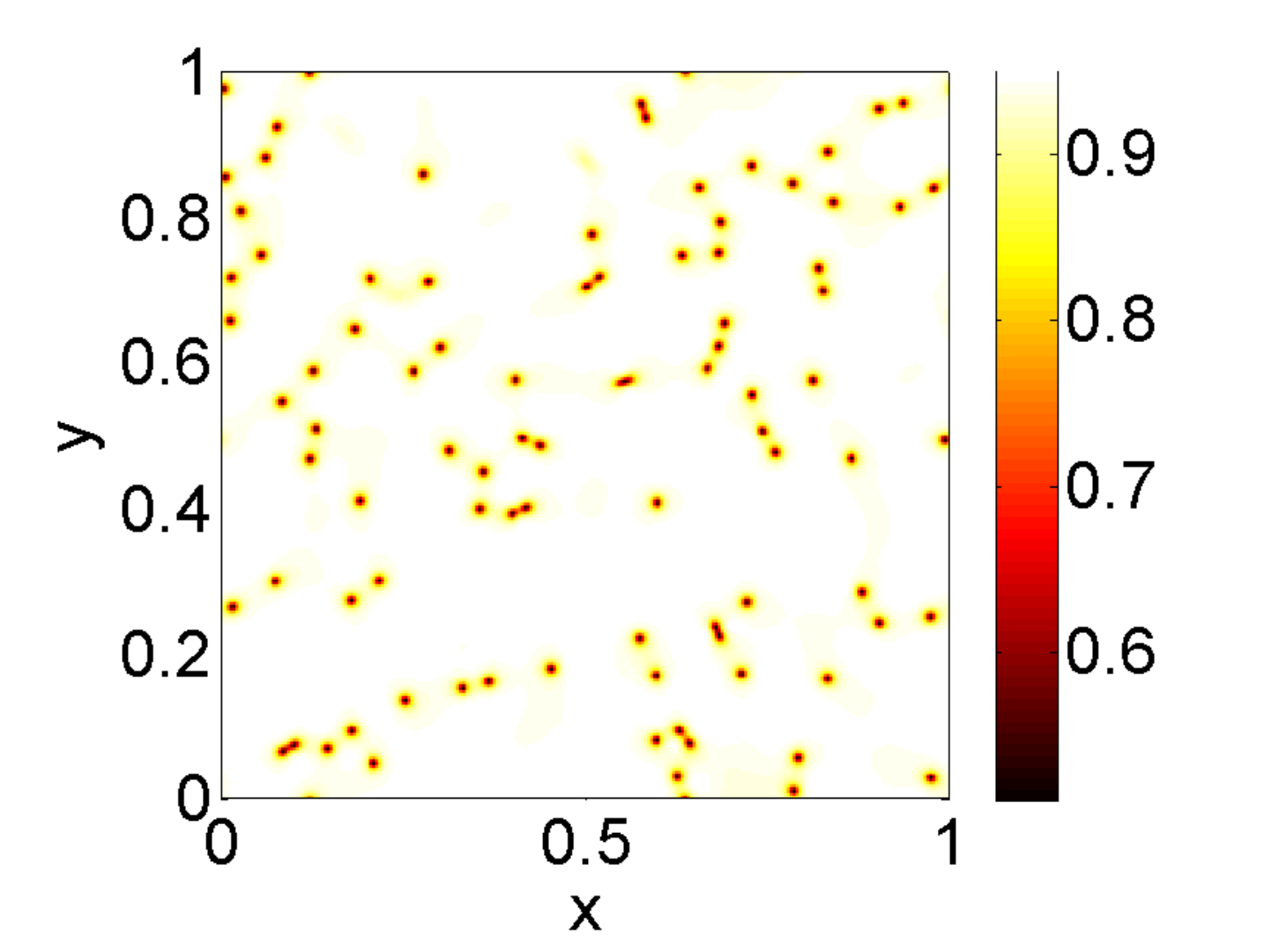}}
		\subfigure[$\,\,t=5000$]{\includegraphics[width=0.32\textwidth]{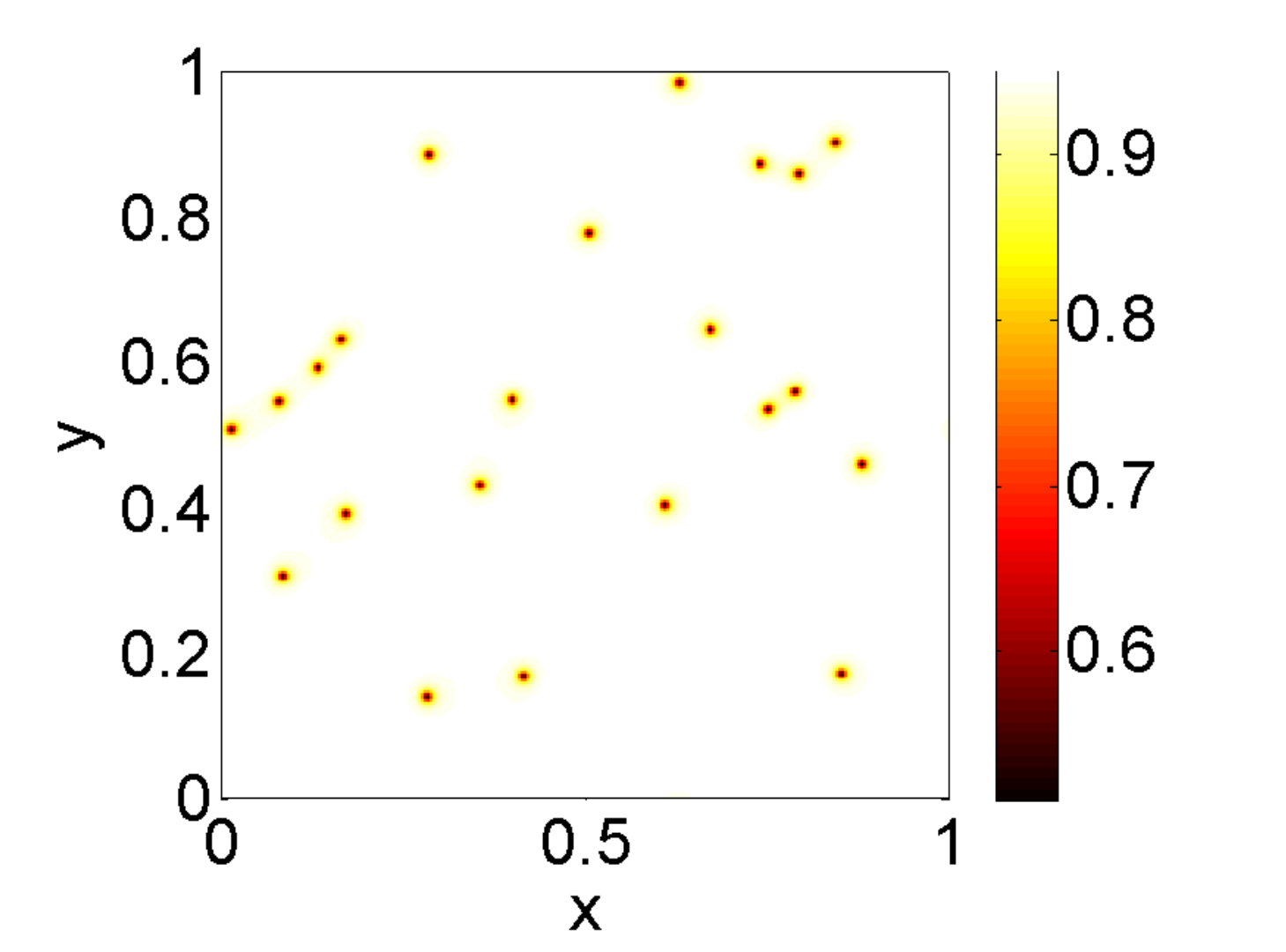}}
		\caption{Snapshots of scalar order parameter at various times.}
	\label{fig:op1}
\end{figure}

\begin{figure}
	\centering
	  \subfigure[]{\includegraphics[width=0.48\textwidth]{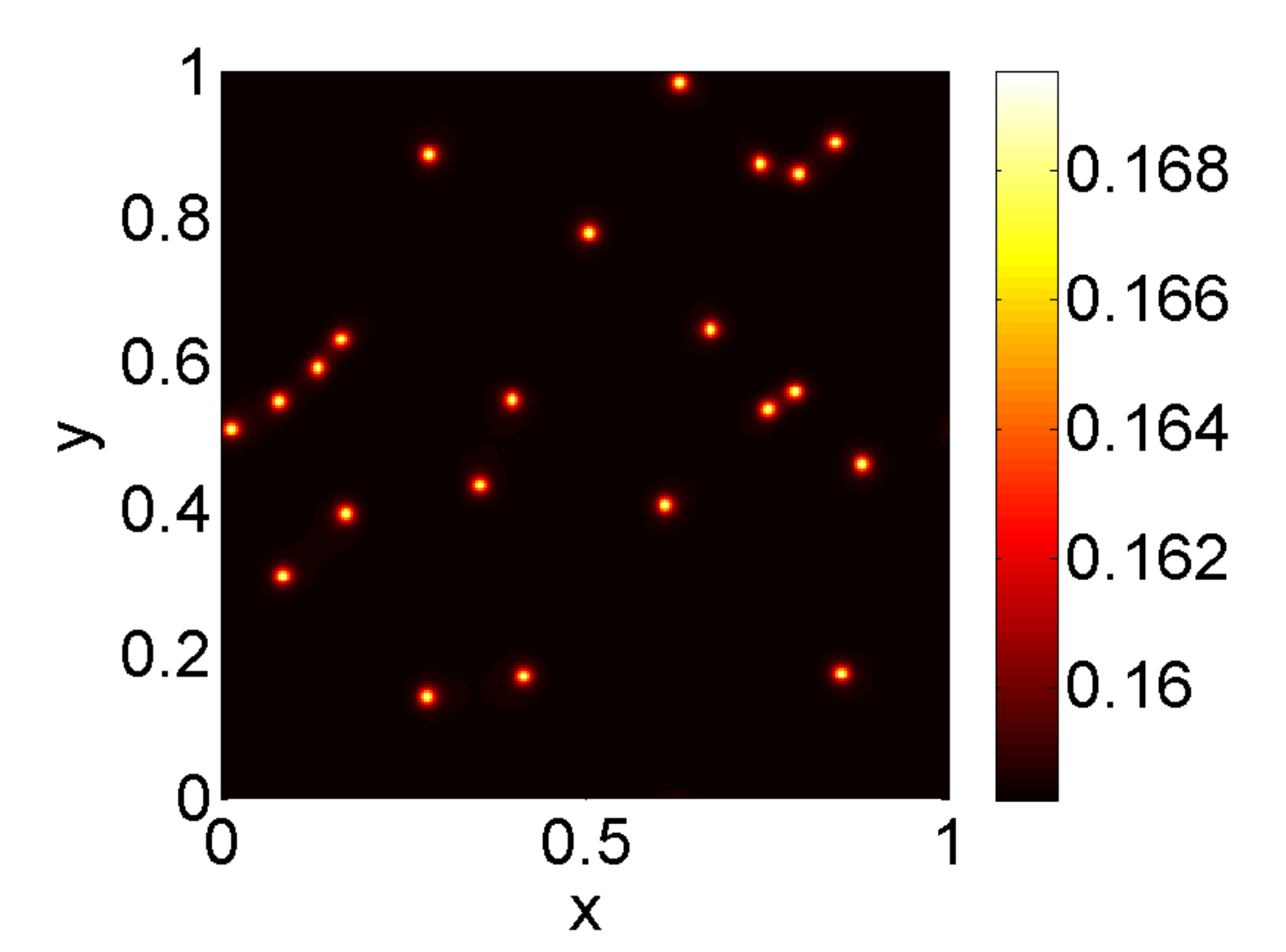}}\\
		\subfigure[]{\includegraphics[width=0.48\textwidth]{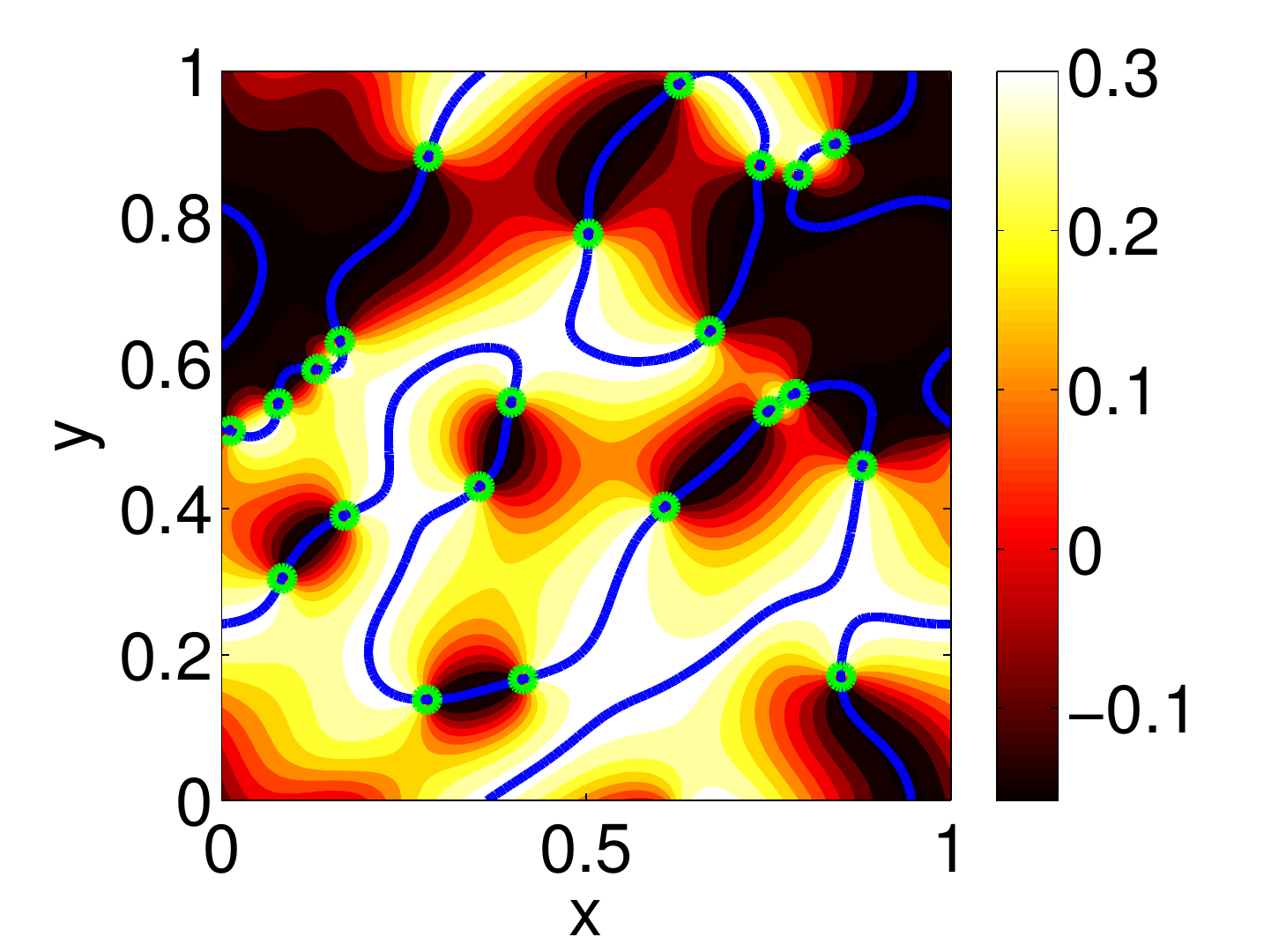}}
		\subfigure[]{\includegraphics[width=0.48\textwidth]{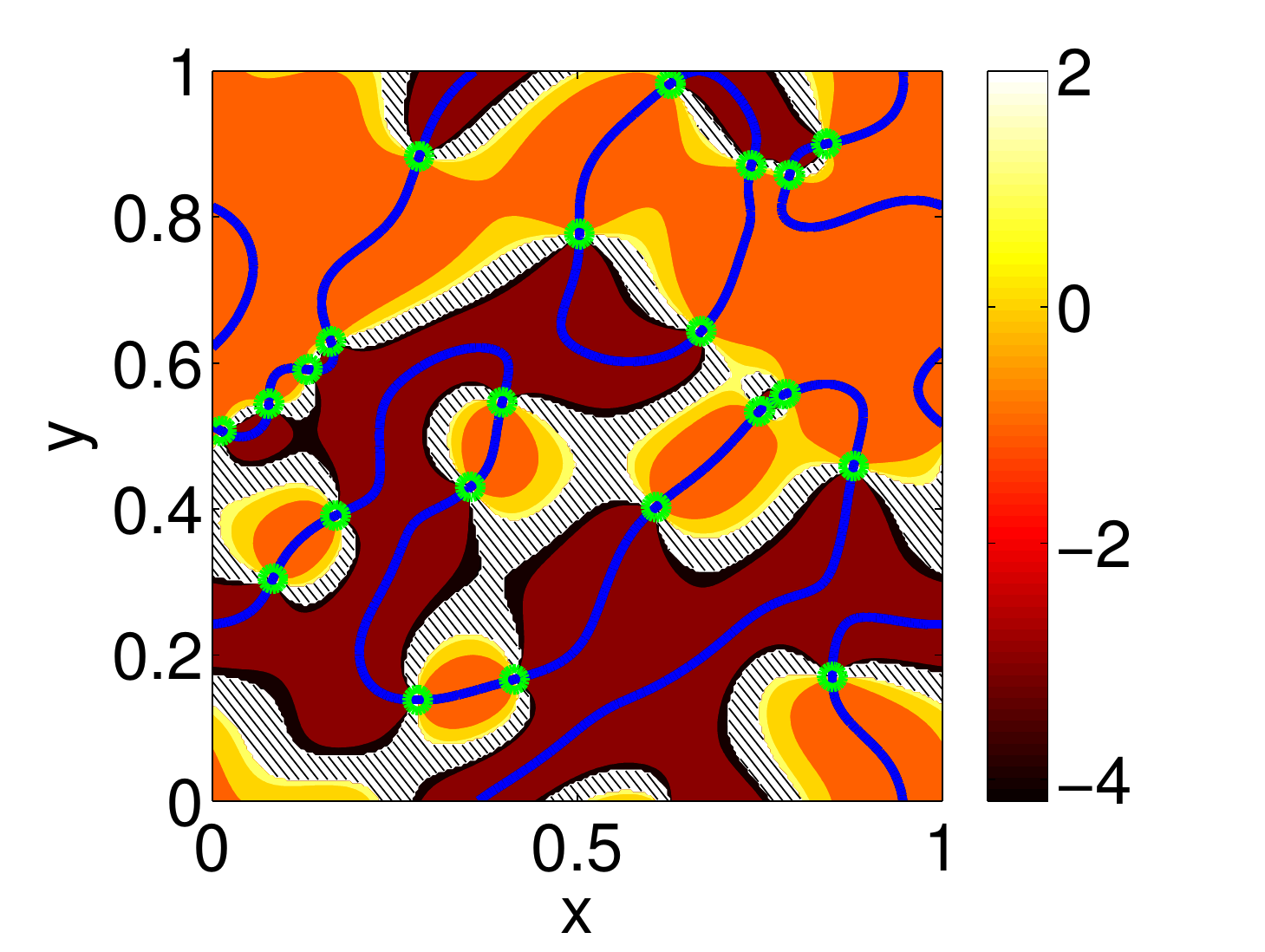}}
		\caption{Fuller characterization of the system at $t=5,000$.
		(a) Plot of $s+q$ at $t=5000$ revealing a bimodal distribution.  The value $s+q\approx A_+$ corresponds to a type-3 stable fixed point while the $s+q\approx 2(A_+-g_2)$ corresponds to a type-1(a) unstable bi-axial fixed point; (b) Plot of $s$ at the same time.  The small circular contours correspond $s+q\approx 2(A_+-g_2)$ (i.e. bi-axial regions) and the larger contours correspond to $r=0$; (c) Plot of $\lambda=s/q$ for $|q|>0.05$ (the hatched regions correspond to $|q|<0.05$.
		}
	\label{fig:study_op1}
\end{figure}

Sample simulation results at various times (uni-axial initial data) are shown in Figure~\ref{fig:op1}.
The scalar order parameter $\myop=\sqrt{6\mathrm{tr}(\mxQ^2)}$ tends to a bimodal distribution with $\myop=6|A_{\pm}|$.  To understand this structure in more depth, $s+q$ is plotted in Figure~\ref{fig:study_op1}(a) at $t=5000$.  Again, $s+q$ reveals a bimodal distribution with $s+q\approx A_+$ throughout the domain and $s+q\approx 2(A_+-g_2)$ in small islands.  By referring to the fixed-point analysis, it can be seen that the small islands correspond to regions where the system locally tends to an unstable bi-axial fixed point while in the rest of the domain the system tends locally to stable fixed points of types 1(a,b) and type 3.  Reasons for the (local) selection of an unstable fixed point are given in what follows.

To understand the evolution of the system more fully, the components of the $Q$-tensor are selected for in-depth study at a particular time in Figure~\ref{fig:study_op1}.  In Figure~\ref{fig:study_op1}(a) a contour plot of $s+q$ is shown again revealing a bimodal distribution with $s+q\approx 2(A_+-g_2)$ in the small island regions and $s+q\approx A_+$ elsewhere.  These again correspond (respectively) to unstable bi-axial fixed points and neutral or fixed points of types 1(a,b), and 3, providing further evidence of this characterization of the late-time evolution in terms of (local) relaxation to fixed points.
In Figure~\ref{fig:study_op1}(b) a plot of $s$ by itself is shown, overlaid with the  zero contour of the scalar field $r$ and a further contour with $s+q\approx 2(A_+-g_2)$.
Along level curves with $r=0$, the system is seen to collapse locally into a type-1(a,b) fixed neutral point with $r=0$.   More precisely, such curves are heteroclinic orbits, where the system transitions between one kind of neutral type-1(a) uniaxial fixed point and another; the crossover between the fixed points corresponds to the unstable biaxial state.  This is consistent with analysis of the steady-state heteroclinic orbits in the literature~\cite{schopohl1987}.  Away from lines with $r=0$, the system collapses into a type-3 fixed point, as can be seen by plotting $\lambda=s/q$ for $|q|>0.05$ (Figure~\ref{fig:study_op1}(c)).  Here $\lambda$ exhibits a high degree of spatial ordering consistent with the notion that the components of the $Q$-tensor form coherent domains (as in Figure~\ref{fig:study_op1}(b)).  Here,  $\lambda\in (-\infty,-2]\cup [-1/2,\infty)$, with $\lambda=-2,-1/2$ corresponding to the coincidence of the type-1(b,c) fixed points and type 3 fixed points at the level cures $r=0$.

\begin{figure}
	\centering
		\subfigure[]{\includegraphics[width=0.48\textwidth]{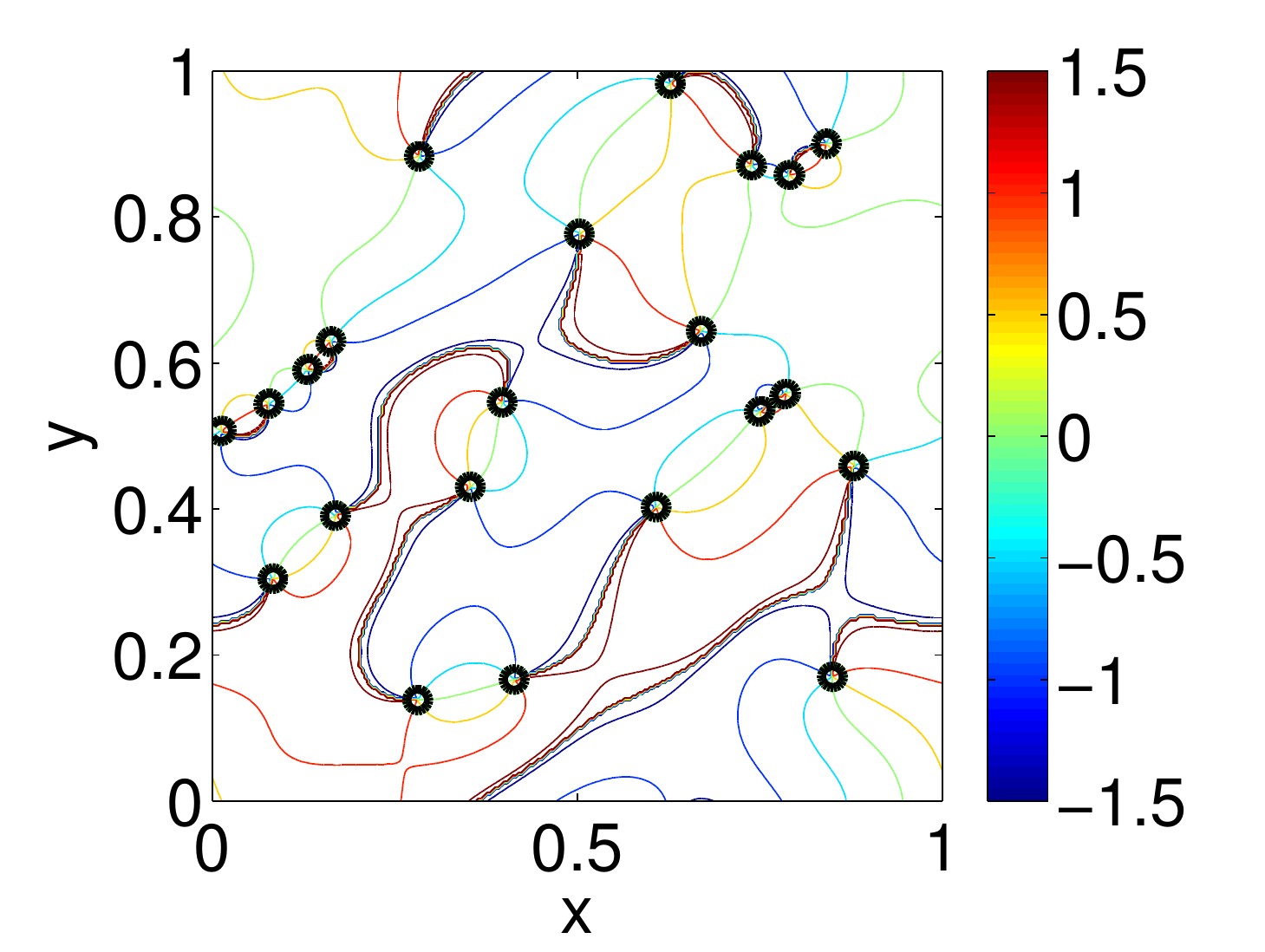}}
		\subfigure[]{\includegraphics[width=0.48\textwidth]{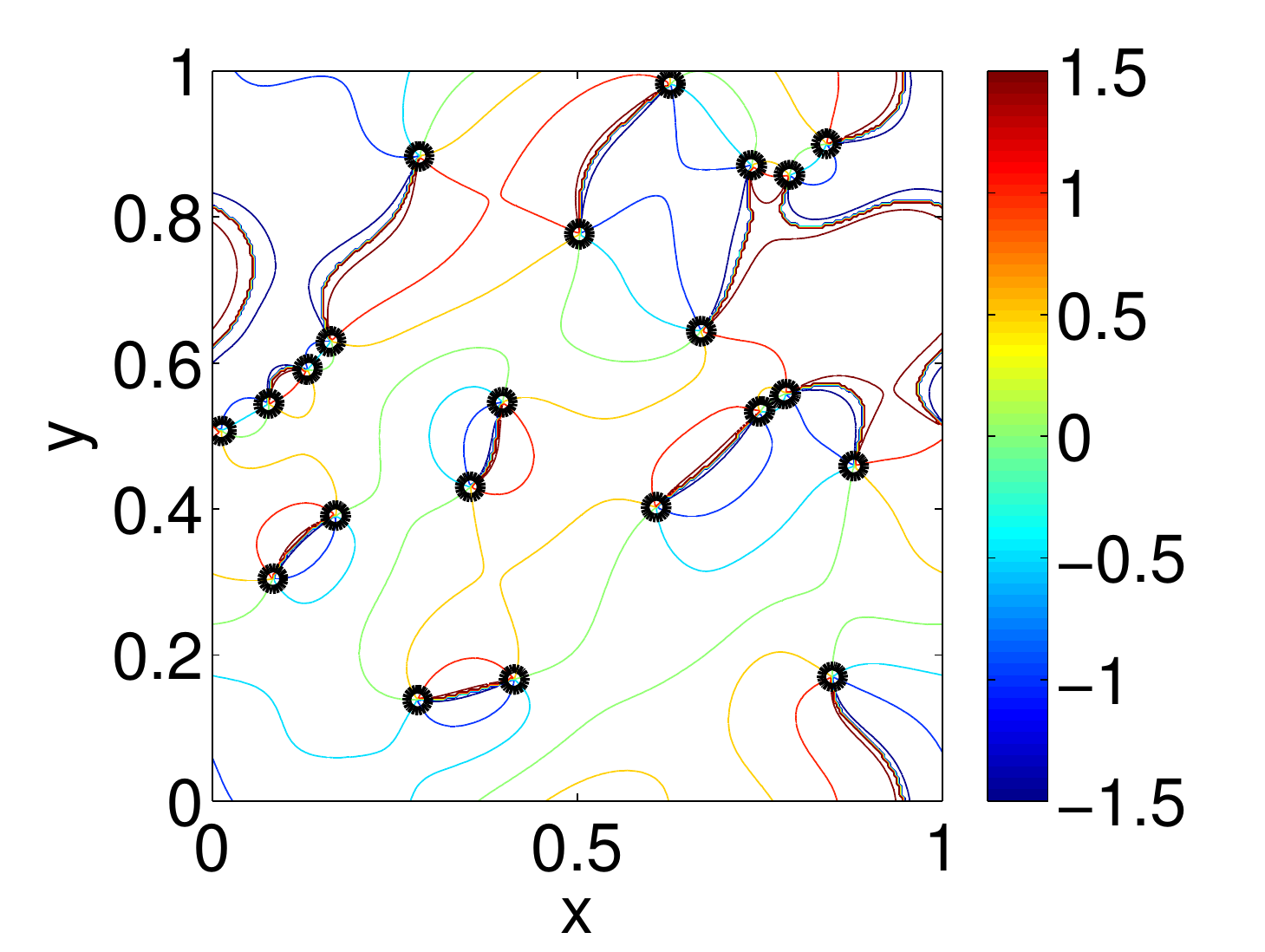}}
		\caption{Characterization of the topology of the director at $t=5,000$:
		 Contours of (a) $\varphi_1$ and (b) $\varphi_2$ showing the jumps (jagged contours) where the angle jumps from $\pm \pi/2$ to $\mp\pi/2$.  
		}
	\label{fig:chi}
\end{figure}

To understand the presence of the type-1(a) unstable bi-axial fixed points in the system at late time, consideration is given to the topology of the eigendirections of the $Q$-tensor.  Specifically, the eigendirections $\bm{n}^{(1,2)}=(\cos\varphi_{1,2},\sin\varphi_{1,2},0)$ are examined and the phases $\varphi_1$ and $\varphi_2$ are plotted Figure~\ref{fig:chi}.
The plot reveals that  $\varphi_{1,2}$ experience jumps of magnitude $\pi$  along curves starting from one of the islands and finishing at another.  Thus, the phase jumps and the corresponding islands of type-1(a) fixed points are properly seen together as line defects and defect cores respectively with a topological charge $\pm 1/2$. 
(Specifically, a defect core can be assigned a topological charge $1/2$ if $\varphi_1$ winds from $-\pi/2$ to $\pi/2$ in an anti-clockwise fashion and $-1/2$ if the winding is an a clockwise fashion.  The value of $1/2$ refers to the fact that precisely two windings are required for $\varphi_1$ to return to its original value, modulo $2\pi$.)
Additionally, the location of the jumps coincides exactly with the $r=0$ contour.  Thus, a state with $r=0$ and $Q$ uni-axial corresponds to a jump in $\varphi_{1,2}$ and hence, a director that takes two opposite values while $r=0$ with $Q$ bi-axial corresponds to a point where the director cannot be assigned any value(s), i.e. a defect core.  These findings are consistent with the literature~\cite{bhattacharjee2008,schopohl1987}.
Since the type-1(a) fixed point with $\myop=6|A_-|$ is unstable, it can be asked why it appears at late time.  The reason is that the system globally conserves the topological charge of the defects.  Thus, (neglecting integer-charged defects, as such defects tend to be absent at late time in planer systems~\cite{bhattacharjee2008,bhattacharjeePhd}) the only way for the system to remove defects and hence move to a more stable state is by the merger of two defect cores of opposite charge.  In this way, two `pins' supporting a domain wall annihilate and the domain expands in size.  

\begin{figure}
	\centering
		\subfigure[$\,\,t=100$]{\includegraphics[width=0.24\textwidth]{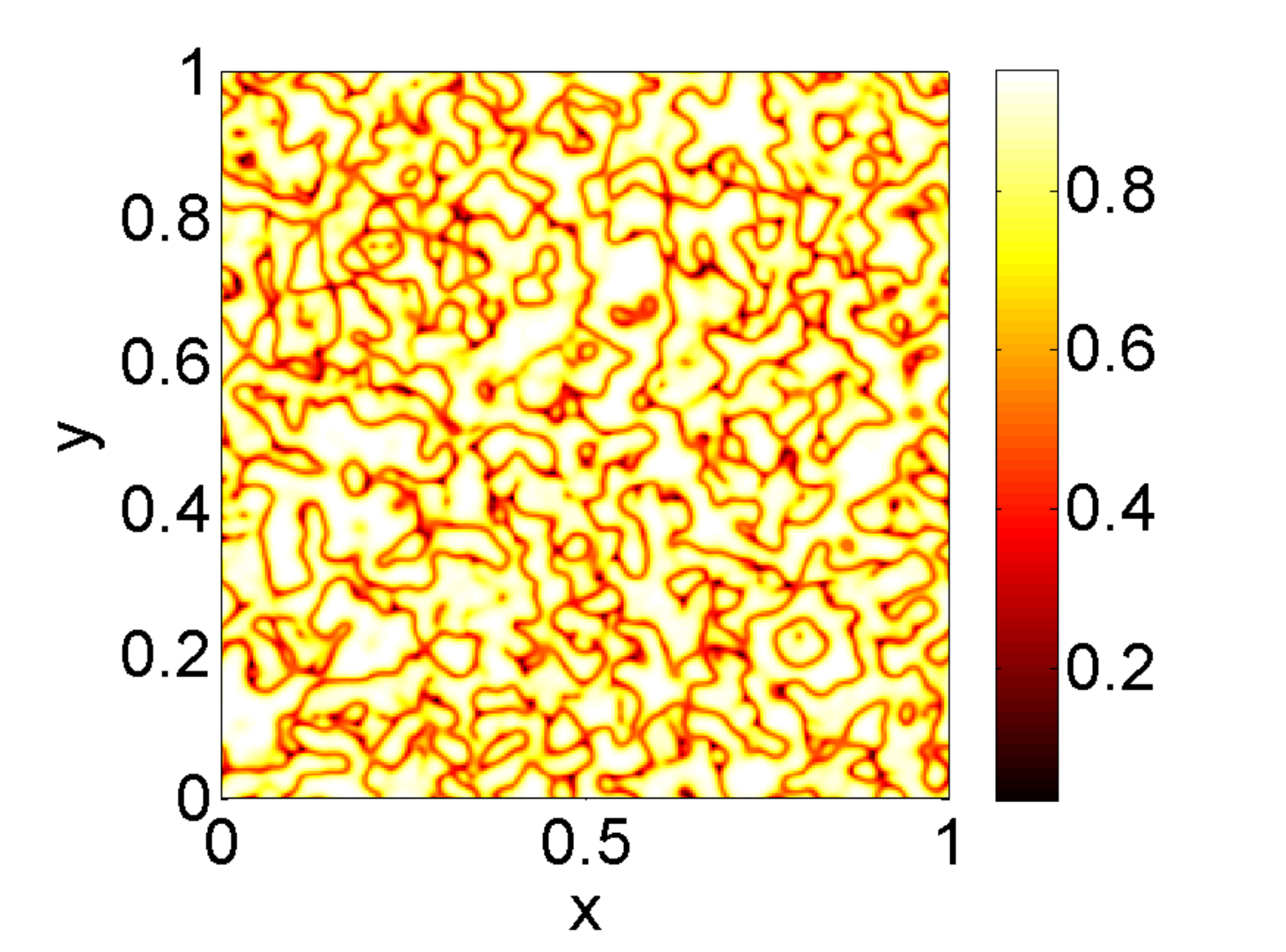}}
		\subfigure[$\,\,t=1000$]{\includegraphics[width=0.24\textwidth]{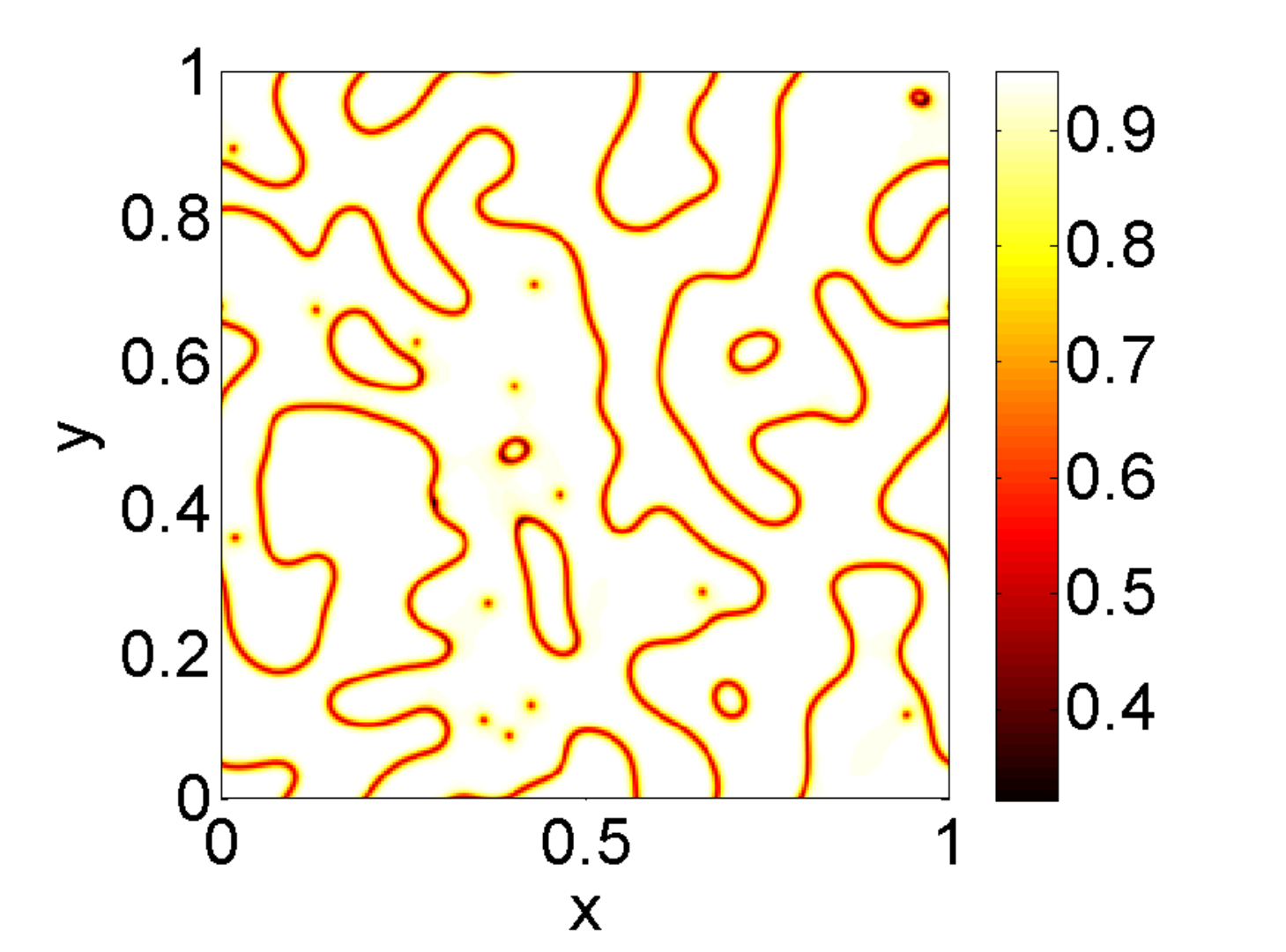}}
		\subfigure[$\,\,t=2000$]{\includegraphics[width=0.24\textwidth]{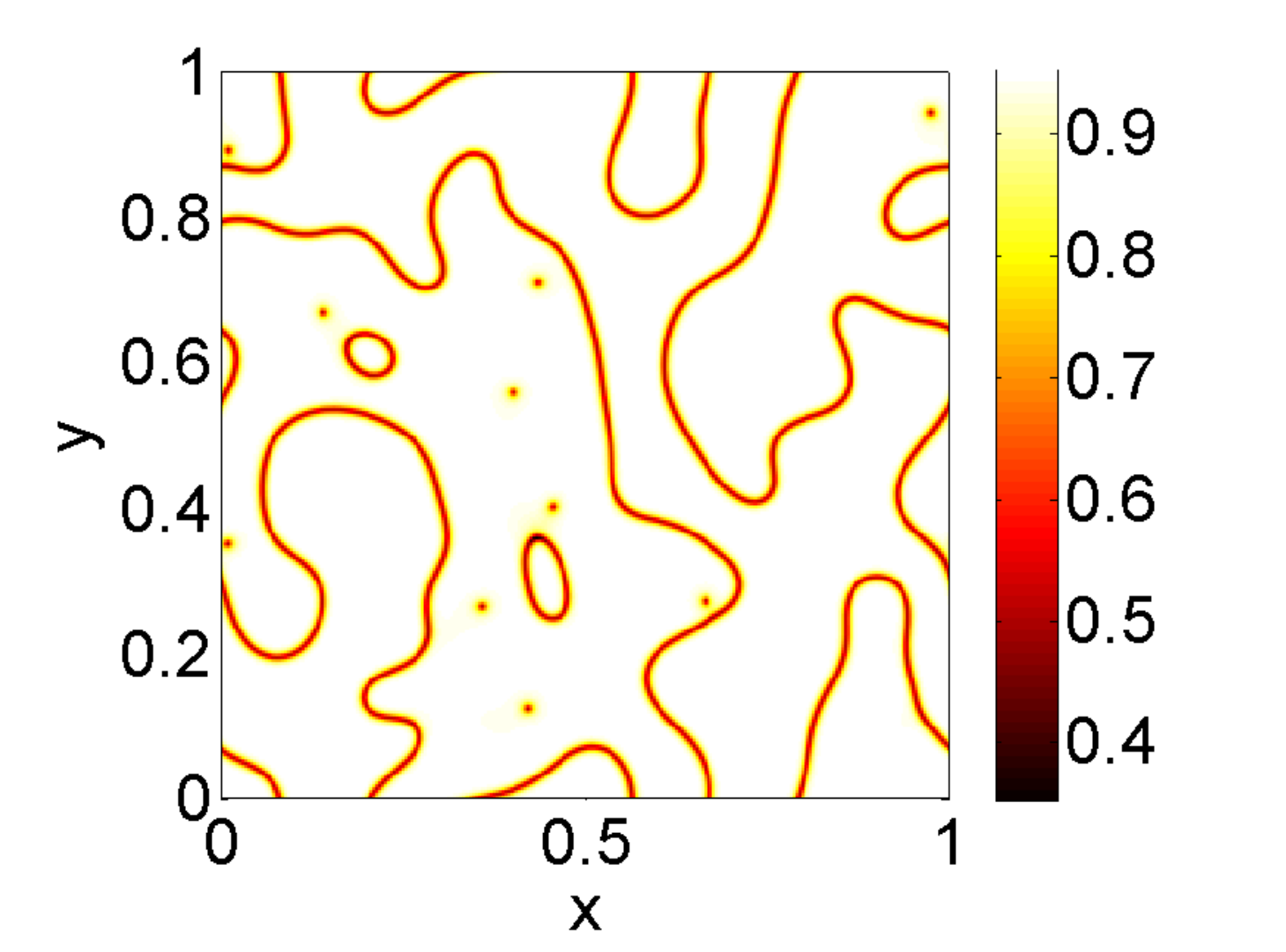}}
		\subfigure[$\,\,t=5000$]{\includegraphics[width=0.24\textwidth]{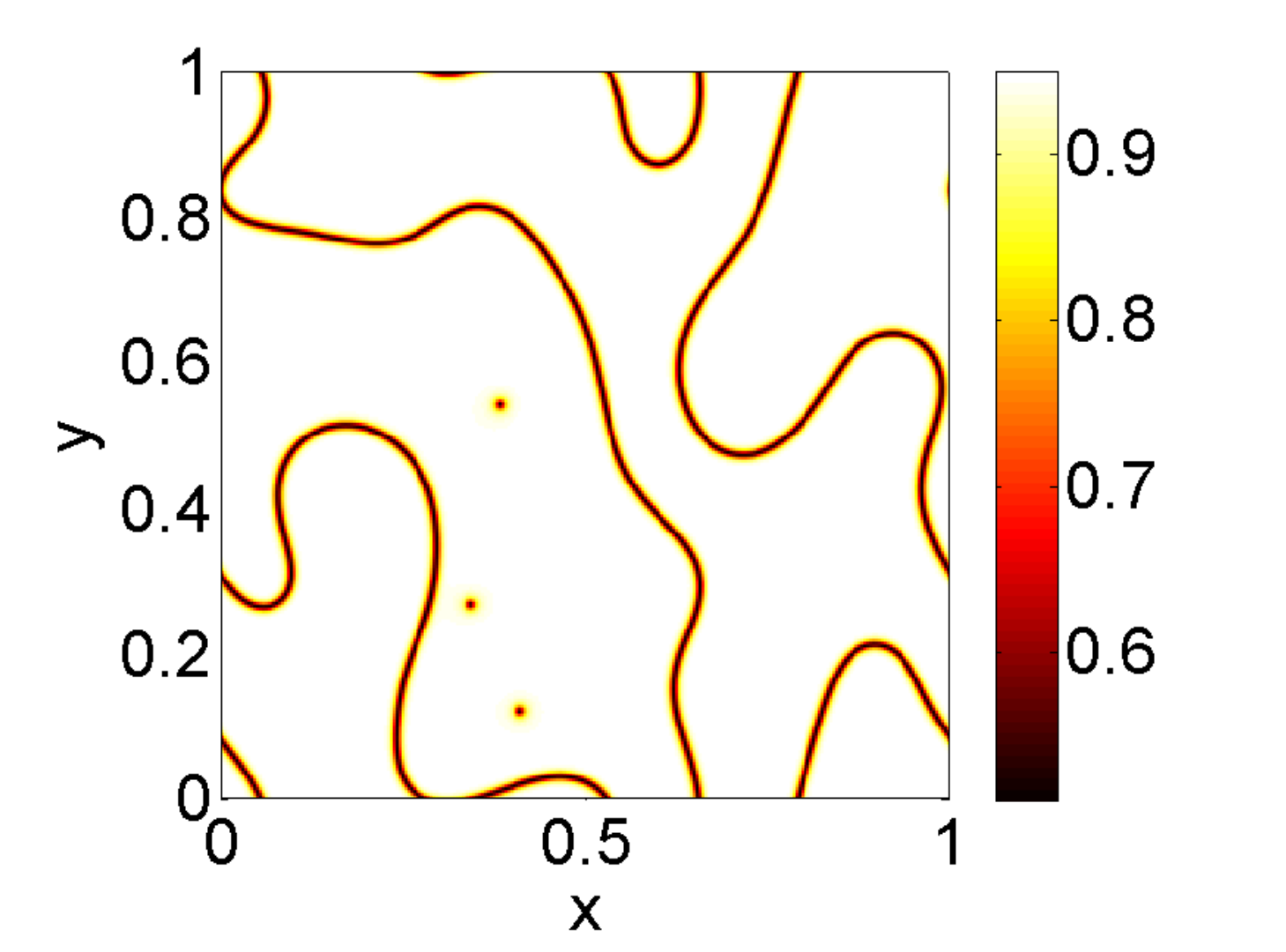}}
		%
		%
				\subfigure[$\,\,t=100$]{\includegraphics[width=0.24\textwidth]{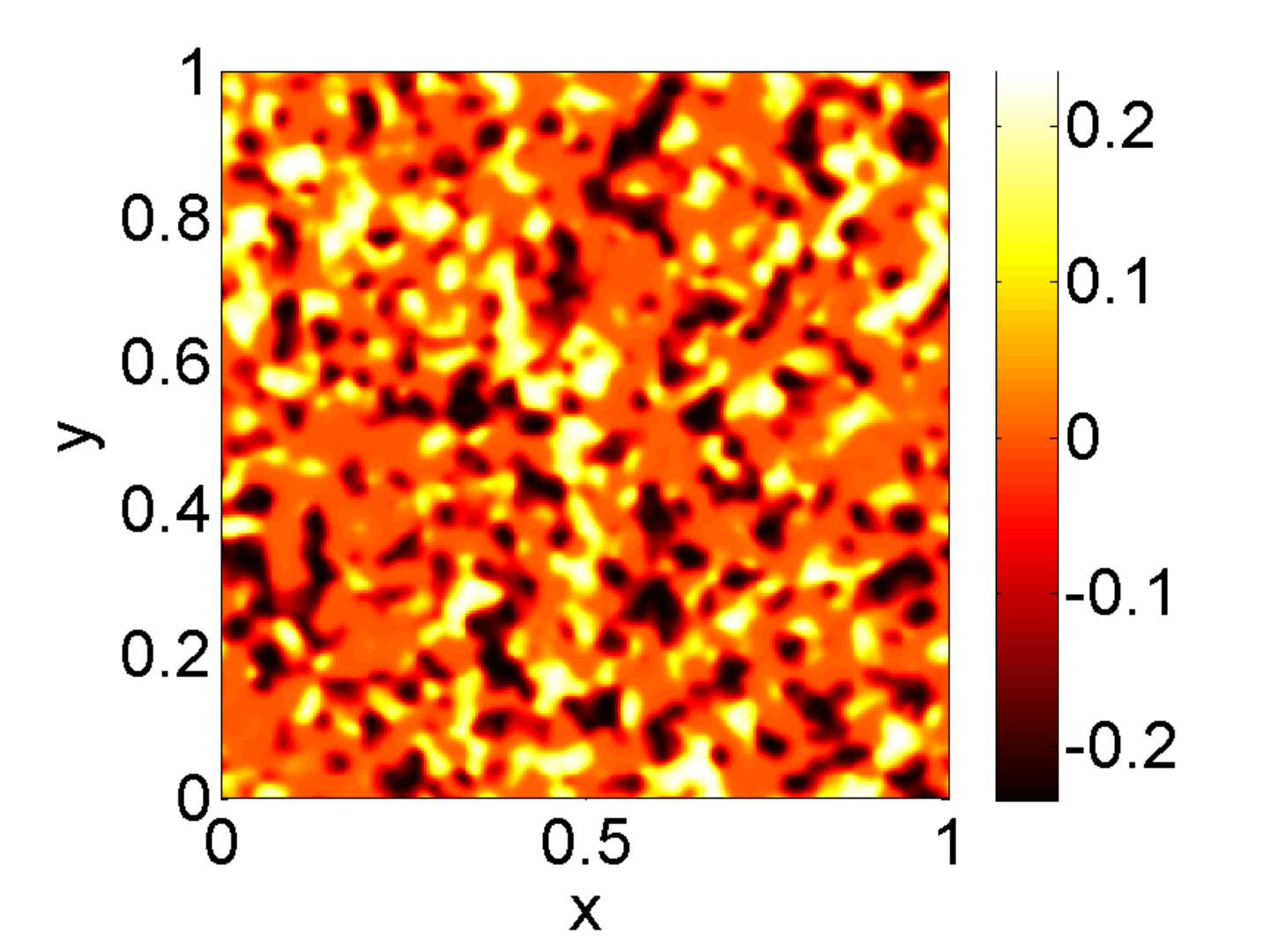}}
		\subfigure[$\,\,t=1000$]{\includegraphics[width=0.24\textwidth]{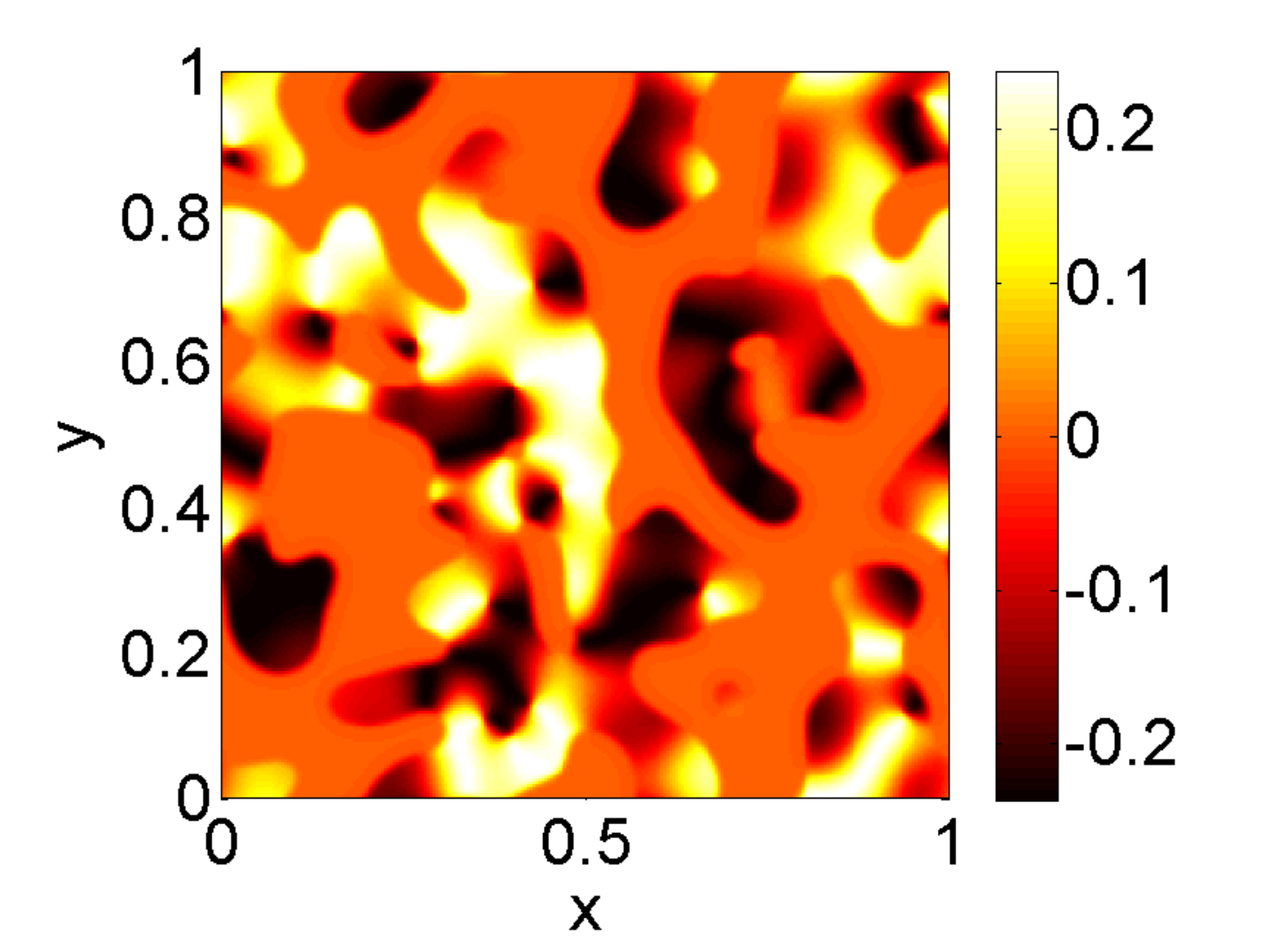}}
		\subfigure[$\,\,t=2000$]{\includegraphics[width=0.24\textwidth]{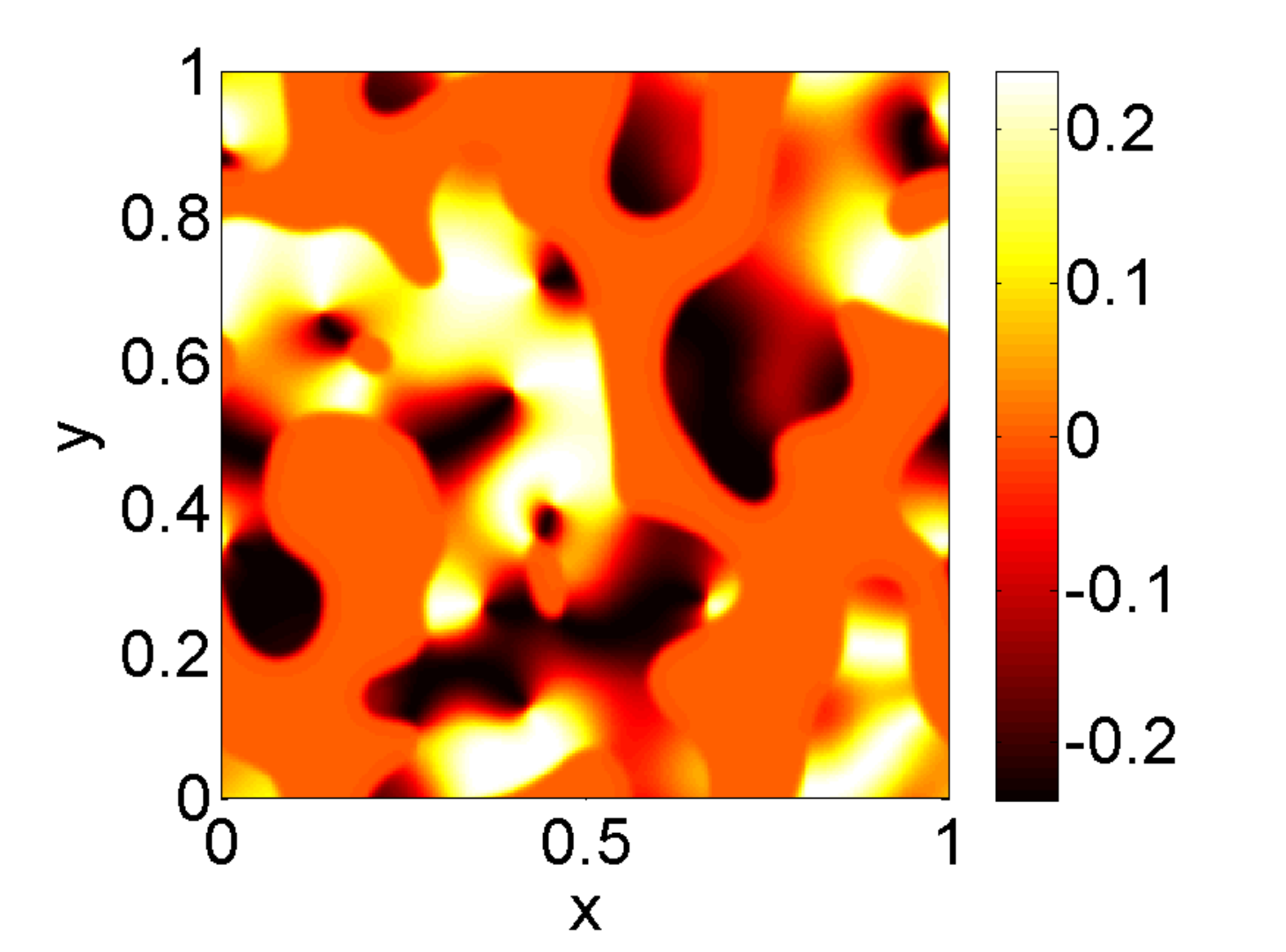}}
		\subfigure[$\,\,t=5000$]{\includegraphics[width=0.24\textwidth]{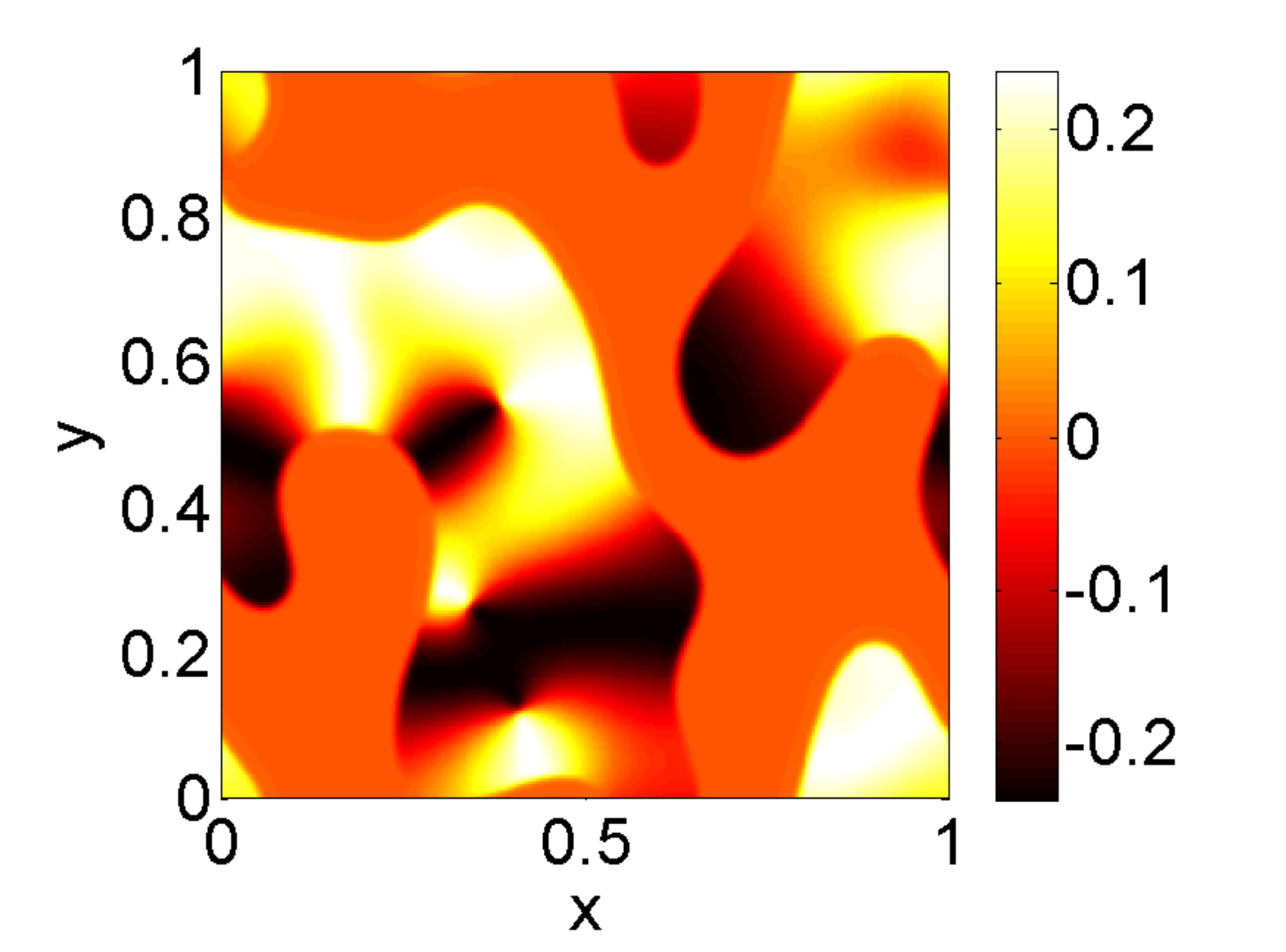}}
		\caption{Top: snapshots of scalar order parameter at various times for the bi-axial initial data.  
		Bottom: corresponding snapshots for $r$.  The domains with $r=0$ correspond to the bi-axial state.}
	\label{fig:op2}
\end{figure}

\begin{figure}
	\centering
		\includegraphics[width=0.5\textwidth]{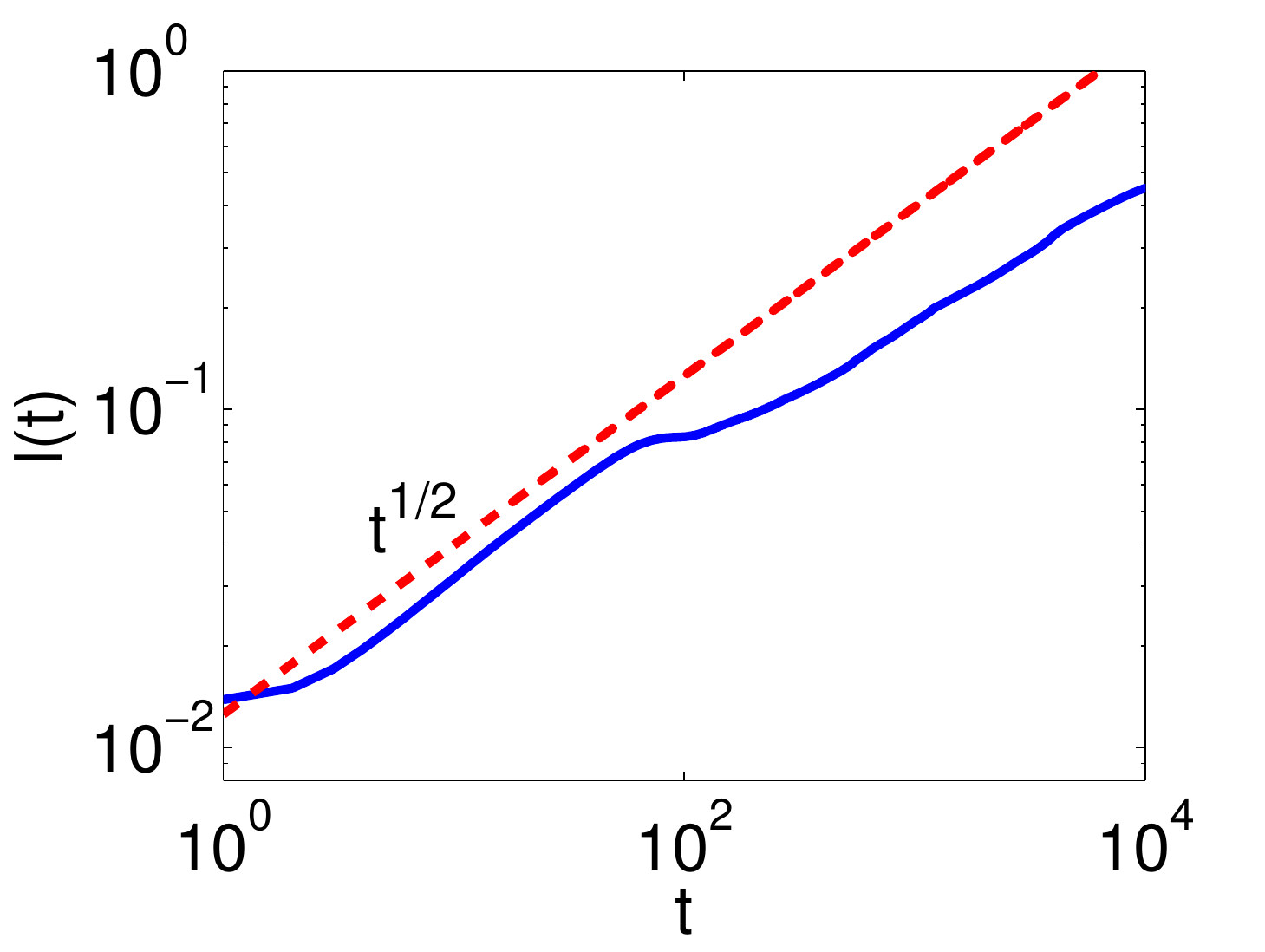}
		\caption{Growth of the domain scale $L(t)$ for the case of biaxial initial data}
	\label{fig:production_Lt}
\end{figure}

Sample simulation results  at various times (bi-axial initial data) are shown in Figure~\ref{fig:op2}.  The system forms domains of two distinct kinds -- regions that approximate a type 1(a) stable biaxial fixed point and further regions that are a patchwork of uniaxial states wherein the domain structure (including the defect structure) closely resembles that already studied in the context of uni-axial initial data.
It is of interest to study the growth of the domains, particularly for the case of biaxial initial data.   Therefore, a measure $L(t)$ of the domain scales is introduced:
\begin{subequations}
\begin{equation}
L(t)=2\pi/k_1,\qquad k_1= \frac{\int|\widehat{C}_{\bm{k}}|^2\mathd^2k}{\int|\bm{k}|^{-1}|\widehat{C}_{\bm{k}}|^2\mathd^2k},
\end{equation}
where $\widehat{C}_{\bm{k}}$ is the Fourier transform of the mean-zero part of the scalar order parameter, i.e.
\begin{equation}
\langle \myop\rangle=\frac{1}{|\Omega|}\int_{\Omega} \myop(\bm{x})\mathd ^2x ,\qquad
\widehat{C}_{\bm{k}}=\int_{\Omega}\mathe^{-\imag \bm{x}\cdot\bm{k}}\left(\myop-\langle \myop\rangle\right)\mathd^2 x.
\end{equation}%
\label{eq:Lt_def}
\end{subequations}%
This agrees with the standard approach of measuring $k_1$ in the literature, i.e. using the spherically-averaged power spectrum of the relevant order parameter as a distribution function and computing the expected value of $|\bm{k}|$ accordingly~\cite{naraigh2007bubbles}.
The results of measuring $L(t)$ according to the above procedure are shown in Figure~\ref{fig:production_Lt}
The lengthscale $L(t)$ demonstrates an initial rapid growth phase $L(t)\sim t^{1/2}$ consistent with diffusive scaling~\cite{bhattacharjee2008} and with existing studies~\cite{bhattacharjee2008,bhattacharjeePhd}.  Thereafter, the growth slows down.  This too is consistent with existing studies~\cite{bhattacharjeePhd} on systems where a homogeneous biaxial state is energetically favourable compared to the corresponding uni-axial state (for the present system, both such states are equally energetically favourable / stable).  However, finite-size effects may also play a role here, as the domains extend across the periodic box at late times in the present study (as in Figure~\ref{fig:op2}(d,h).  Nevertheless, the clear initial trend consistent with diffusive scaling is unambiguous, and the system therefore coarsens in a well-defined manner.  Consideration is now given to the question of whether this coarsening can be arrested by the application of an external shear flow.

\section{Chaotic advection}
\label{sec:flow}

In this section, the effect of chaotic shear flow via passive advection is modelled using a variety of algebraically-defined velocity fields chosen in part for their simplicity but also because they mimic chaotic flows observed in the laboratory~\cite{naraigh2007bubbles,solomon2003lagrangian}.   In the first instance the 
quasi-periodic constant-phase sine flow is introduced; the other flow models are introduced at various points throughout the section with a view to comparing the robustness of the results to the choice of model flow.
A third motivation for considering such simple model flows is the resulting reduction in complexity, which enables us to focus on several key aspects whereby the presence of a flow field modifies the $Q$-tensor dynamics.  As such, for a generic two-dimensional flow, Equation~\eqref{eq:qtensor} reduces in complexity to the following set of coupled equations:
\begin{subequations}
\begin{align}
\frac{\partial q}{\partial t}&+\bm{v}\cdot\nabla q-2\Omega_{12} r=\epsilon^2 \nabla^2 q \nonumber\\
&+g_1(1-\theta) q + 3g_2 (q^2+r^2)-(q^2+r^2+s^2+qs)q-2g_2(q^2+r^2+s^2+qs),\\
\frac{\partial r}{\partial t}&+\bm{v}\cdot\nabla r-\Omega_{12}(s-q)+\mytu D_{12}=\epsilon^2  \nabla^2r \nonumber\\
&+g_1(1-\theta) r + 3g_2 (q+s)r-(q^2+r^2+s^2+qs)r,\\
\frac{\partial s}{\partial t}&+\bm{v}\cdot\nabla s+2\Omega_{12} r=\epsilon^2 \nabla^2s \nonumber\\
&+g_1(1-\theta) s + 3g_2 (r^2+s^2)-(q^2+r^2+s^2+qs)s-2g_2(q^2+r^2+s^2+qs),
\end{align}
\label{eq:qtensor_2d}%
\end{subequations}%
where $\Omega_{12}=(1/2)(\partial u/\partial y-\partial v/\partial x)$ and $D_{12}=(1/2)(\partial u/\partial y+\partial v/\partial x)$, and where the other components of the antisymmetric and symmetric rate-of-strain tensors are zero for the sine flow.
Throughout this section, equations~\eqref{eq:qtensor_2d} are solved numerically using the pseudo-spectral method already described in Section~\ref{sec:fixed}.

The first considered model flow is the time-periodic constant-phase sine flow with period $\tau$ is defined such that, at time $t$, the velocity field is given by
\begin{subequations}
\begin{eqnarray}
u&=&A\sin\left(k_0 y\right),\qquad   0\leq \text{mod}(t,\tau)<\tfrac{1}{2}\tau,\\
v&=&A\sin\left(k_0 x\right),\qquad      \tfrac{1}{2}\tau\leq \text{mod}(t,\tau)<\tau,
\end{eqnarray}%
\label{eq:flow_def_2d}%
\end{subequations}%
where the velocity components not listed are zero and where $t$ is written uniquely as $t=\myq\tau+\mu$ for $\myq$ zero or a positive integer, with $0\leq \mu<\tau$, and hence $\mathrm{mod}(t,\tau):=\mu$.  The phases of the sine functions in Equation~\eqref{eq:flow_def_2d} are assumed here to be constant and equal to zero, although  consider a modified sine flow is also considered
 wherein the phases are renewed periodically~\cite{lattice_PH1} (the random-phase sine flow).  In both scenarios, the purpose of the model sine flow is to mimic high-Prandtl number turbulence~\cite{naraigh2007bubbles}.   For the present purposes, the chosen parameter values are $A=0.8$ fixed (for both the constant-phase and random-phase sine flows), with $\tau$ varying as indicated below.  Also for the present purposes, the phases in the random-phase verison of the sine flow are updated precisely once per quasi-period $\tau$.

\begin{minipage}{0.95\linewidth}
      \centering
      \begin{minipage}{0.37\linewidth}
			\begin{center}
      \includegraphics*[width=1\textwidth,viewport=30 10 460 380]{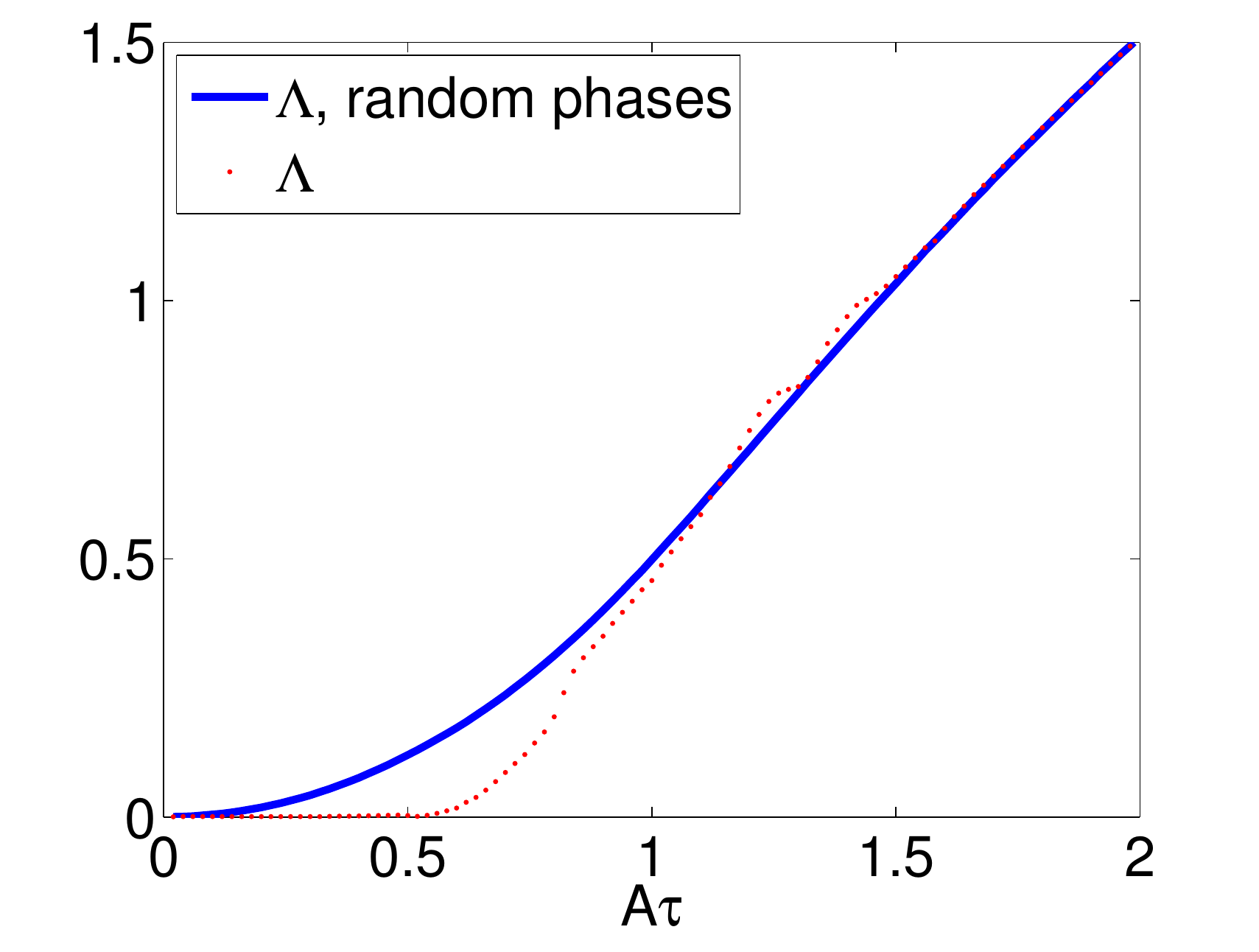}
			\end{center}
      \end{minipage}
			%
      \hspace{0.01\linewidth}
			%
      \begin{minipage}{0.58\linewidth}
			\begin{center}
\includegraphics*[width=0.4\textwidth,viewport=100 50 350 290]{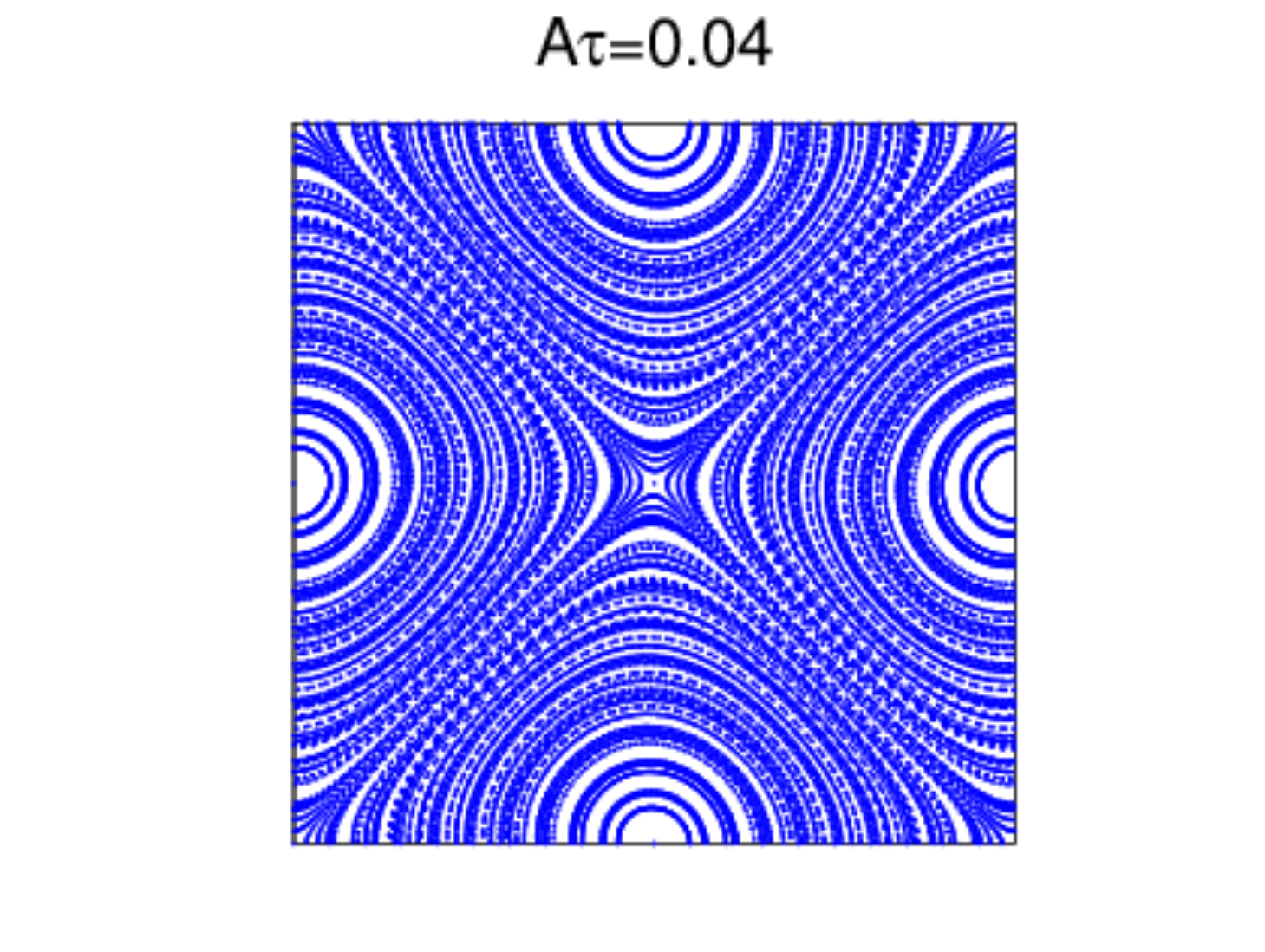}
\includegraphics*[width=0.4\textwidth,viewport=100 50 350 290]{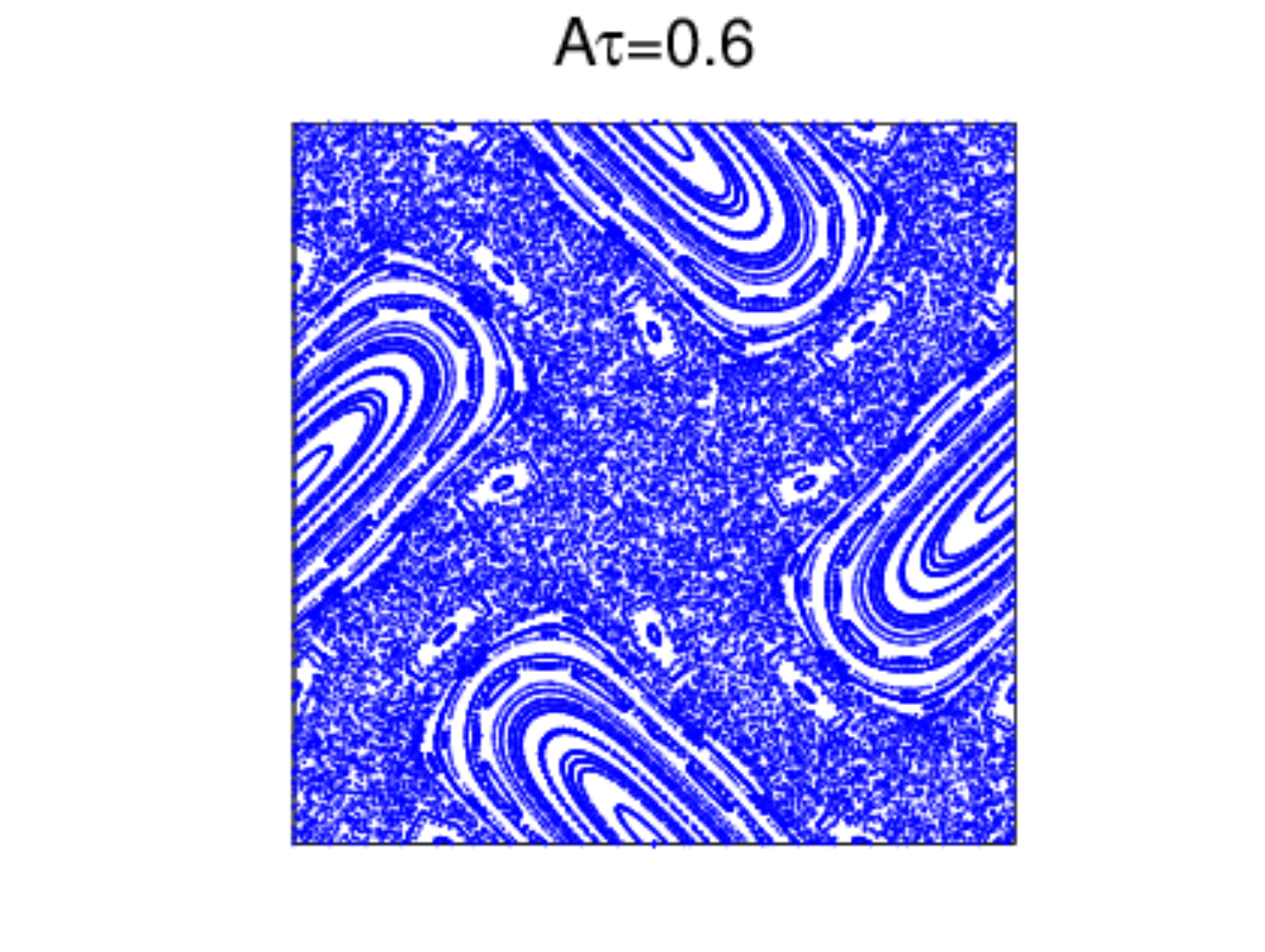}\\
\hspace{-0.1in}(a)$\phantom{aaaaaaaaaaaaaaa}$(b)\\		
\includegraphics*[width=0.4\textwidth,viewport=100 50 350 290]{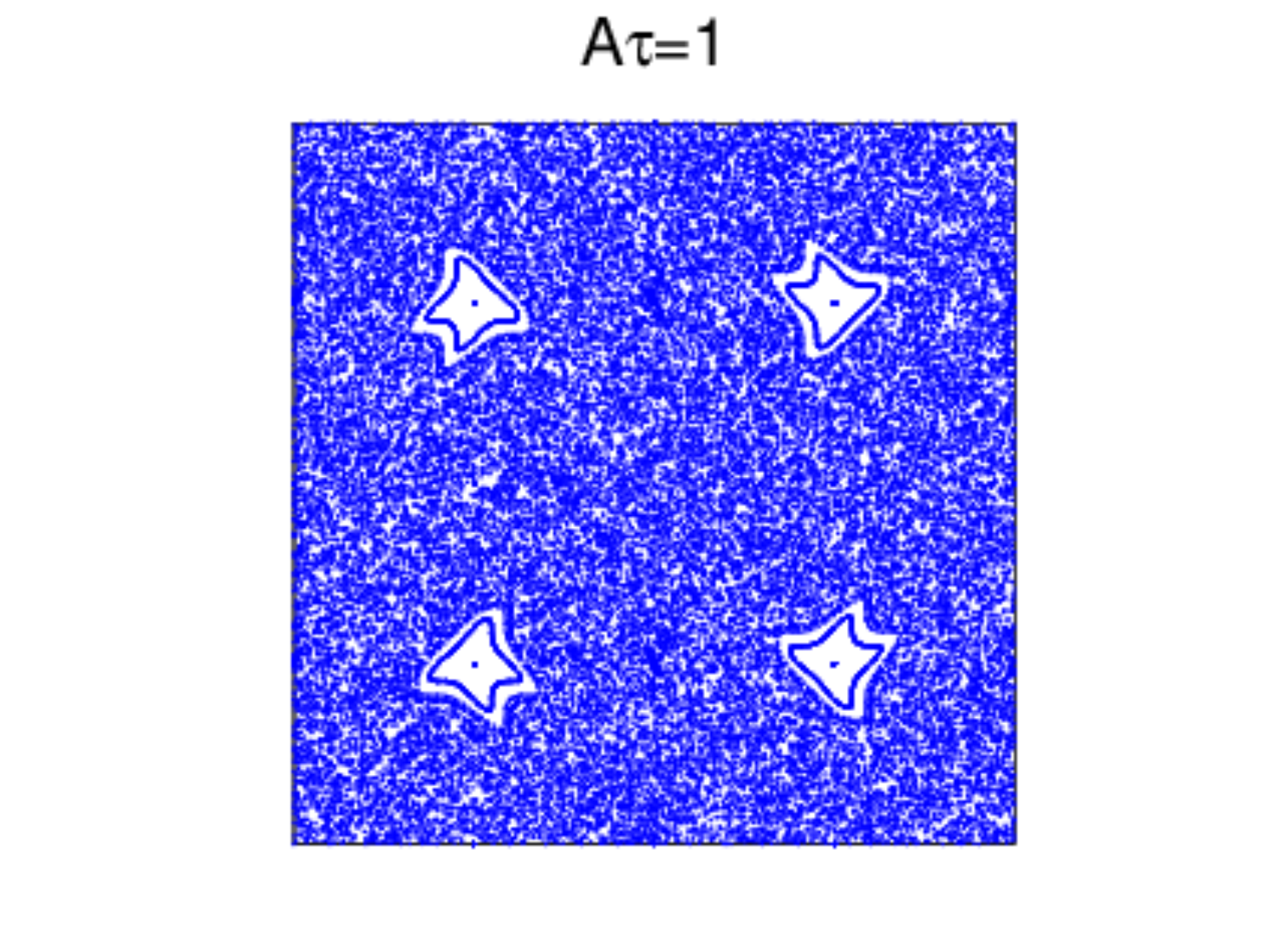}
\includegraphics*[width=0.4\textwidth,viewport=100 50 350 290]{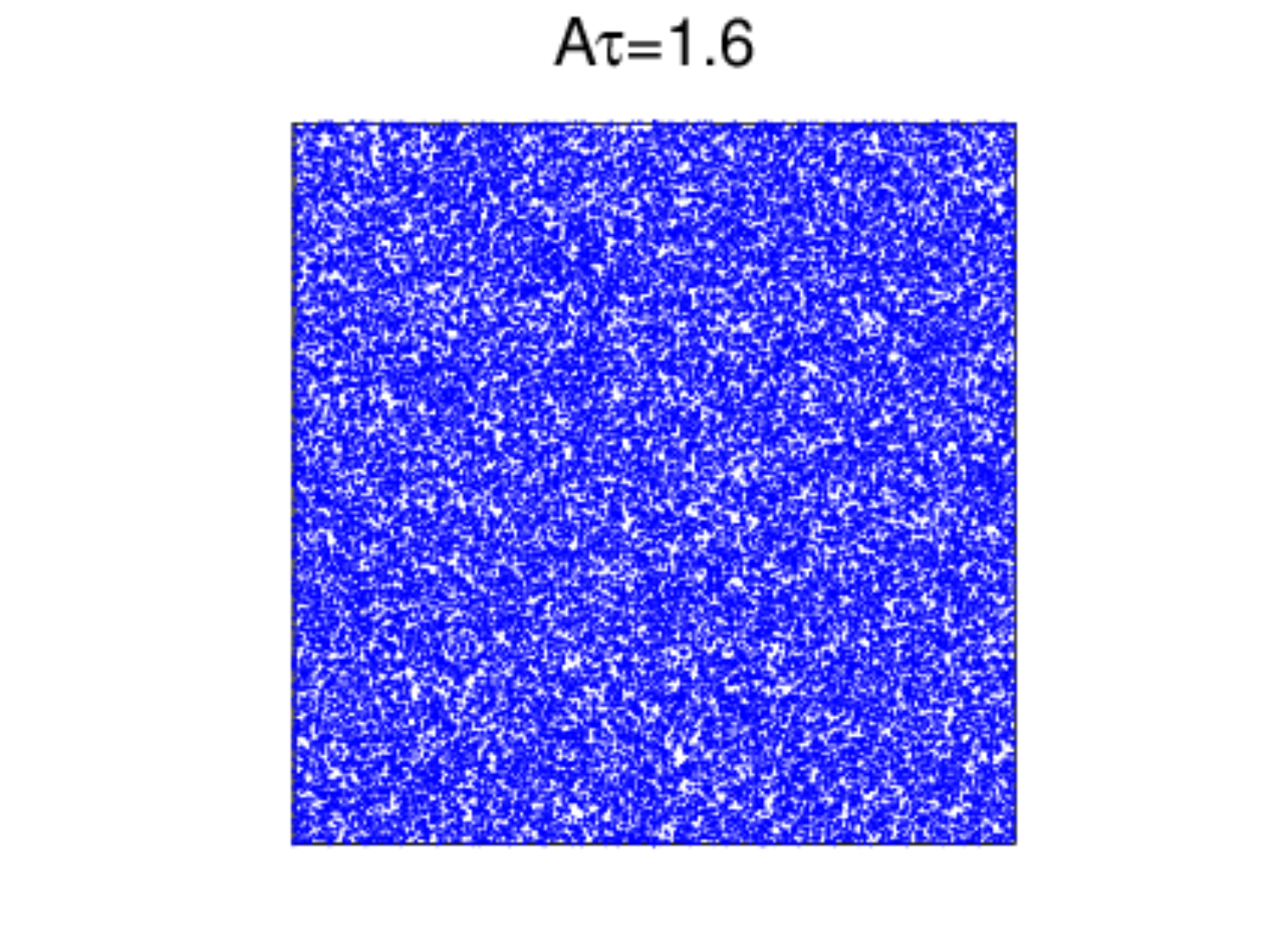}\\
\hspace{-0.1in}(c)$\phantom{aaaaaaaaaaaaaaa}$(d)
      \end{center}
      \end{minipage}
			%
			%
			\vspace{-12pt}
			\begin{figure}[H]
			\includegraphics*[width=0.0000000000001\textwidth,viewport=0 0 1 1]{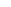}
			\caption{(Left) The average Lyapunov exponent for the constant-phase sine flow in Equation~\eqref{eq:flow_def_2d} (dotted line).  Shown for comparison is the average Lyapunov exponent for a similar sine flow wherein the $u$- and $v$- phases are renewed with separate independent random values once per flow period $\tau$. (Right)  Lagrangian trajectories for the constant-phase sine flow.  
Panel (a): $A\tau=0.04$; (b): $A\tau=0.6$; (c): $A\tau=1$; (d): $A\tau=1.6$.
Regular regions are visible where the trajectories are periodic.  As $A\tau$ increases the regular regions decrease such that by $A\tau=1.6$ the entire flow domain possesses chaotic trajectories.  The horizontal axis is the $x$-axis and the vertical axis is the $y$-axis.  The coordinates range from $(x,y)=0$ in the bottom left-hand corner to $(x,y)=(1,1)$ in the top right-hand corner.}
			\label{fig:mylep}
			\end{figure}
\vspace{8pt}
\end{minipage}
To characterize the flow in Equation~\eqref{eq:flow_def_2d} and its random-phase variant in a succinct fashion, a measure of the mean strain rate associated with the flow is introduced, denoted here by $\mylep$.
The chosen measure takes account of of the flow amplitude $A$, the flow lengthscale $2\pi/k_0$, and the flow timescale $\tau$, and is therefore identified with the average value of the maximal Lyapunov exponent of the flow, computed in a standard fashion~\cite{phdlennon}. 

The results are shown in Figure~\ref{fig:mylep}, for which the constant-phase sine flow~\eqref{eq:flow_def_2d} is first of all discussed.  The figure reveals that
the mean Lyaponov exponent (averaged over the entire domain) is positive for $A\tau\apprge 0.5$, which indicates that the sine flow in this regime is chaotic~\cite{lattice_PH1,Berthier2001}.  A characteristic of such flows is that neighboring parcels of fluid follow trajectories that separate exponentially at a distance that grows  as $\mathe^{\mylep t}$. 
Below the value $A\tau\approx 0.5$, the Lyaponov exponent is zero, meaning the flow is regular: particles remain trapped in well-defined regions of the flow,  and the flow is laminar effectively laminar.  
Care is needed in characterizing the flow based on the average Lyapunov exponent, as such an averaged approach hides fine details.  Indeed, the  cutoff at $A\tau\approx 0.5$ is by no means clear-cut for the constant-phase sine flow, since a small positive values of $\mylep$ at $A\tau\approx 0.5$ corresponds to a flow with inhomogeneous characteristics, i.e. chaotic regions where the local Lyaponov exponent is positive and regular regions where it is zero.   This is is illustrated in the inset panels in Figure~\ref{fig:mylep} where $\mathd\vecx /\mathd t=\bm{v}(\vecx,t)$ for an ensemble of particles: for $A\tau=0.04$ the trajectories are entirely regular, for $A\tau=0.6$--$1$ the trajectories are regular in certain regions and chaotic in others, while for $A\tau=1.6$ the trajectories are chaotic throughout the flow domain.  
In contrast, for the random-phase sine flow, the mean Lyapunov exponent is positive for all values of the parameter $A\tau$, meaning the flow is more chaotic than its constant-phase counterpart.  Intuitively, this makes sense, as the periodic randomization of the phases breaks up the regular flow regions apparent in the constant-phase case in Figure~\ref{fig:mylep}(a)--(c), such that the trajectories resemble Figure~\ref{fig:mylep}(d), regardless of the value of $A\tau$.

\subsection{Results -- no tumbling}

\begin{figure}
	\centering
		\subfigure[$\,\,A\tau=0.04$]{\includegraphics[width=0.32\textwidth]{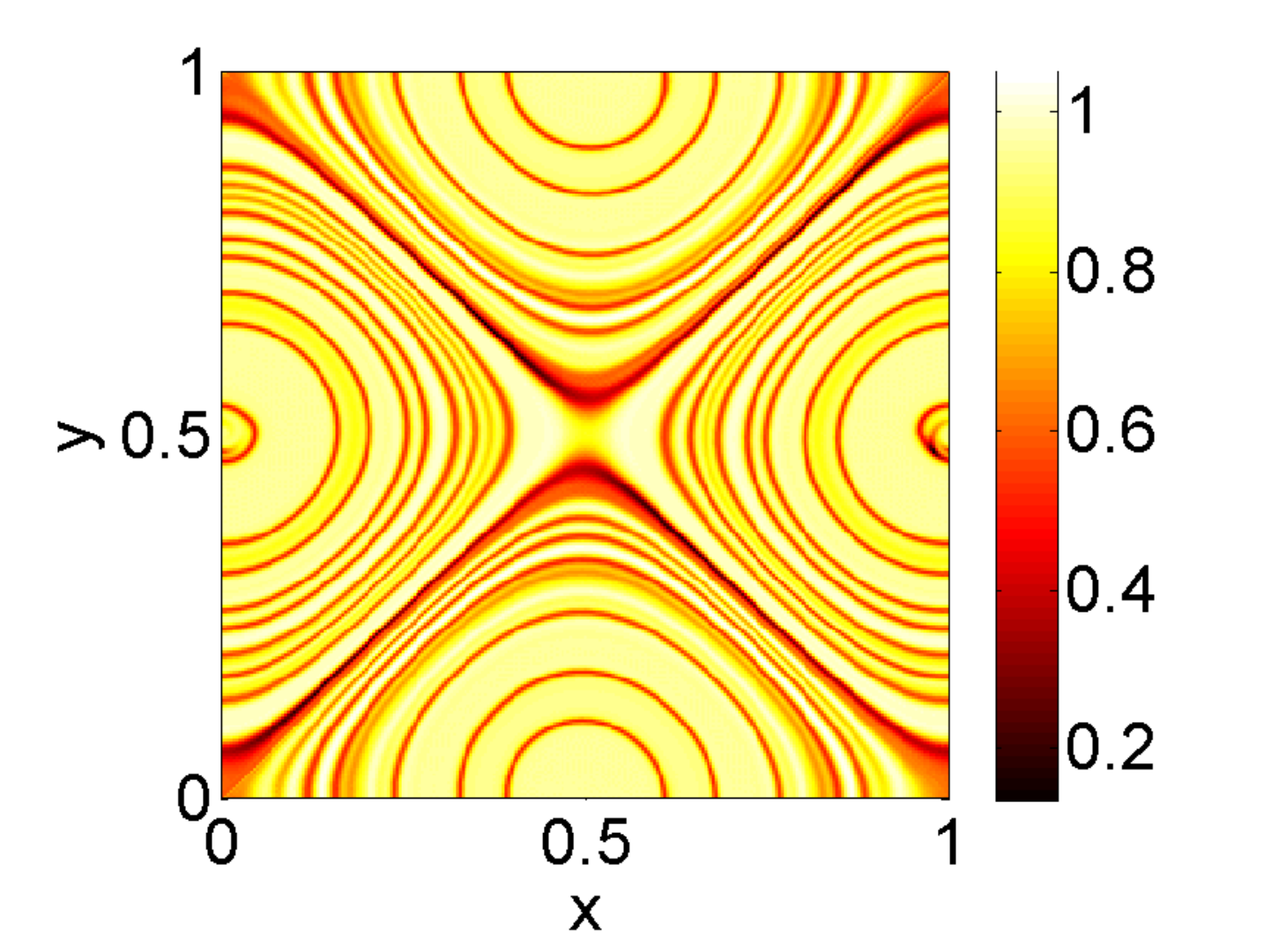}}
		\subfigure[$\,\,A\tau=0.6$]{\includegraphics[width=0.32\textwidth]{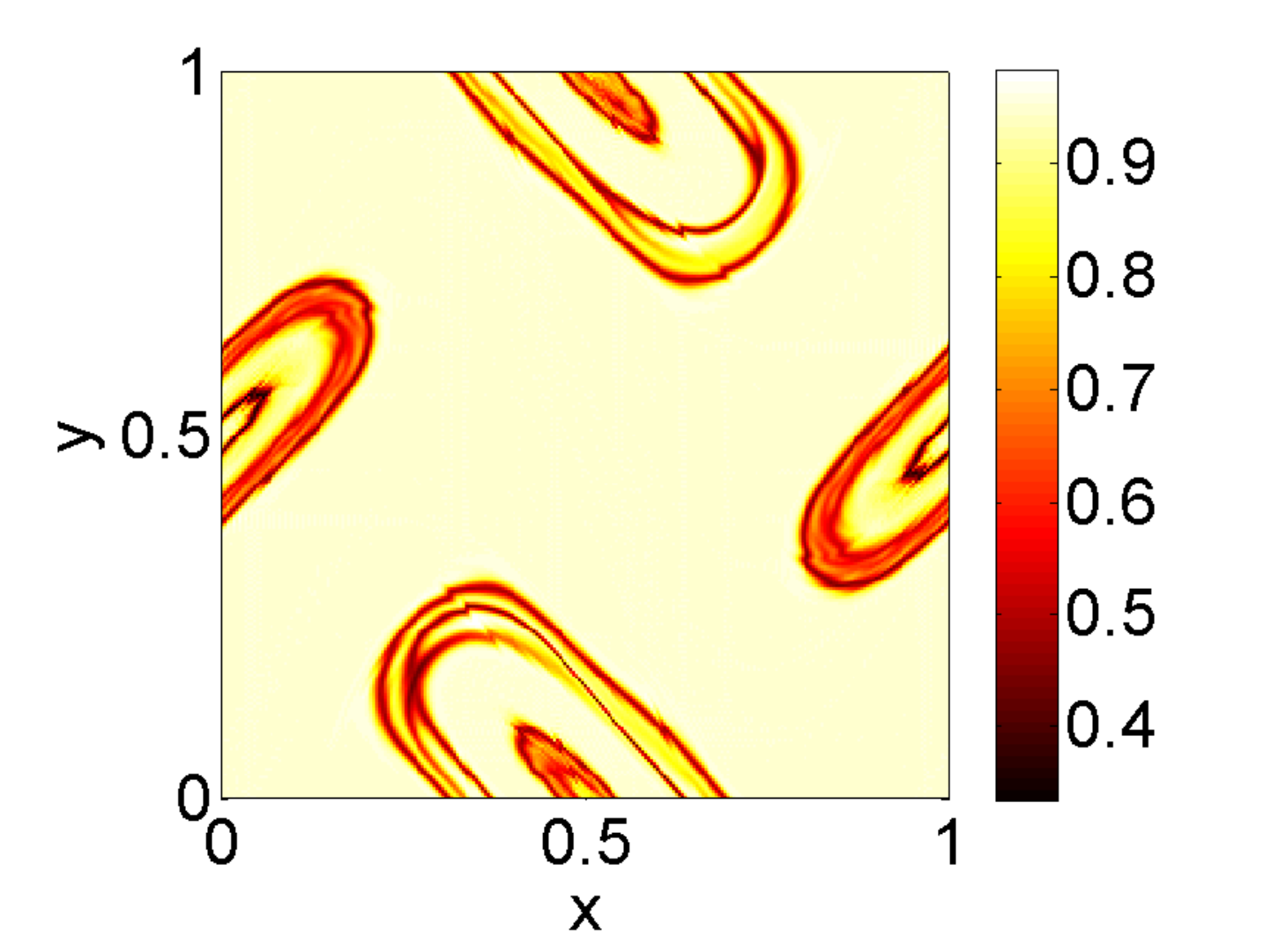}}
		\subfigure[$\,\,A\tau=1.6$]{\includegraphics[width=0.32\textwidth]{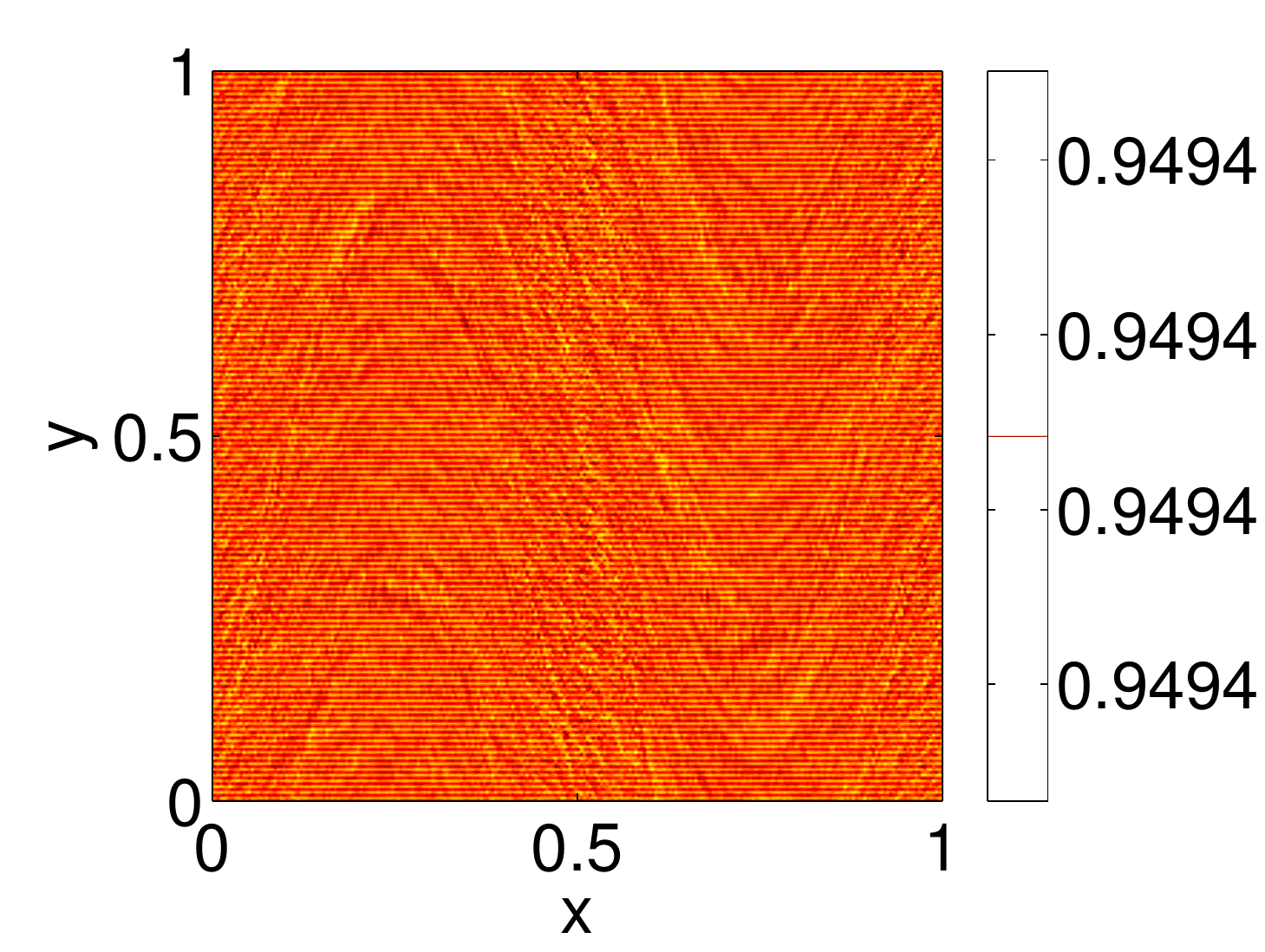}}\\
		\subfigure[$\,\,A\tau=0.04$]{\includegraphics[width=0.32\textwidth]{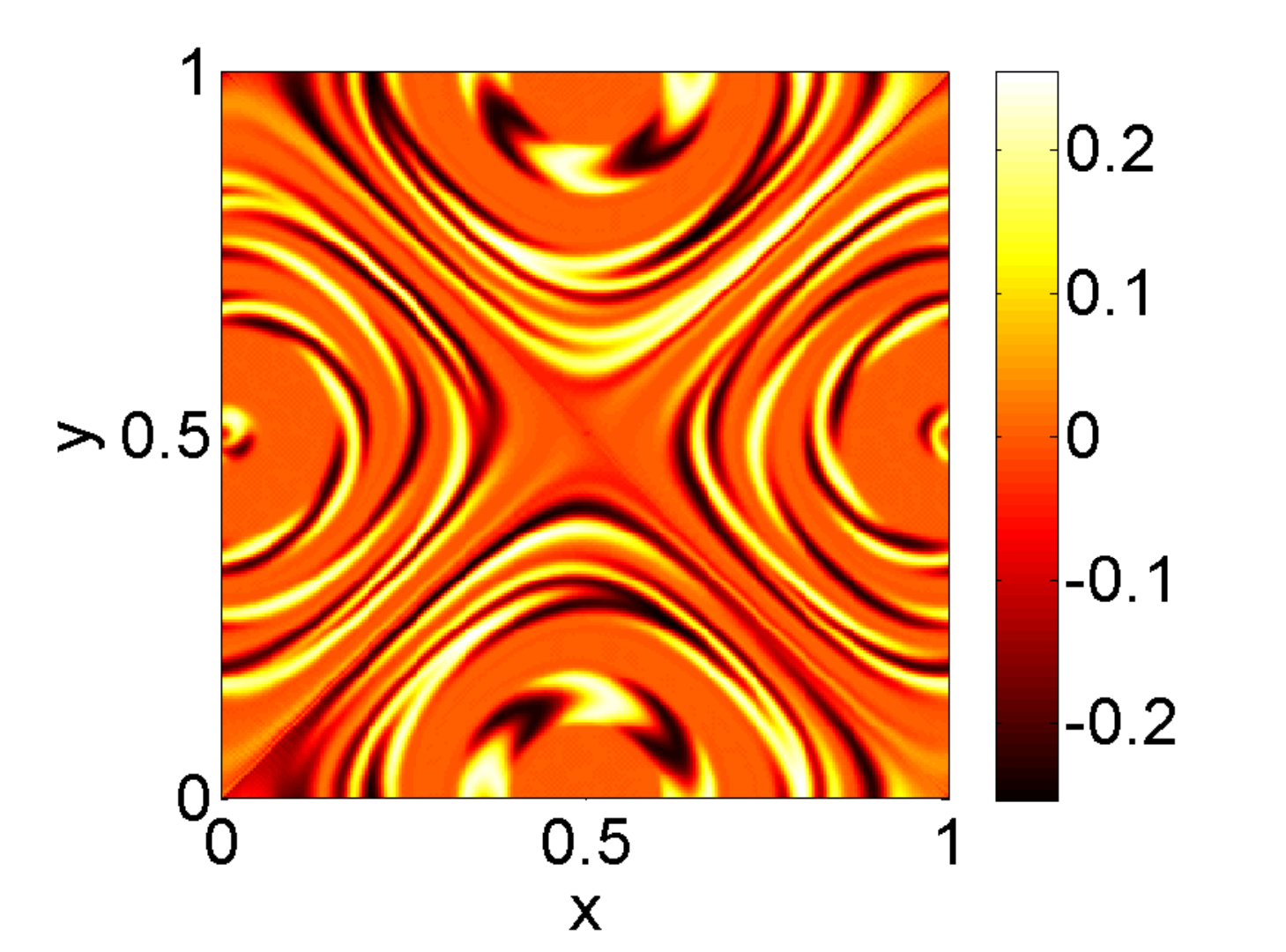}}
		\subfigure[$\,\,A\tau=0.6$]{\includegraphics[width=0.32\textwidth]{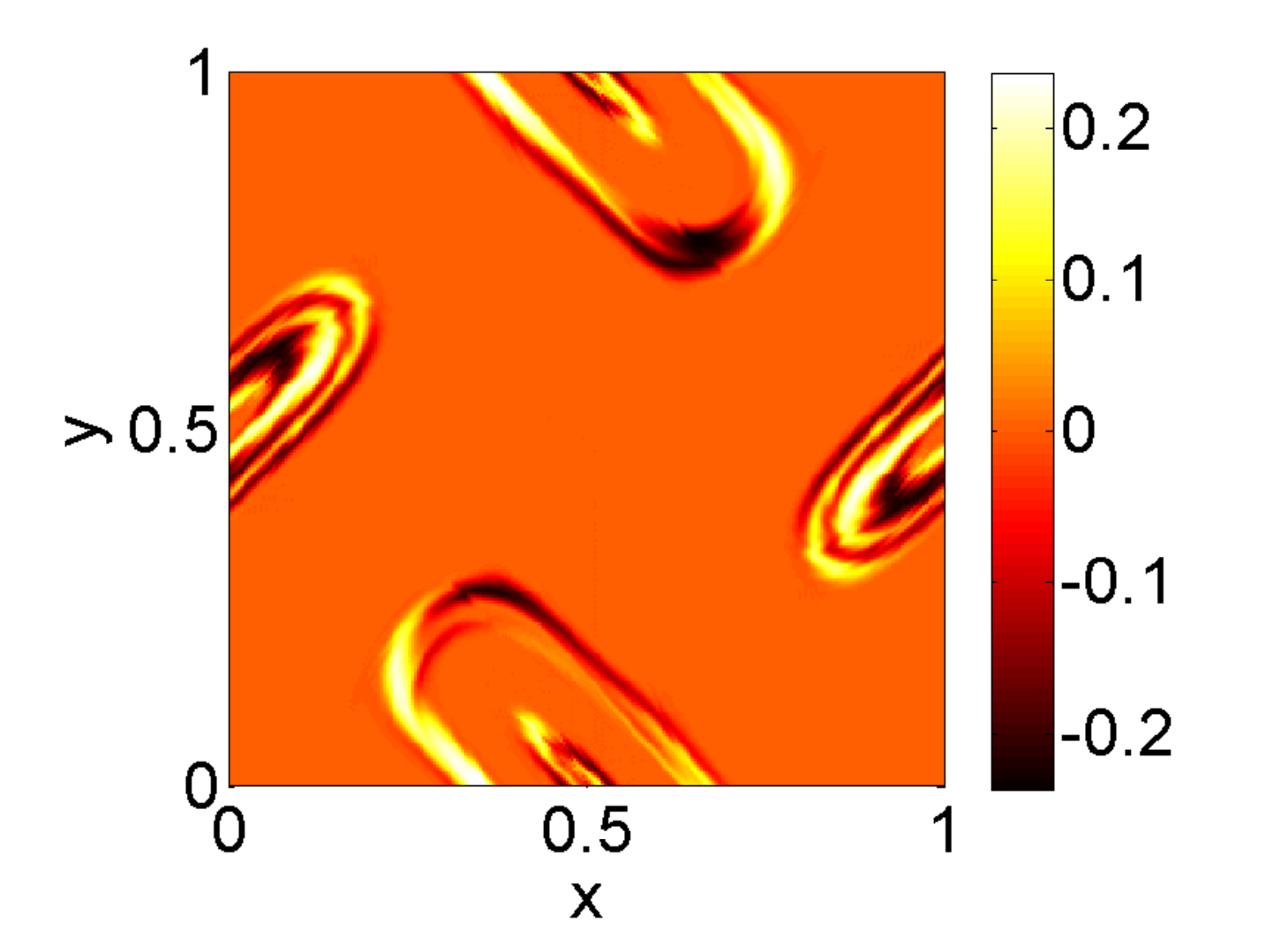}}
		\subfigure[$\,\,A\tau=1.6$]{\includegraphics[width=0.32\textwidth]{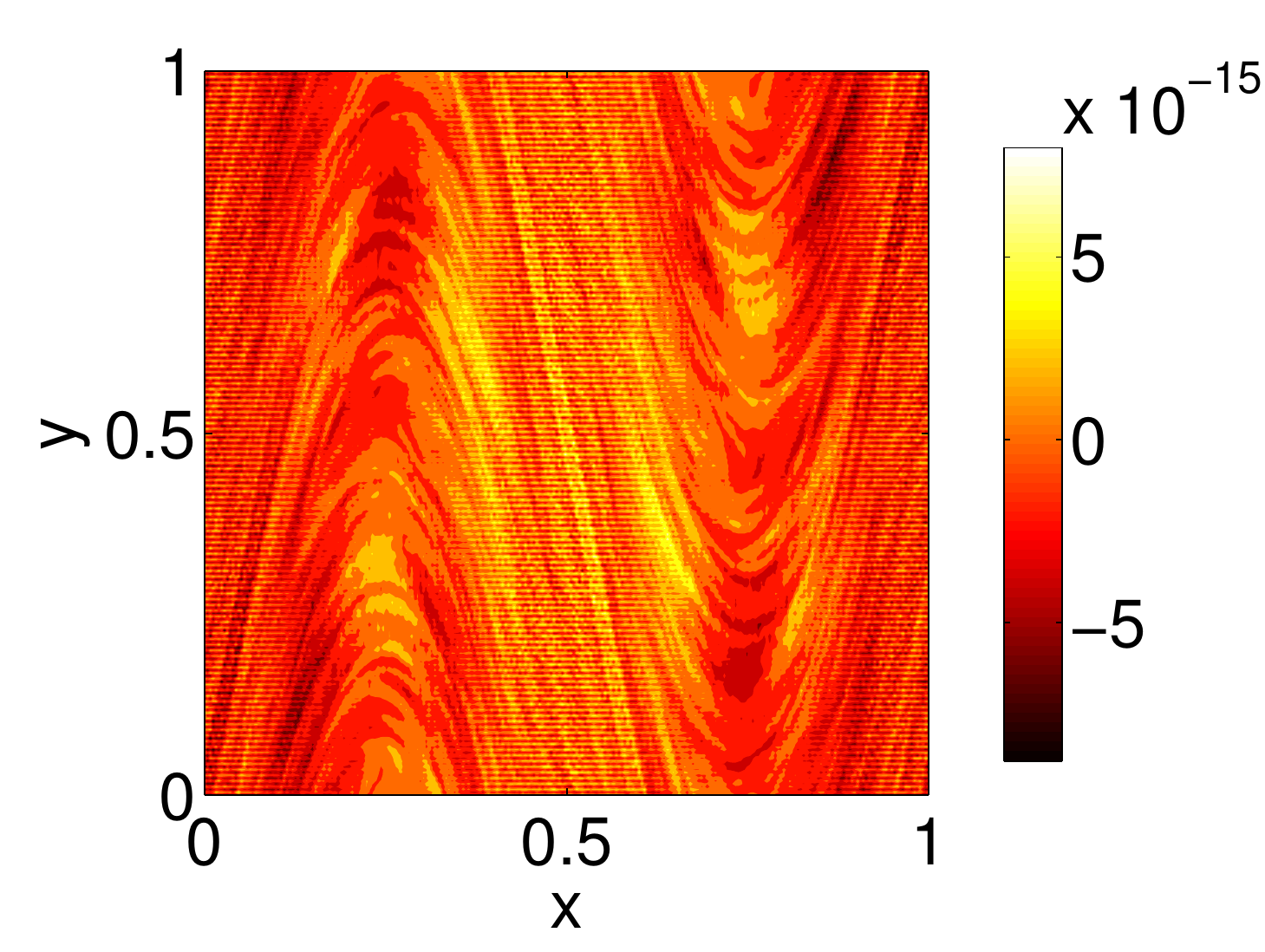}}
		\caption{Across the top: snapshots of scalar order parameter at various times for various values of $A\tau$, with $\mytu=0$. Across the bottom: corresponding snapshots of $r$.  Snapshots in the first two columns are taken at $t=320$.  The third column (figures (c) and (f)) concerns $A\tau=1.6$, for which the snapshots are taken at $t=320$; these are included here to demonstrate the relaxation to a uniform state for the large values of $A\tau$.  The compressed colour bars in these figures is a consequence of the rapid relaxation to the uniform steady state.}
	\label{fig:flow_no_tumble1}
\end{figure}

A first set of results is shown in Figure~\ref{fig:flow_no_tumble1} for the constant-phase sine flow with no tumbling ($\mytu=0$).
For $A\tau=0.04$ the order parameter is trapped inside regions whose boundaries are described by the separatrices $y=x$ and $y=1-x$, corresponding to the directions of maximum stretching of the sine flow.  Inside these regions, the $Q$-tensor is stretched and folded in a uniform fashion by the application of the sine flow, leading to thin annular-shaped nematic domains where the $Q$-tensor locally relaxes to values related either a bi-axial or a uni-axial fixed point.   The natural formation of the nematic domains is thereby disrupted, as the imposition of the flow separatrices sets an upper limit of the domain size: in short, the nematic domains (whether uniaxial or biaxial) are `frozen-in' to the flow structure.  For $A\tau=0.6$ the growth of the domains is again disrupted, and the chaotic regions of the flow extend to larger parts of the domain.  Inside the regular regions, the $Q$-tensor again locally relaxes to fixed-point values as previously, whereas in the chaotic regions, the $Q$-tensor relaxes to a uniform state corresponding to a stable bi-axial fixed point, as evidenced by the extended domain wherein $r=0$ in the corresponding plot of the $r$-component of the $Q$-tensor.  
These trends continue with increasing $A\tau$: the `islands' in which the $Q$-tensor assumes a mixture of uni-axial and bi-axial fixed-point values shrink while the surrounding `sea' of biaxial fixed-point values expands.
 Beyond $A\tau=1$, the islands disappear altogether the $Q$-tensor relaxes to a uniform state given by the stable bi-axial fixed point (panels (c), (f) in Figure~\ref{fig:flow_no_tumble1}).

\begin{figure}
	\centering
		\subfigure[]{\includegraphics[width=0.45\textwidth]{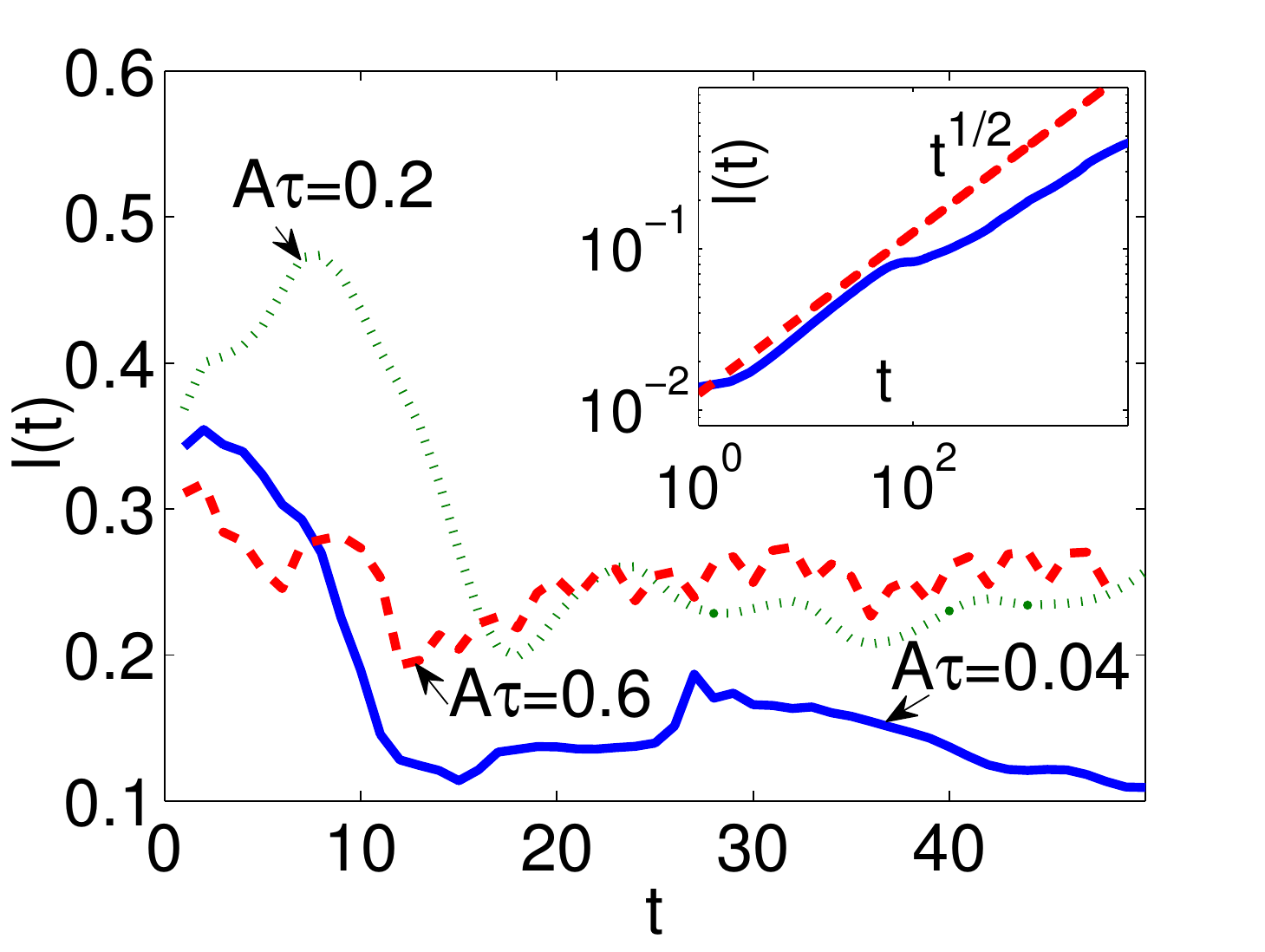}}
		\subfigure[]{\includegraphics[width=0.46\textwidth]{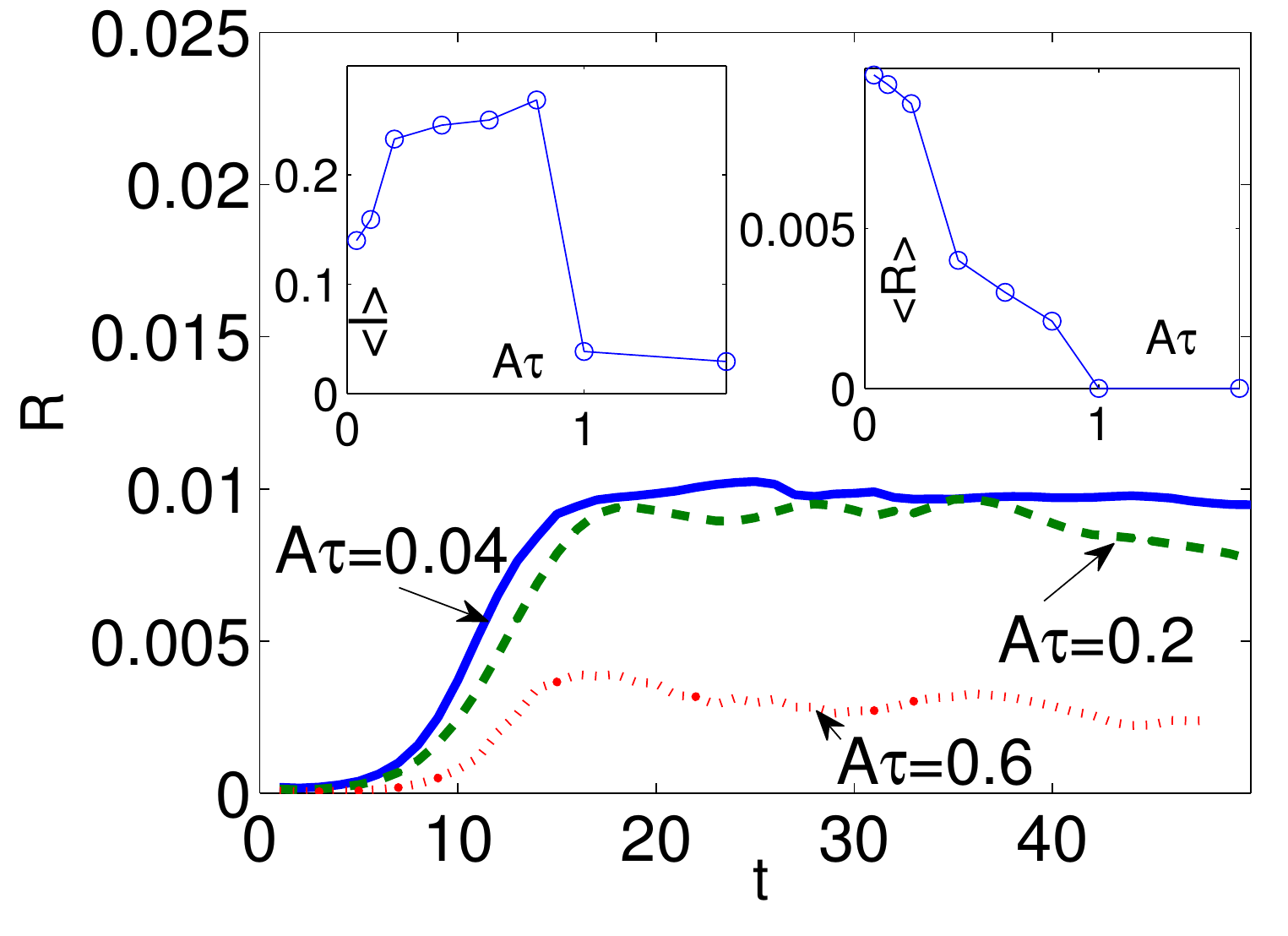}}
		\caption{(a) Time evolution of the domain scale $L(t)$ for various values of $A\tau$.  The inset shows the time-averaged values of $L(t)$ for a much larger range of $A\tau$-values, with angle brackets denoting a time average.  The time-averages are taken over intervals where the $Q$-tensor dynamics are in a statistically steady state.  (b)  The same, for  $R:=L_x^{-1}L_y^{-1}\iint r^2\,\mathd x\,\mathd y$}
	\label{fig:Lt_no_tumbling}
\end{figure}

The above results can be summarized succinctly by examining $L(t)$ as given by Equation~\eqref{eq:Lt_def} (Figure~\ref{fig:Lt_no_tumbling}(a)).  
For $A\tau\apprle 0.8$, $L(t)$ saturates at a value $L(t)\approx 0.25$, consistent with coarsening arrest governed by the `freezing in' of the nematic domains into regular flow regions bounded by well-defined flow separatrices.  Beyond $A\tau=0.8$, the islands of regular flow shrink and the chaotic regions expand, leading to a corresponding shrinking of the nematic domains, such that by $A\tau=1$ a uniform bi-axial state is favoured.  Correspondingly, the average value of $r^2$ (averaged over the spatial domain) is plotted in Figure~\ref{fig:Lt_no_tumbling}(b) as a function of time.
According to Figure~\ref{fig:flow_no_tumble1}, the mean square value of $r$ can be used as a measure of bi-axiality, with $r=0$ corresponding to a biaxial fixed point.  Thus, it can be seen quantitatively from the figure that increasing $A\tau$ increases the degree to which the system relaxes to a bi-axial fixed point.  It is noted that the time series in Figure~\ref{fig:Lt_no_tumbling} show only a portion of the results obtained: the simulations have been continued out to $t\approx 2,000$ and the steady state observed in the figure persists at all such late times.

\subsection{Results -- with tumbling}

\begin{figure}[htb]
	\centering
		\subfigure[$\,\,A\tau=0.04$]{\includegraphics[width=0.32\textwidth]{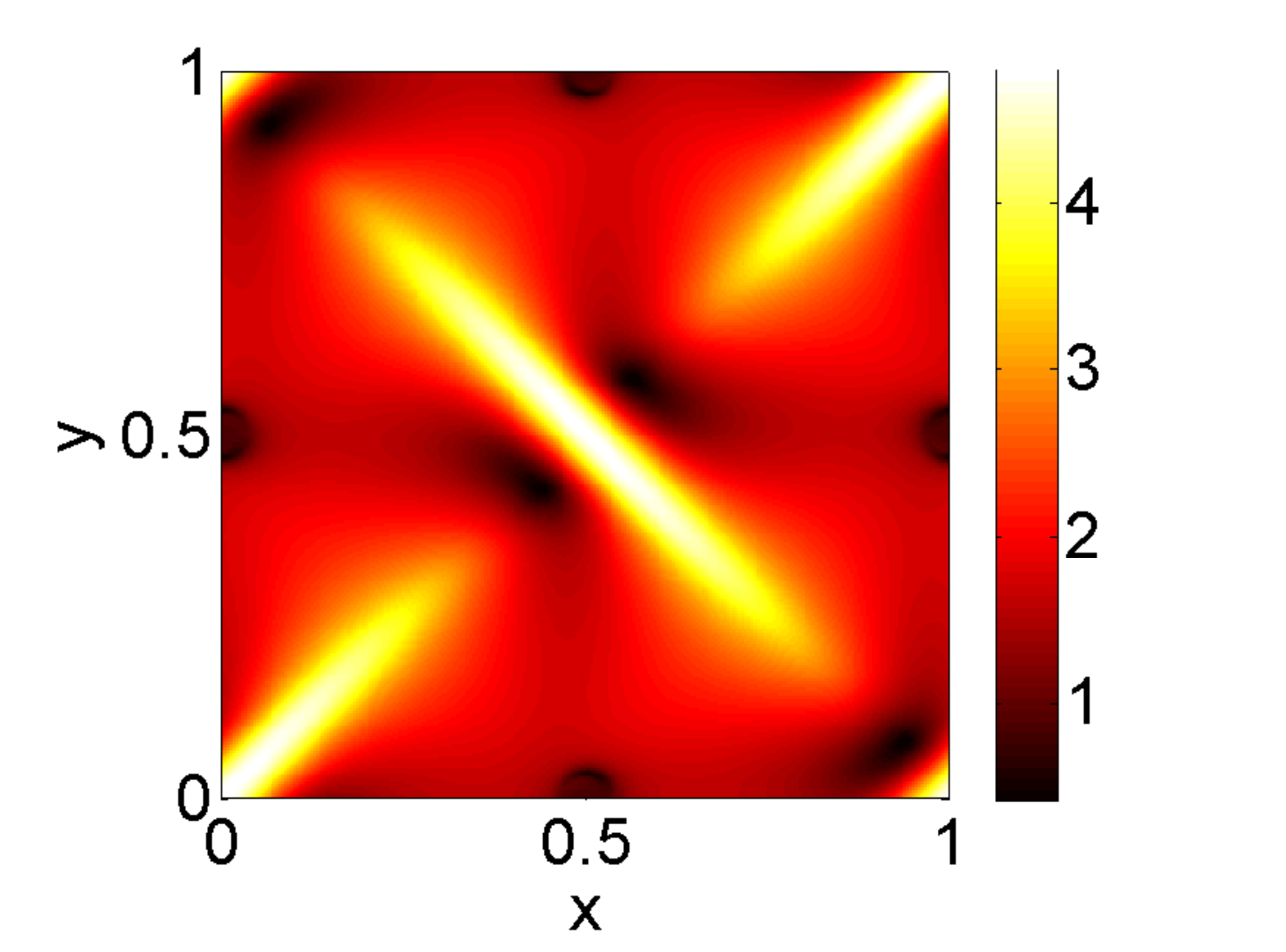}}
		\subfigure[$\,\,A\tau=0.8$]{\includegraphics[width=0.32\textwidth]{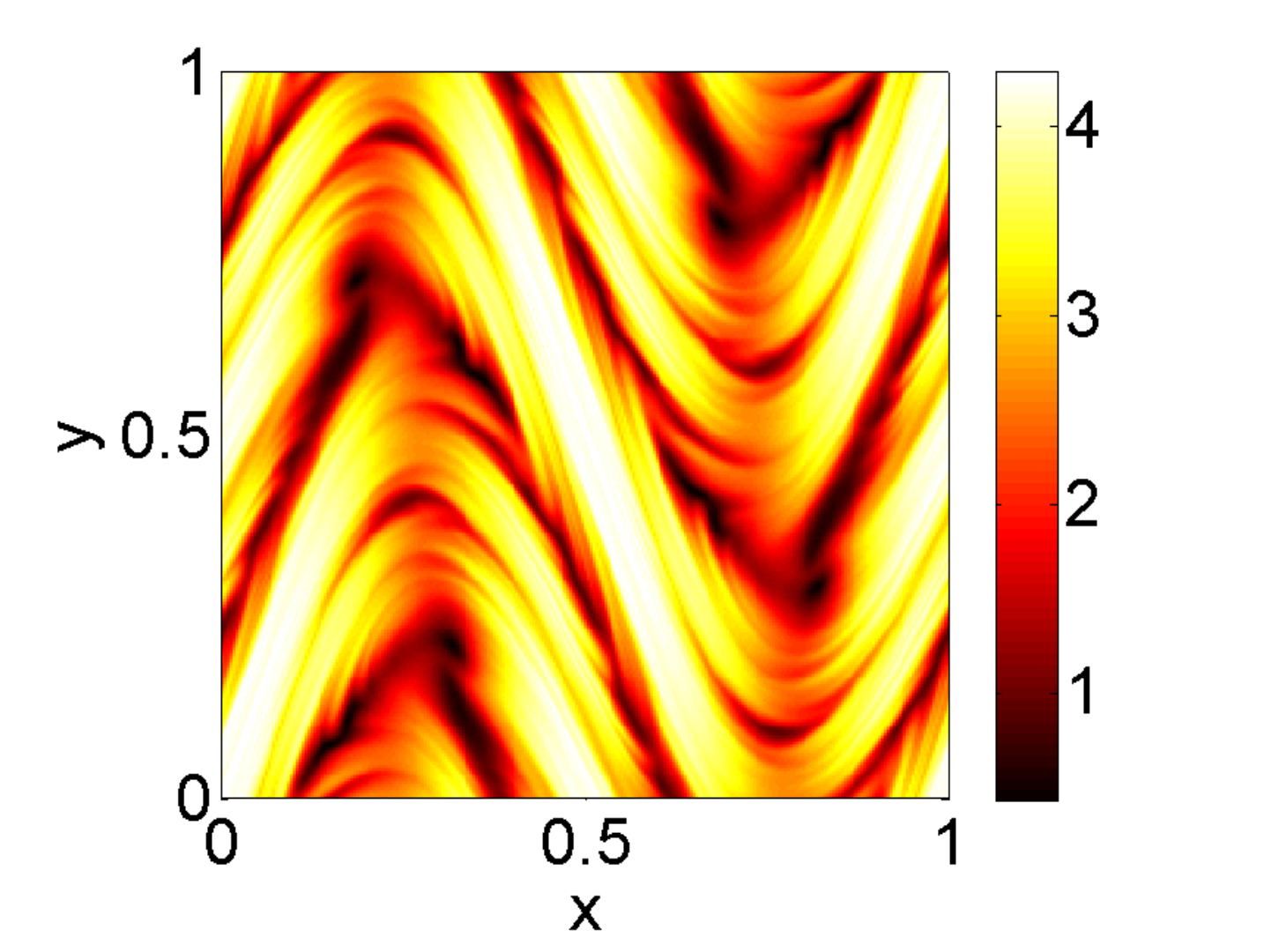}}
				\subfigure[$\,\,A\tau=1.6$]{\includegraphics[width=0.32\textwidth]{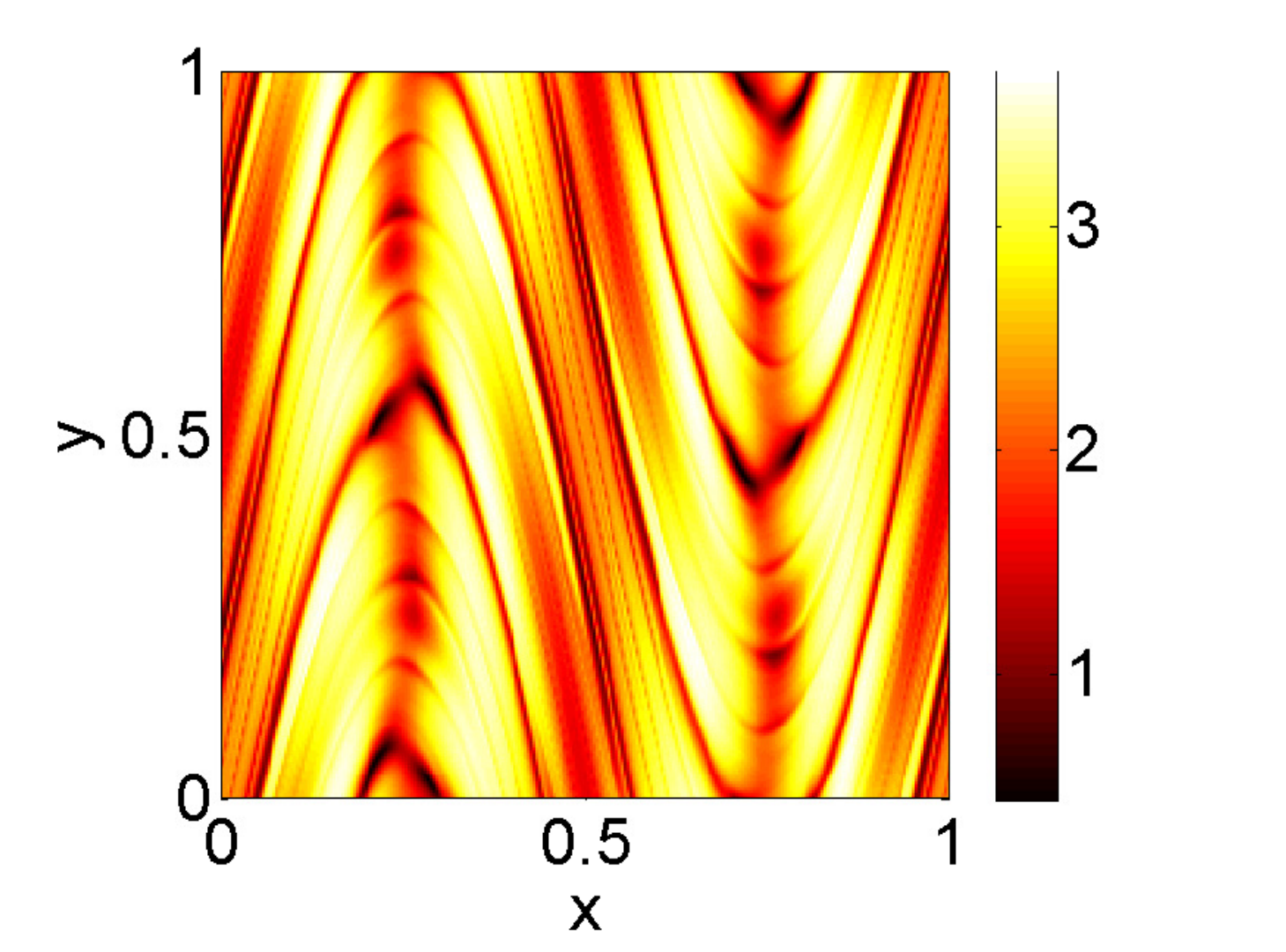}}\\
		\subfigure[$\,\,A\tau=0.04$]{\includegraphics[width=0.32\textwidth]{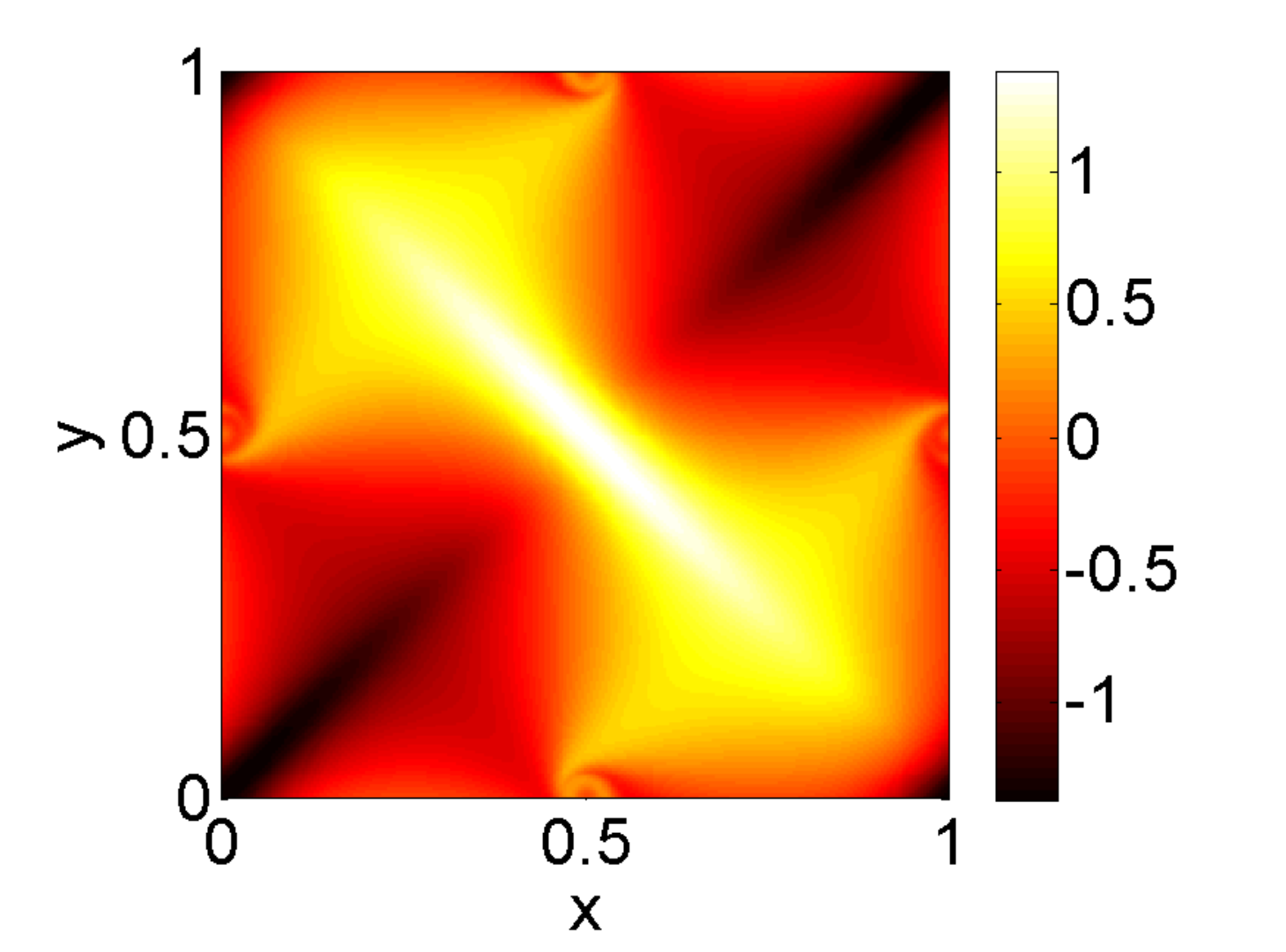}}
		\subfigure[$\,\,A\tau=0.8$]{\includegraphics[width=0.32\textwidth]{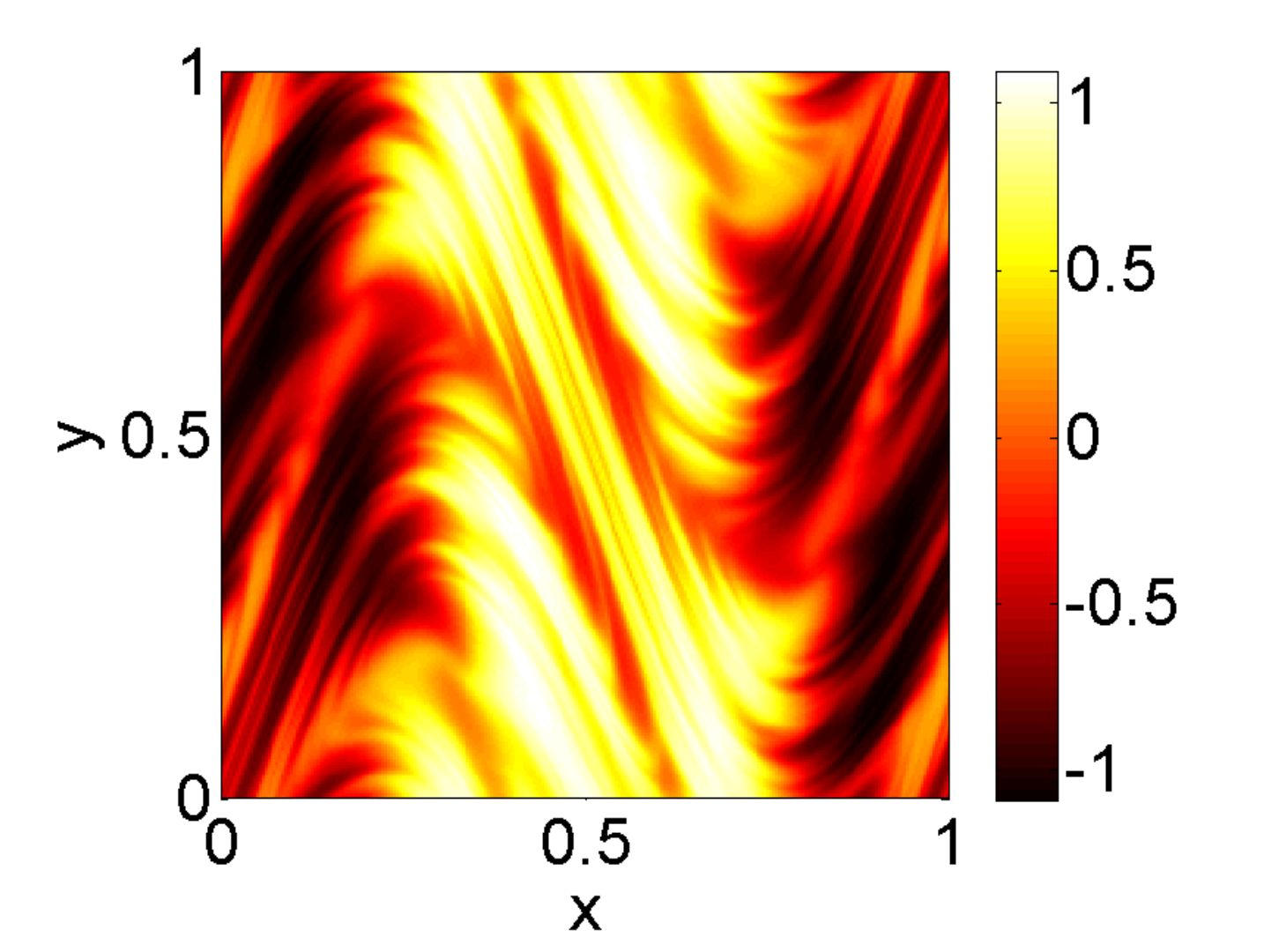}}
				\subfigure[$\,\,A\tau=1.6$]{\includegraphics[width=0.32\textwidth]{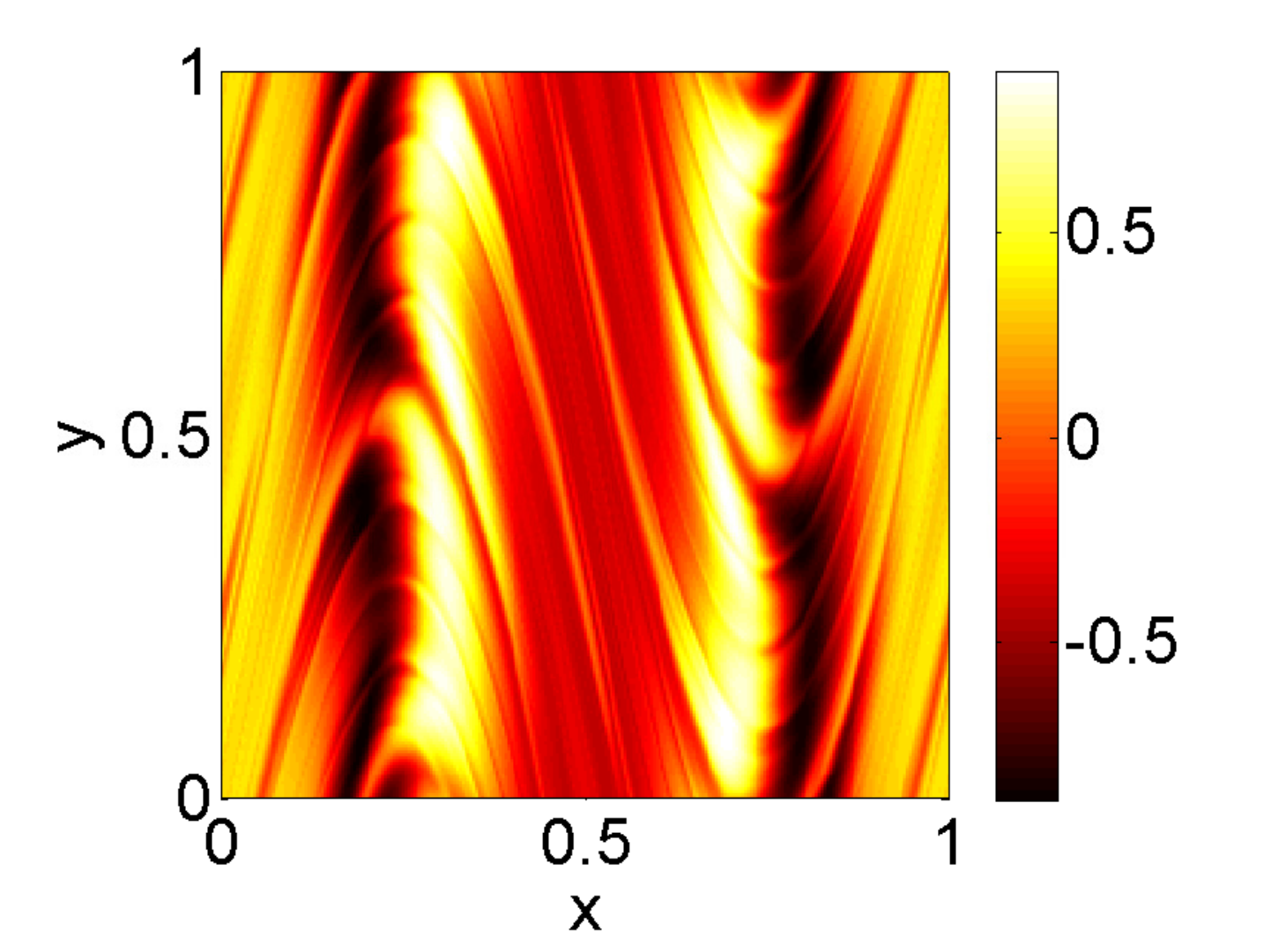}}\\
		\caption{Across the top: snapshots of scalar order parameter at various times for various values of $A\tau$, with $\mytu=1$. Across the bottom: corresponding snapshots of $r$.  The snapshots are illustrative and are taken at different times: the snapshots at $A\tau=0.04,0.8,1.6$ are taken at $t=120,200,6$ respectively.  }
	\label{fig:flow_tumble1}
\end{figure}
A second set of results for the constant-phase sine flow examined with $\mytu=1$.  Since $\partial u/\partial x=\partial v/\partial y=0$ for the sine flow, the imposition of tumbling in the $Q$-tensor dynamics amounts only  to the addition of a source term $D_{12}$ on the left-hand side of the $r$-equation, as can be seen in Equation~\eqref{eq:qtensor_2d}.  Snapshots of the resulting $Q$-tensor properties are shown in Figure~\ref{fig:flow_tumble1}.
\begin{figure}[htb]
	\centering
		\includegraphics[width=0.6\textwidth]{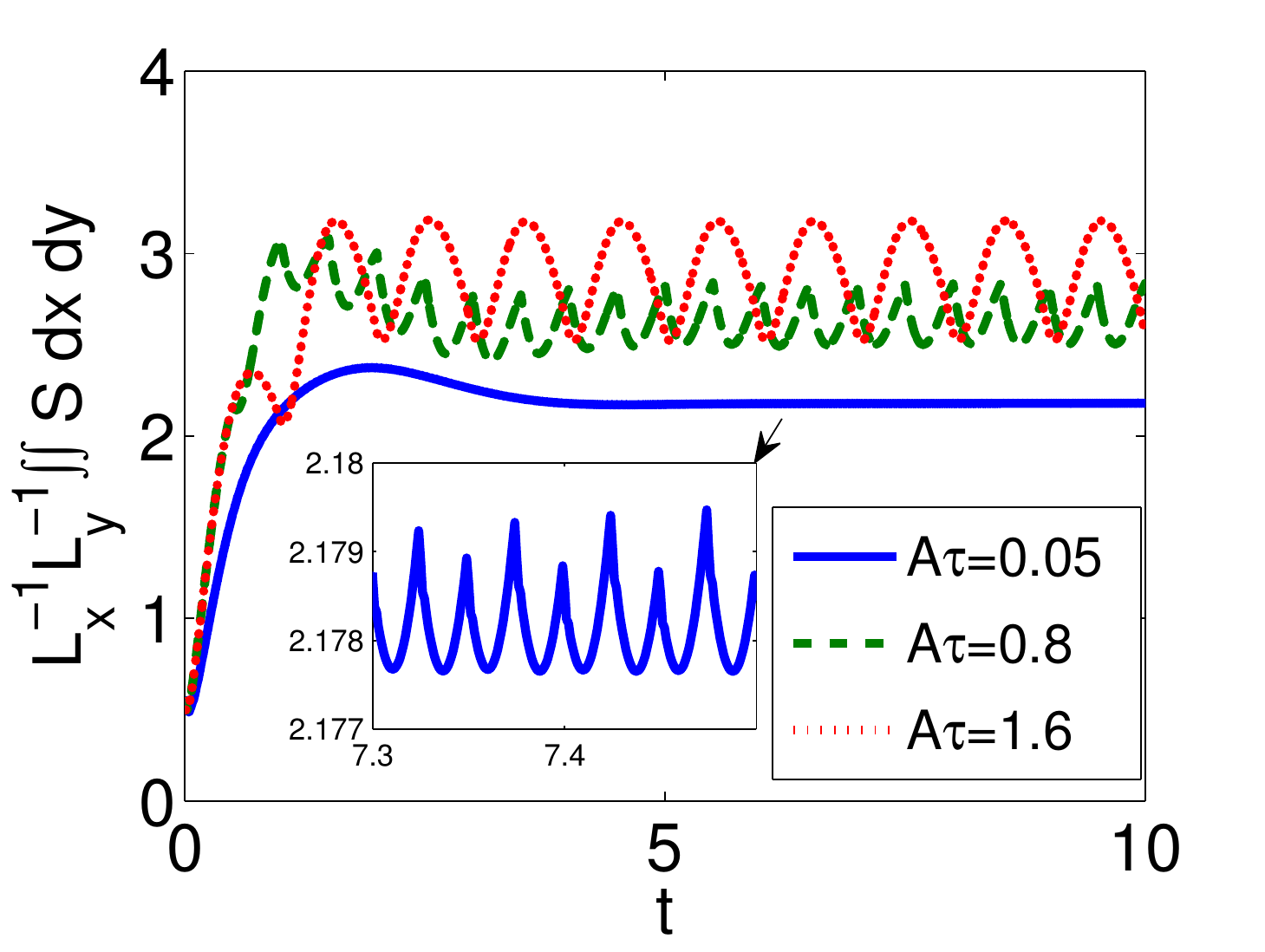}
		\caption{The spatially-averaged order parameter as a function of time for various values of $A\tau$ -- constant-phase sine flow with tumbling, $\mytu=1$. }
	\label{fig:flow_tumble2}
\end{figure}
Corresponding spatially-averaged quantities are shown in Figure~\ref{fig:flow_tumble2} as a function of time.  These reveal that the system settles down to a periodic state whose period is the same (in a first approximation) to the period of the forcing term.  
%
%
%
Furthermore, for the regular flow regime $A\tau = 0.04$, the scalar order parameter assumes the same regular spatial structure as the imposed flow field.  For the chaotic regimes $A\tau=0.8$ and $A\tau=1.6$,  the scalar order parameter again assumes a spatial structure imposed by the flow; in these cases however the resulting pattern is chaotic as opposed to laminar.  In each considered case, the dynamics can be understood simply as a response to the forcing due to the inhomogeneity in Equation~\eqref{eq:qtensor_2d}.

\subsection{Results -- other model flows}

\begin{figure}
	\centering
		\includegraphics[width=0.6\textwidth]{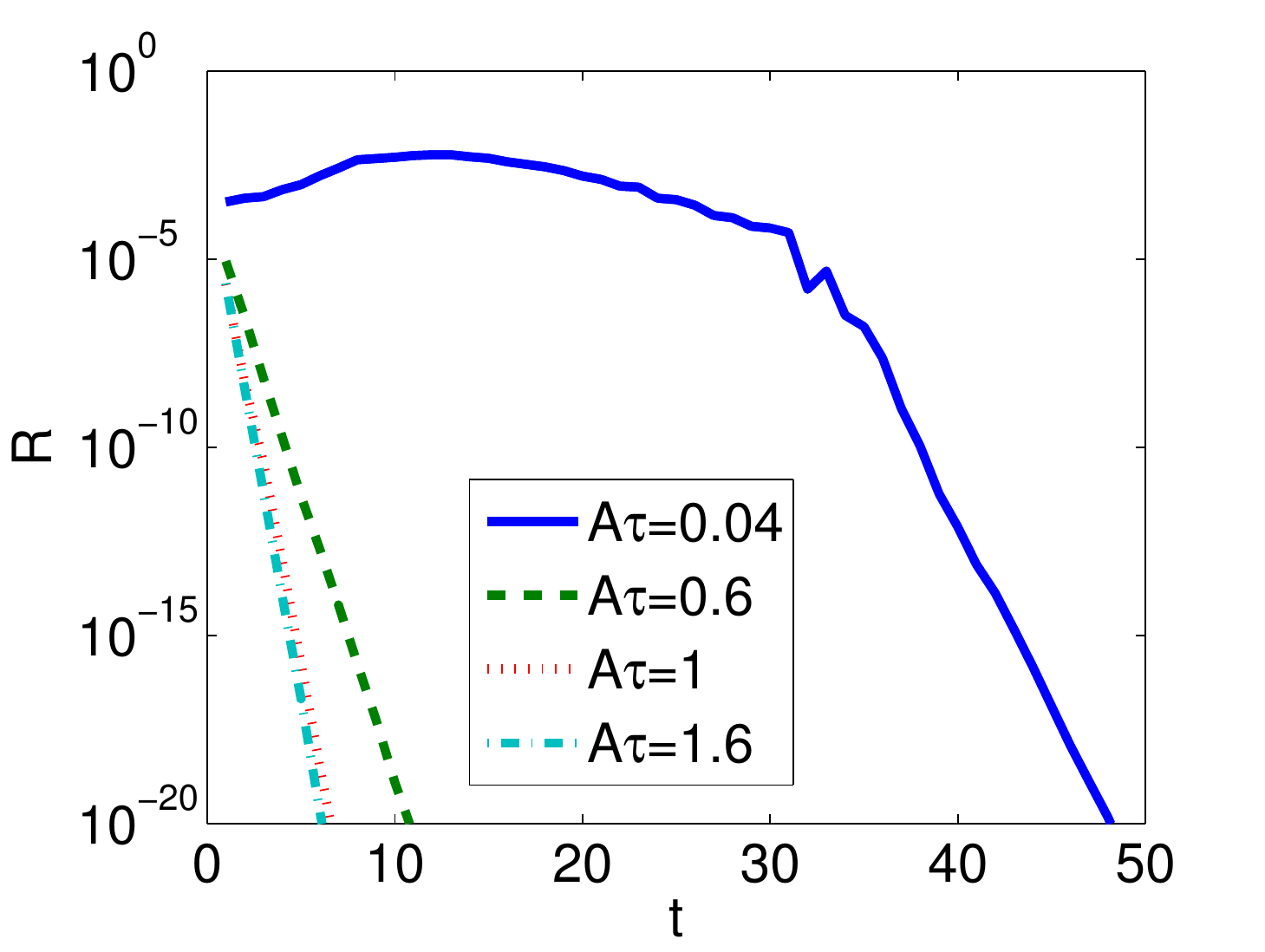}
		\caption{Time evolution of $R(t)$ for various values of $A\tau$, for the random-phase sine flow.}
	\label{fig:Rt_random}
\end{figure}

\begin{figure}
	\centering
		\subfigure[$\,\,t=50$]{\includegraphics[width=0.32\textwidth]{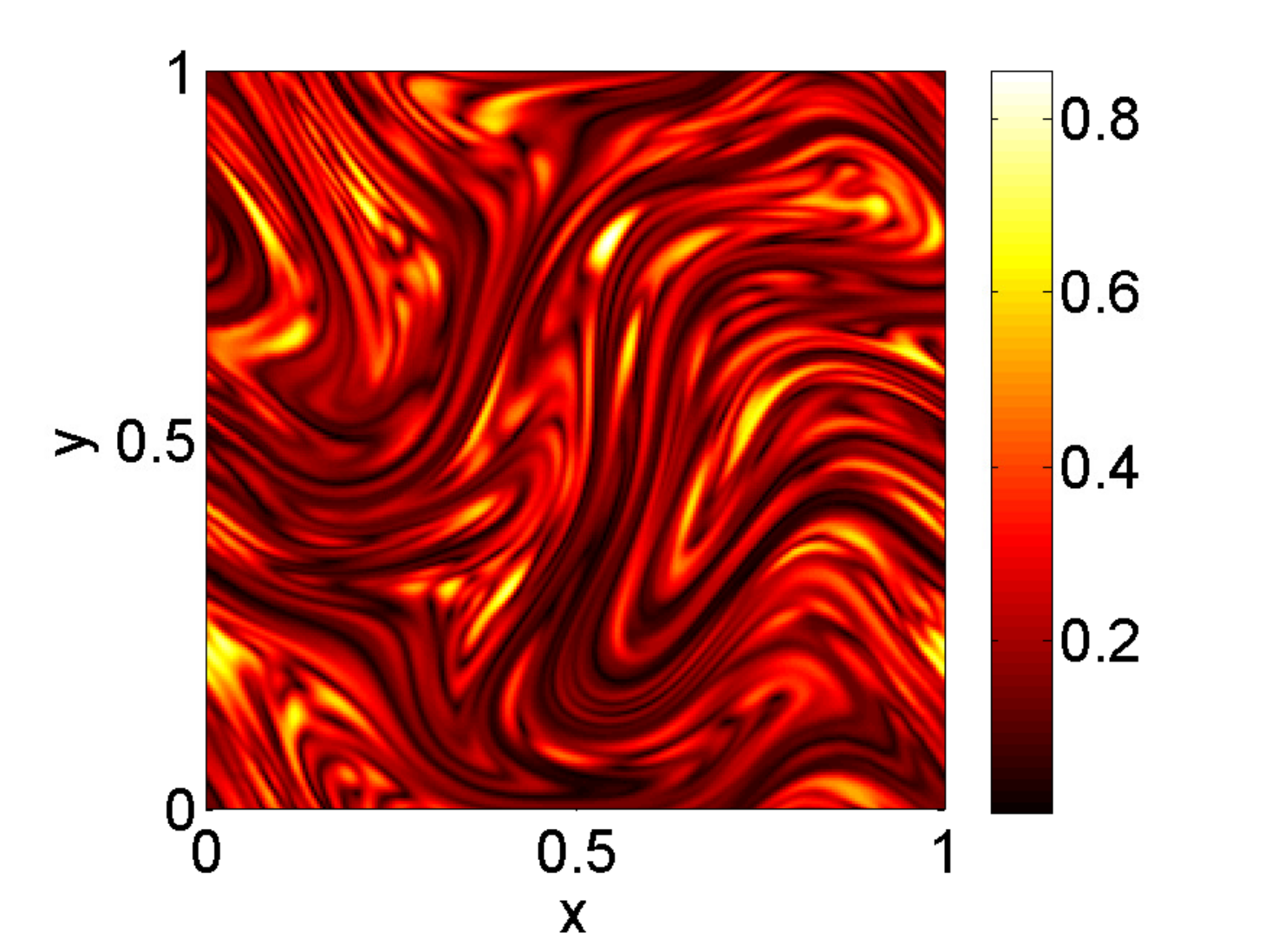}}
		\subfigure[$\,\,t=200$]{\includegraphics[width=0.32\textwidth]{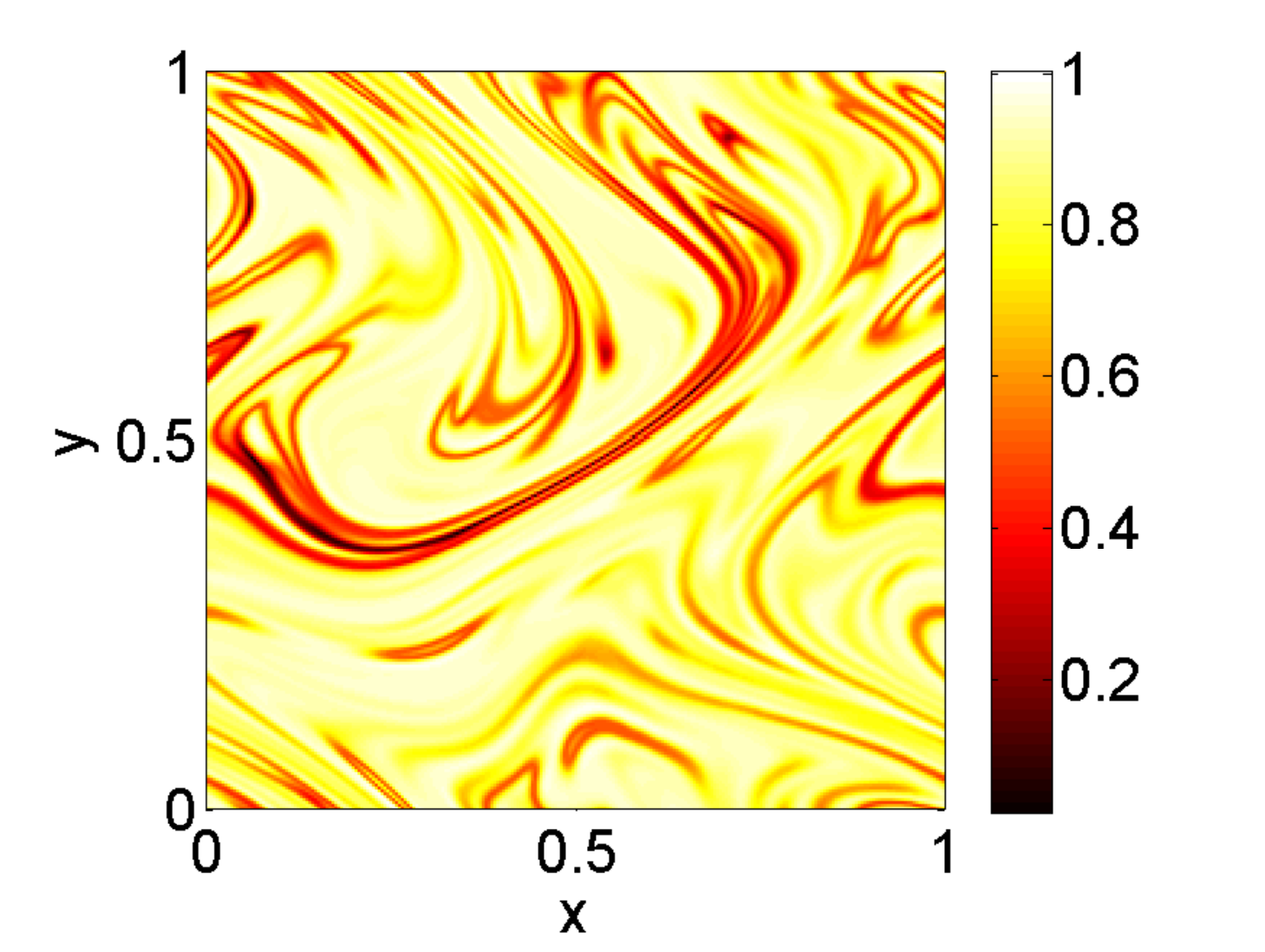}}
		\subfigure[$\,\,t=400$]{\includegraphics[width=0.32\textwidth]{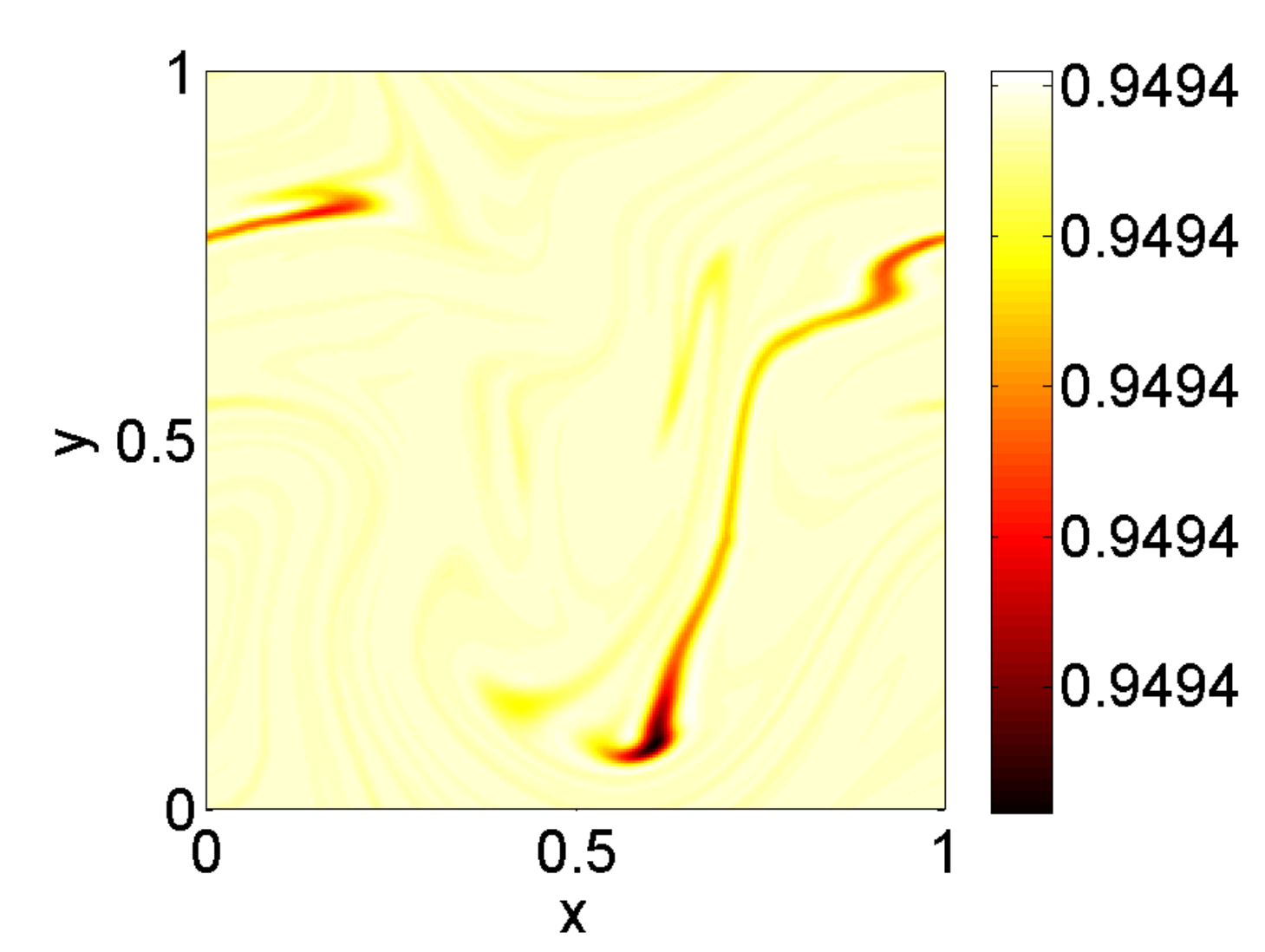}}\\
		\subfigure[$\,\,t=50$]{\includegraphics[width=0.32\textwidth]{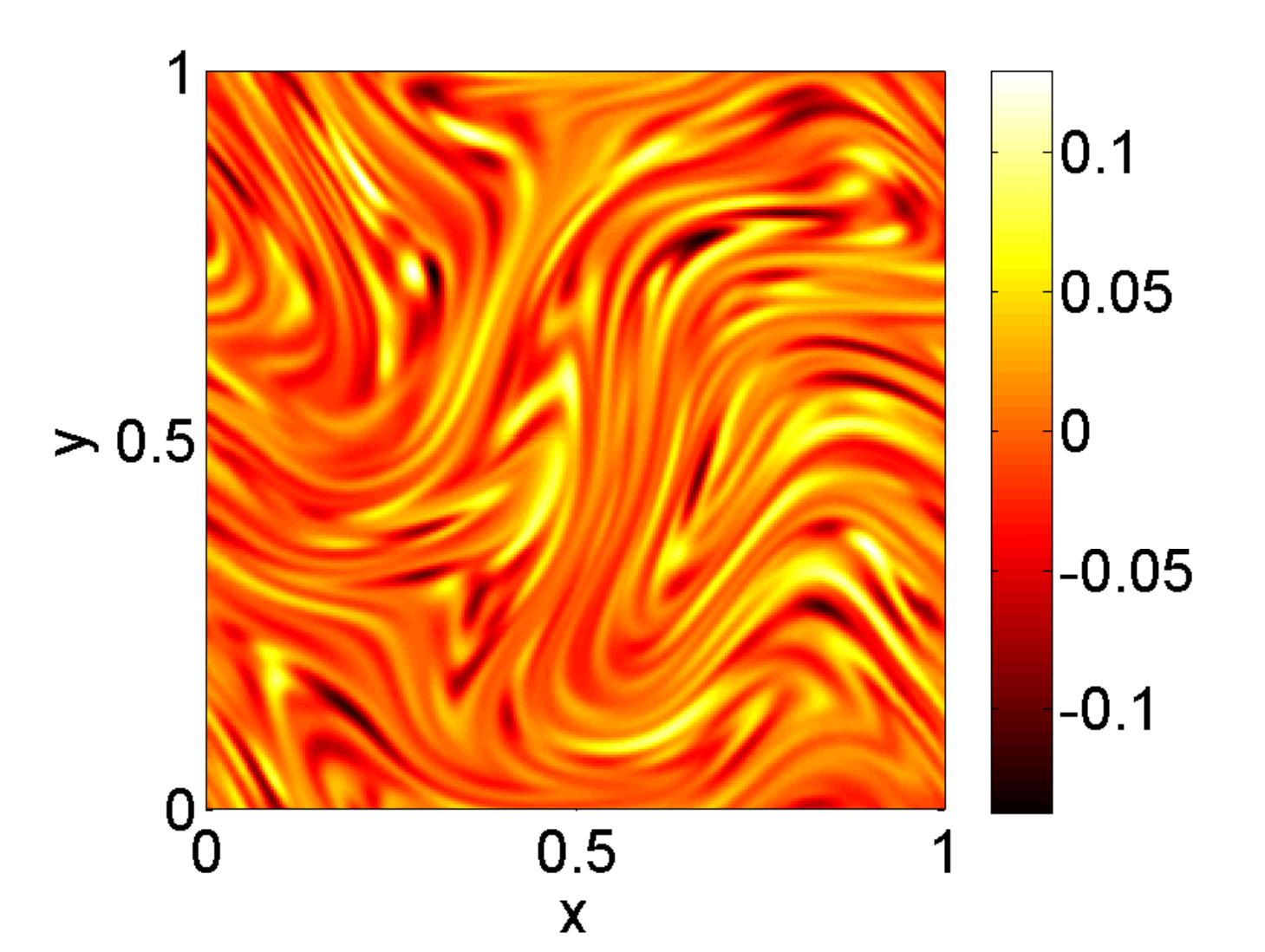}}
		\subfigure[$\,\,t=200$]{\includegraphics[width=0.32\textwidth]{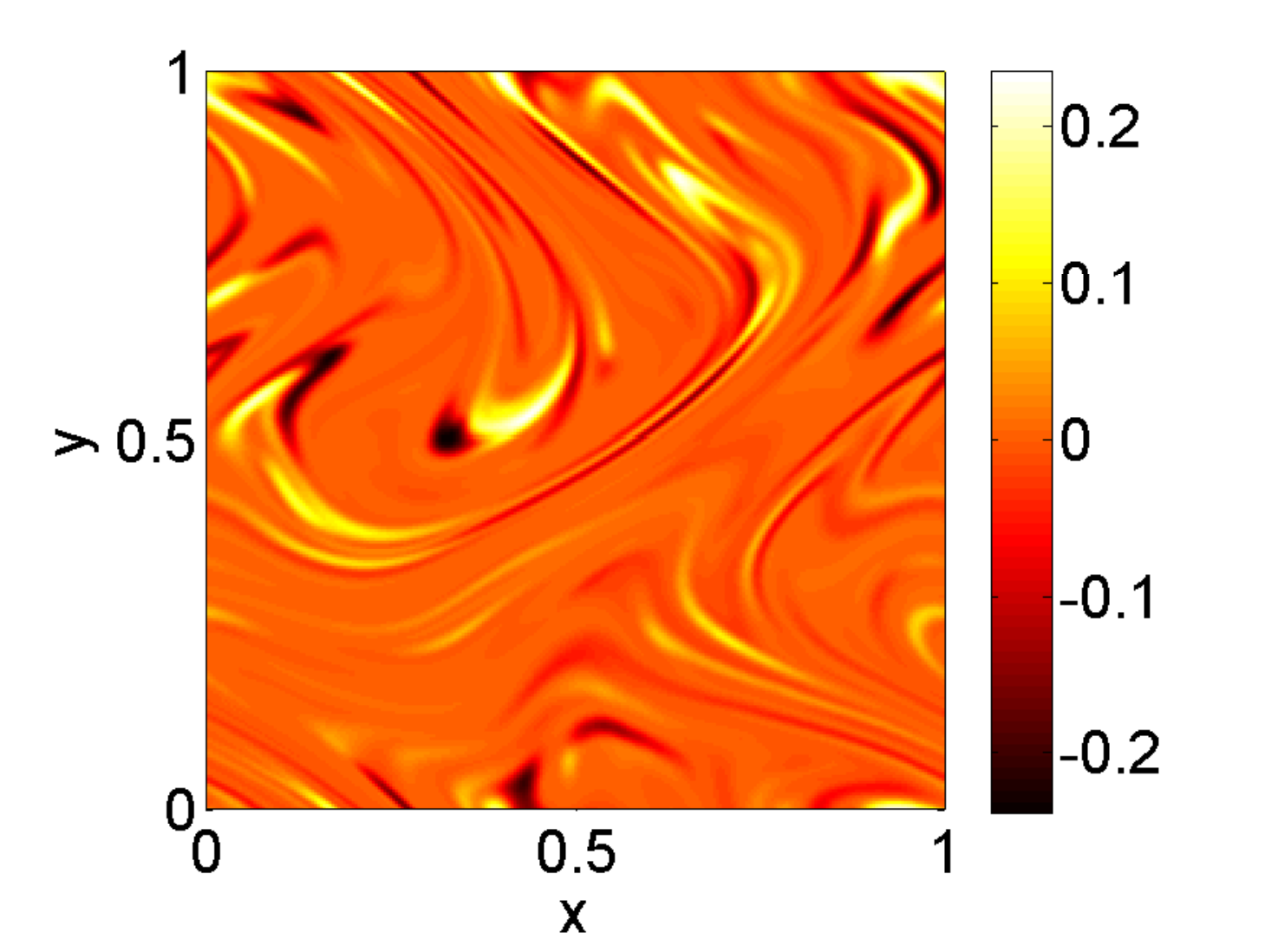}}
		\subfigure[$\,\,t=400$]{\includegraphics[width=0.32\textwidth]{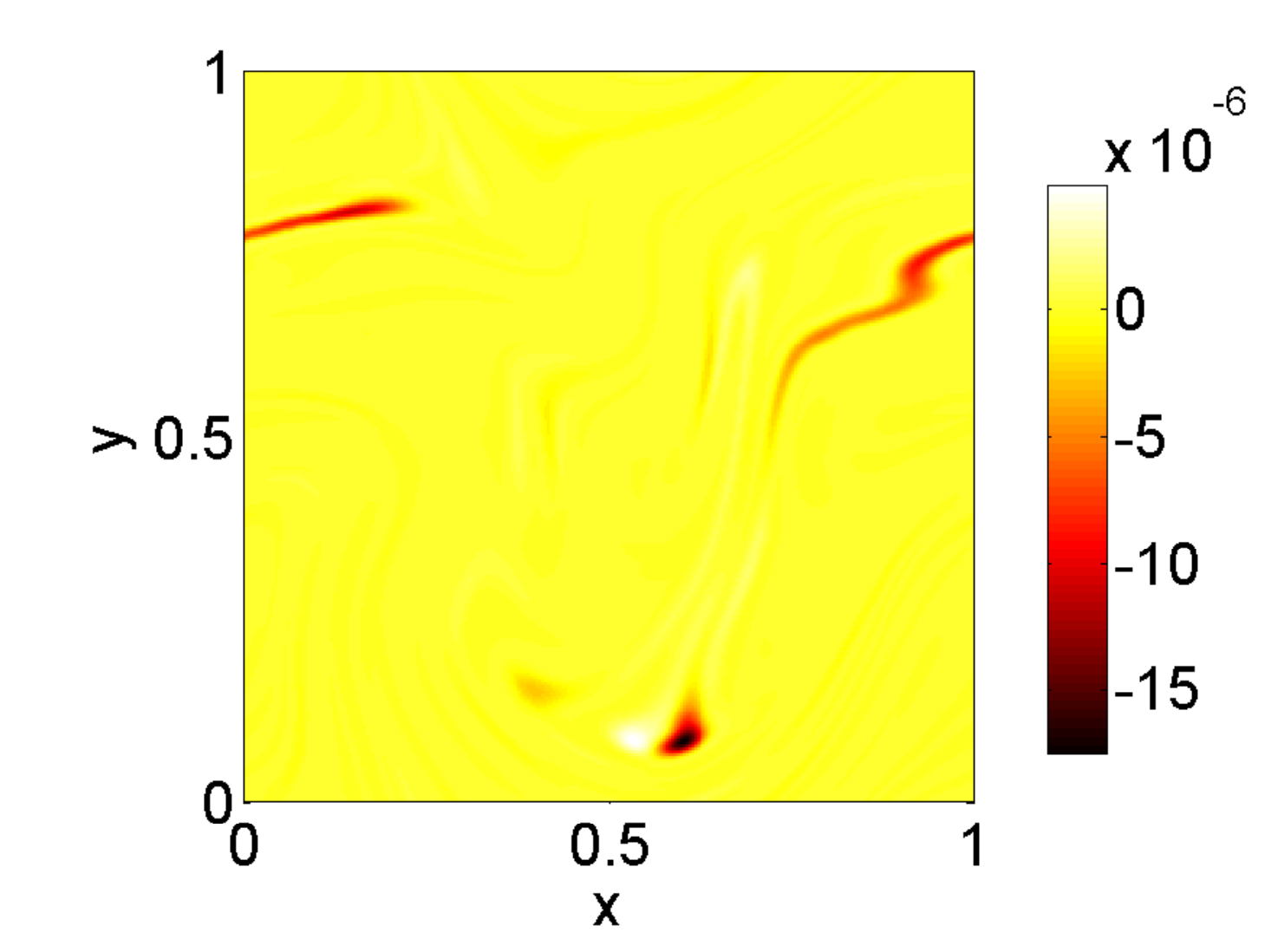}}
		\caption{Across the top: snapshots of scalar order parameter at various times for $A\tau=0.04$ and $\mytu=0$, for the random-phase sine flow.  Across the bottom: corresponding snapshots of $r$.    The compressed colour bars in panel~(c) is a consequence of the rapid relaxation to the uniform steady state.}
	\label{fig:randomphase_r_op}
\end{figure}

Consideration is also given to other model flows, starting with the random-phase sine flow.
The effect on the $Q$-tensor of stirring by the random-phase sine flow is 
 encapsulated in Figure~\ref{fig:Rt_random}, which shows a time series of the spatially-averaged $r$ field $R(t):=L_x^{-1}L_y^{-1}\iint r^2\mathd x\mathd y$ for various values of $A\tau$.
It can be seen that $R(t)$ decays exponentially to zero for all chosen values of $A\tau$, indicating that the biaxial fixed point is favoured in all scenarios.  Snapshots of the scalar order parameter and the $r$-component of the $Q$-tensor confirm that the system does indeed relax to a uniform state corresponding to the biaxial fixed point (Figure~\ref{fig:randomphase_r_op}).

\begin{figure}
	\centering
		\includegraphics[width=0.6\textwidth]{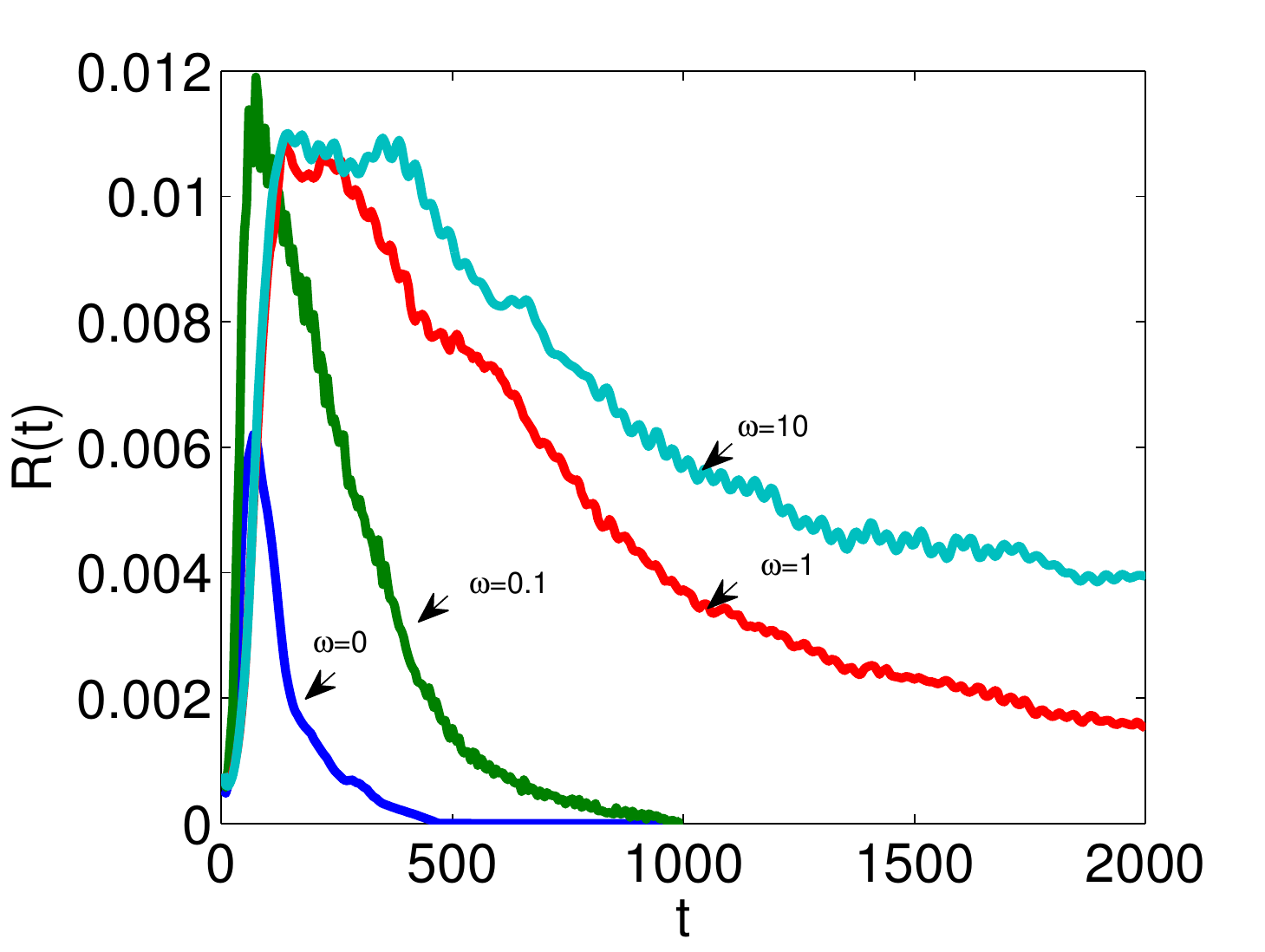}
		\caption{Time evolution of $R(t)$ for various values of $\omega$, for the oscillating cell flow whose streamfunction is given by Equation~\eqref{eq:mypsi}.  Flow parameters: $A_0=0.05/(2\pi)$, $A_1=\pi$, $k_x=6\pi/L_x$, $k_y=2\pi/L_y$, with $L_x=L_y=1$.   The simulations again use periodic boundary conditions in both spatial directions.}
	\label{fig:Rt_omega_all}
\end{figure}

\begin{figure}
	\centering
		%
		%
				\subfigure[$\,\,\omega=0,t=400$]{\includegraphics[width=0.22\textwidth]{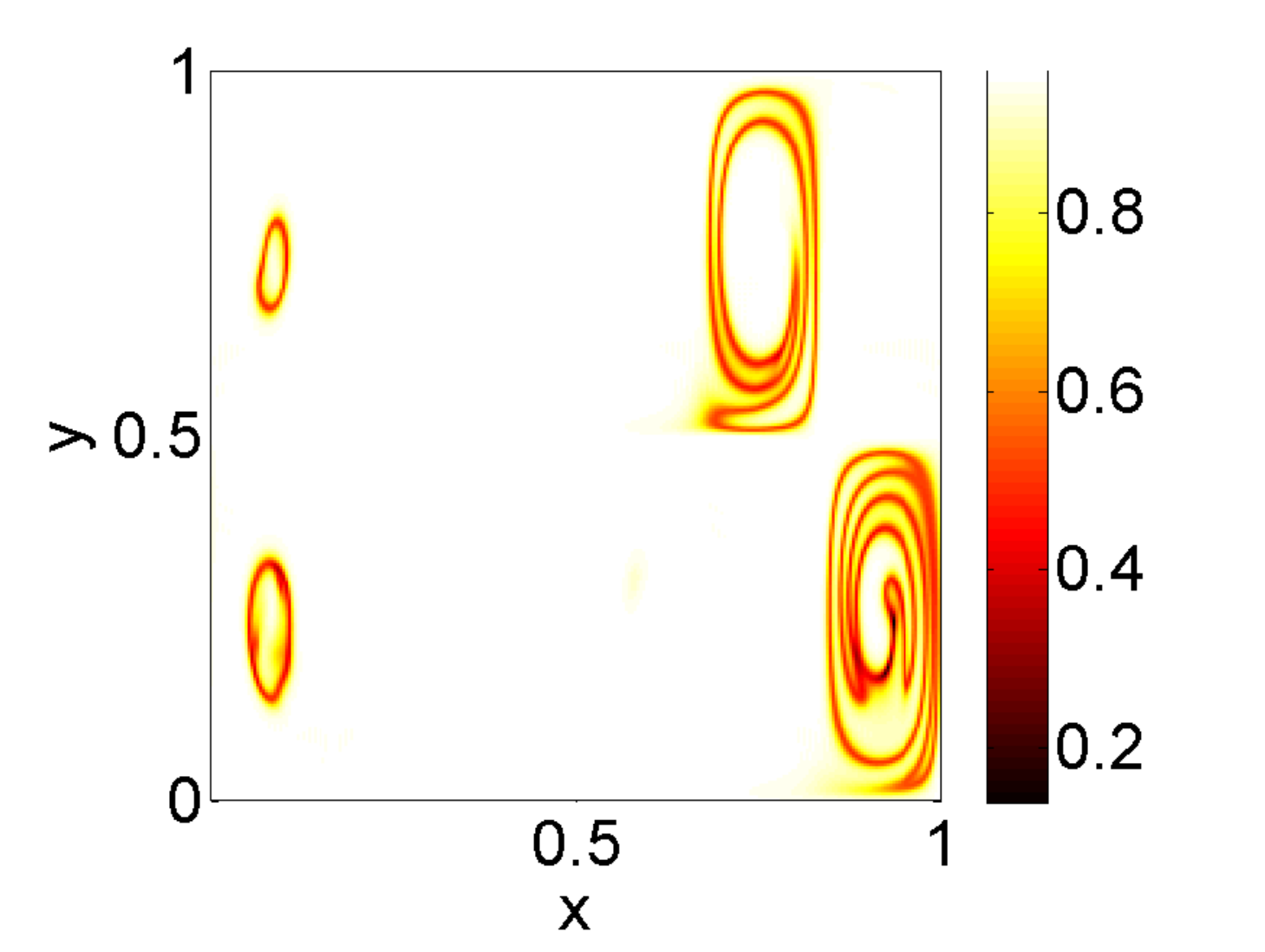}}
		\subfigure[$\,\,\omega=0.1,t=800$]{\includegraphics[width=0.22\textwidth]{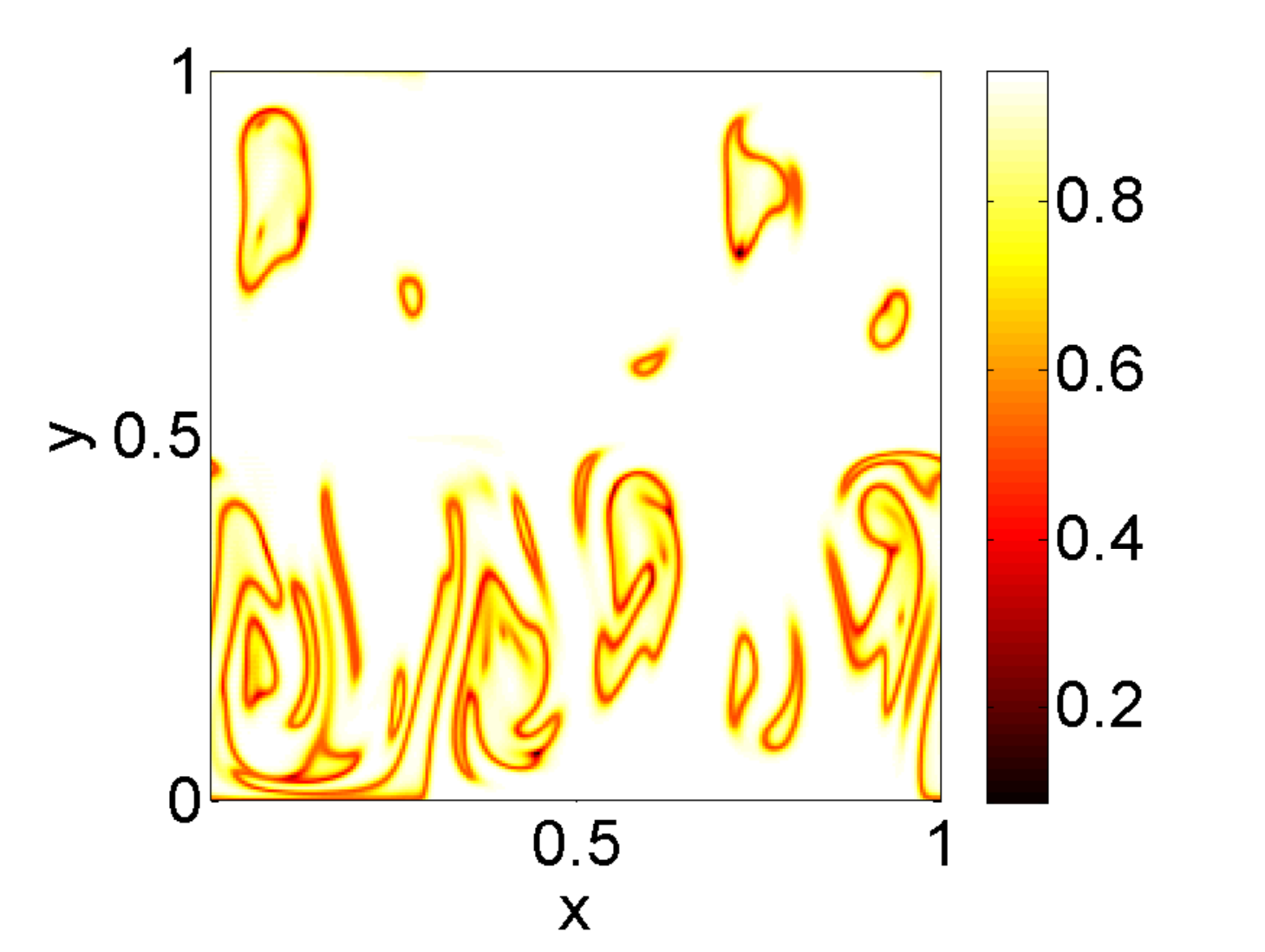}}
		\subfigure[$\,\,\omega=1,t=1500$]{\includegraphics[width=0.22\textwidth]{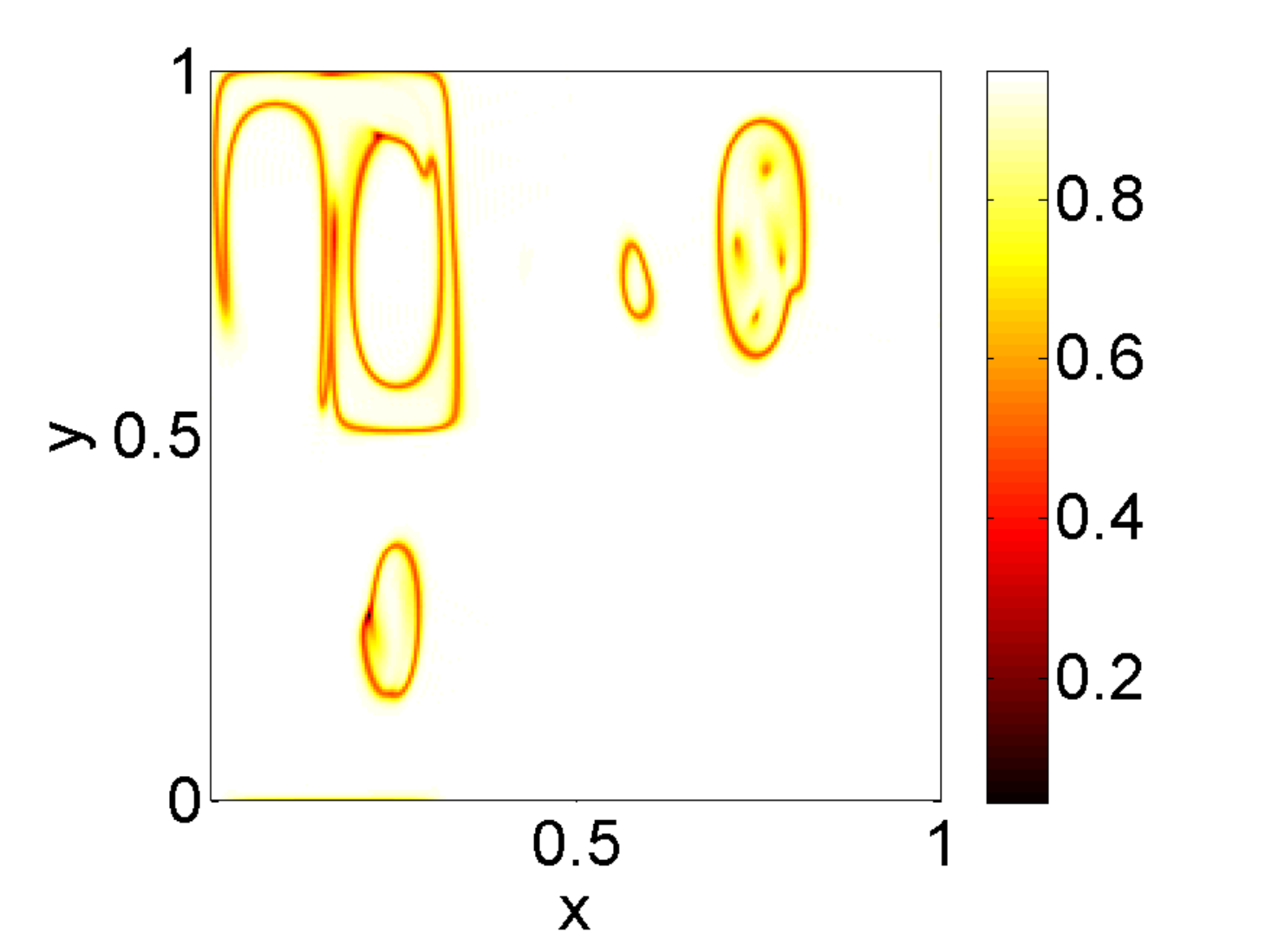}}
		\subfigure[$\,\,\omega=10,t=1500$]{\includegraphics[width=0.22\textwidth]{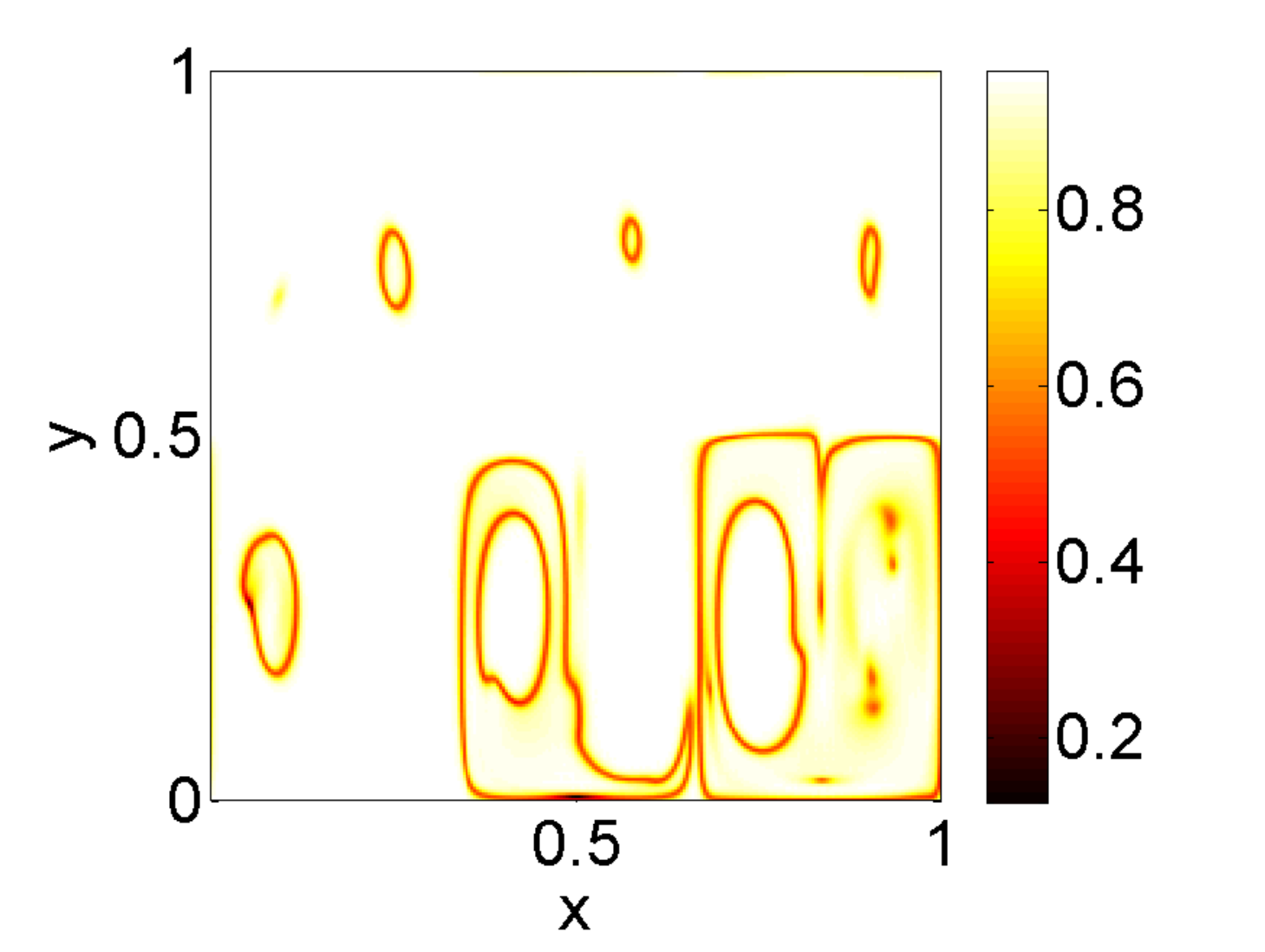}}\\
		\subfigure[$\,\,\,\omega=0,t=400$]{\includegraphics[width=0.22\textwidth]{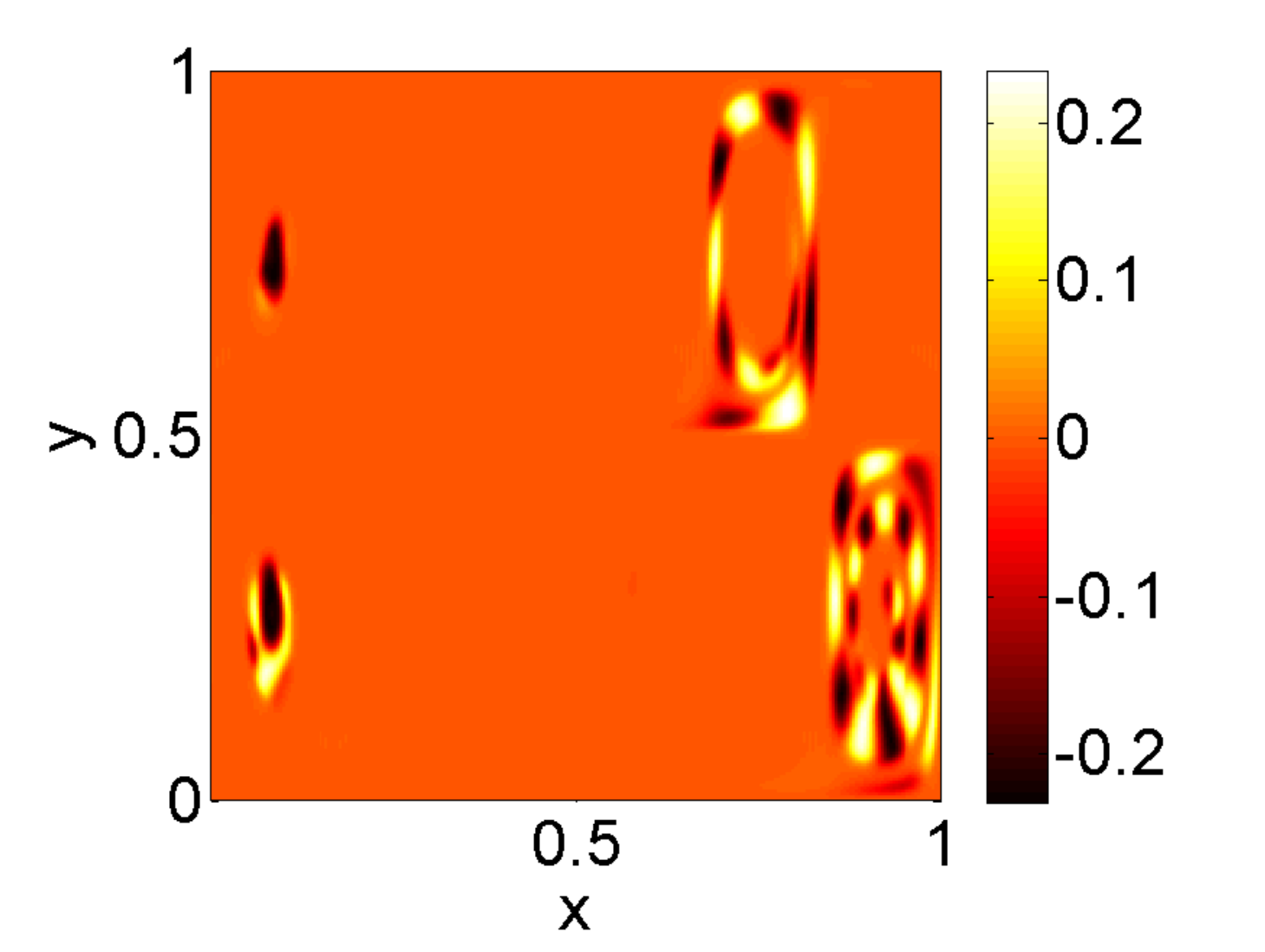}}
		\subfigure[$\,\,\omega=0.1,t=800$]{\includegraphics[width=0.22\textwidth]{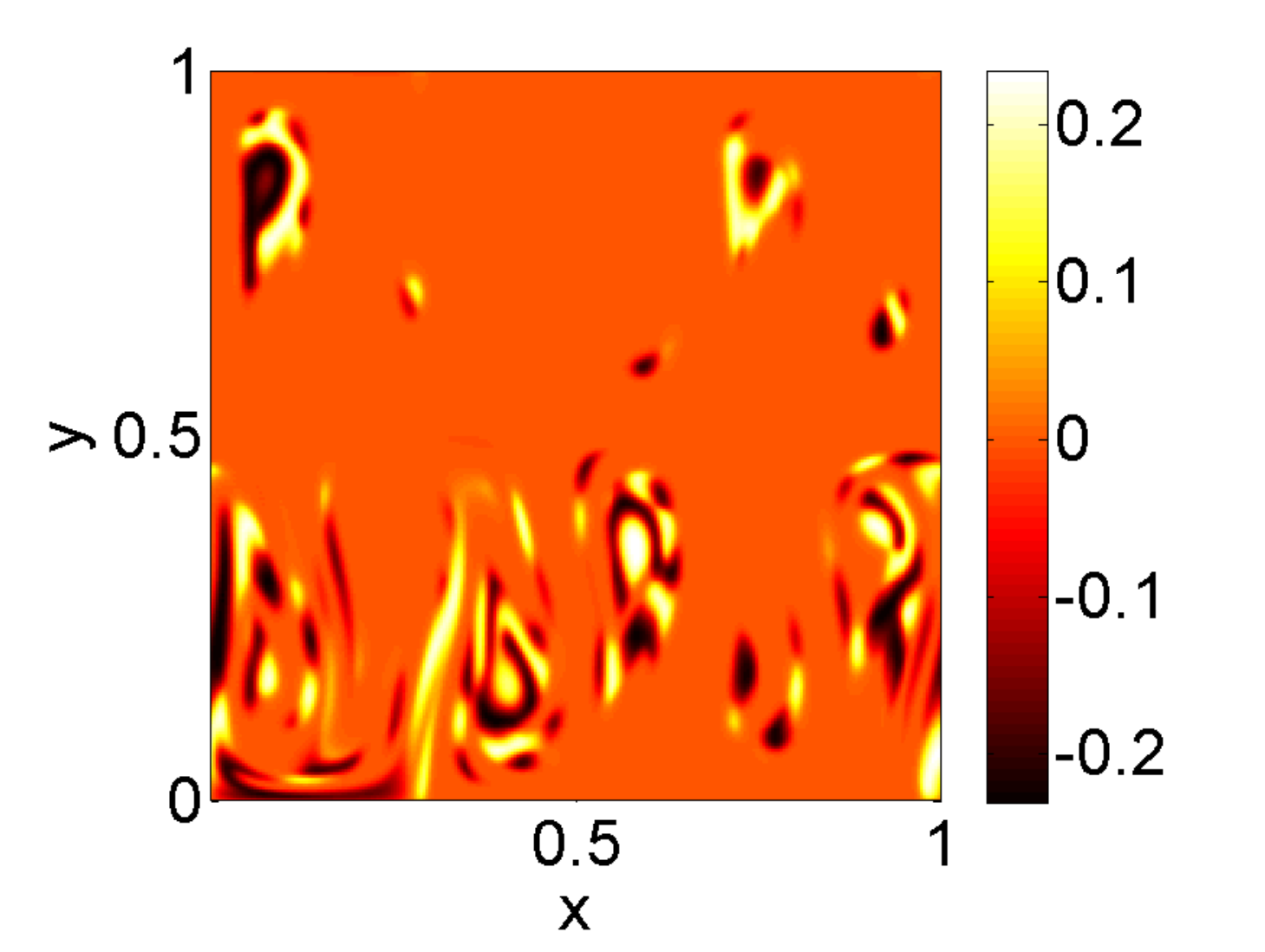}}
		\subfigure[$\,\,\omega=1,t=1500$]{\includegraphics[width=0.22\textwidth]{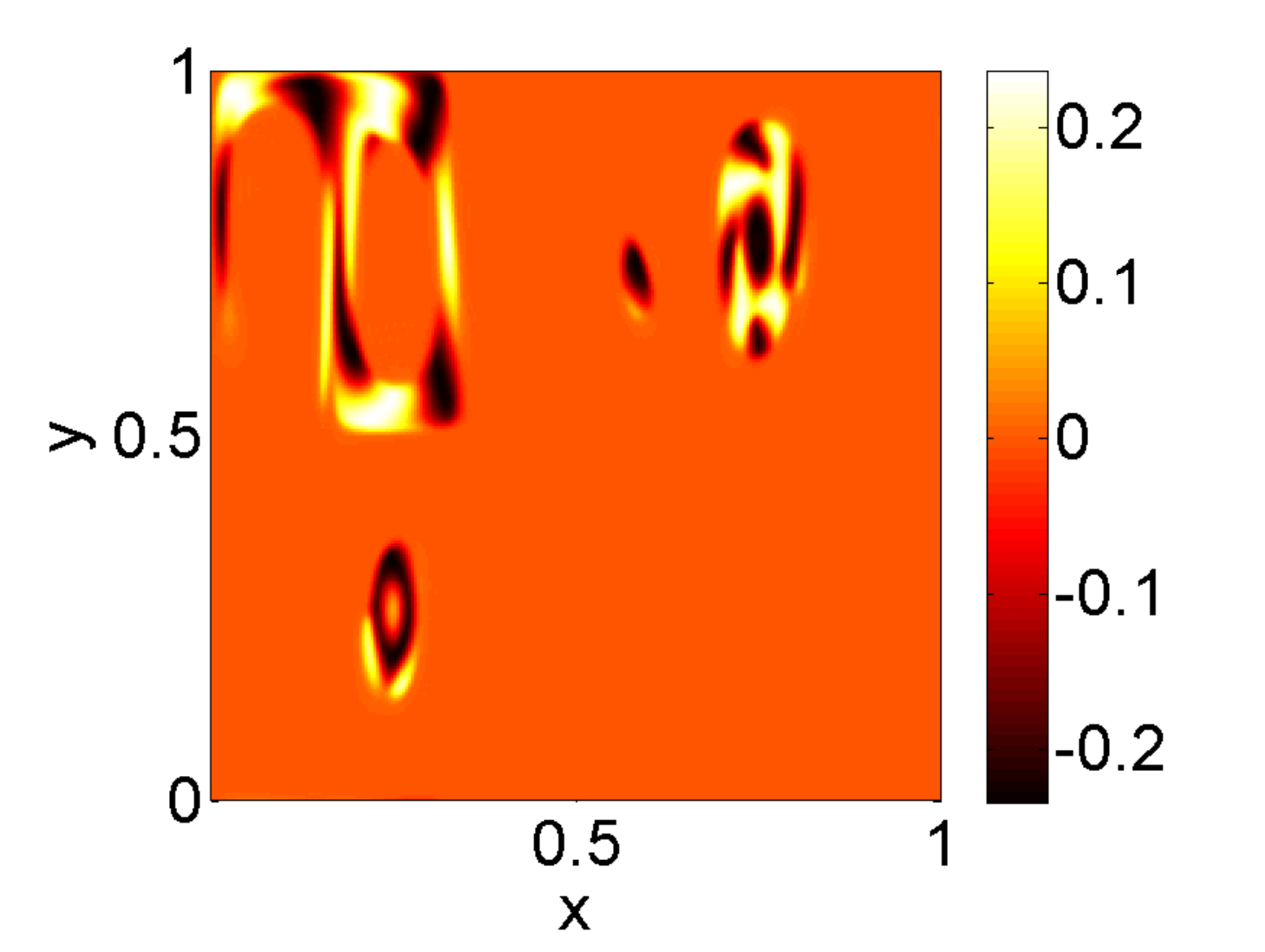}}
		\subfigure[$\,\,\omega=10,t=1500$]{\includegraphics[width=0.22\textwidth]{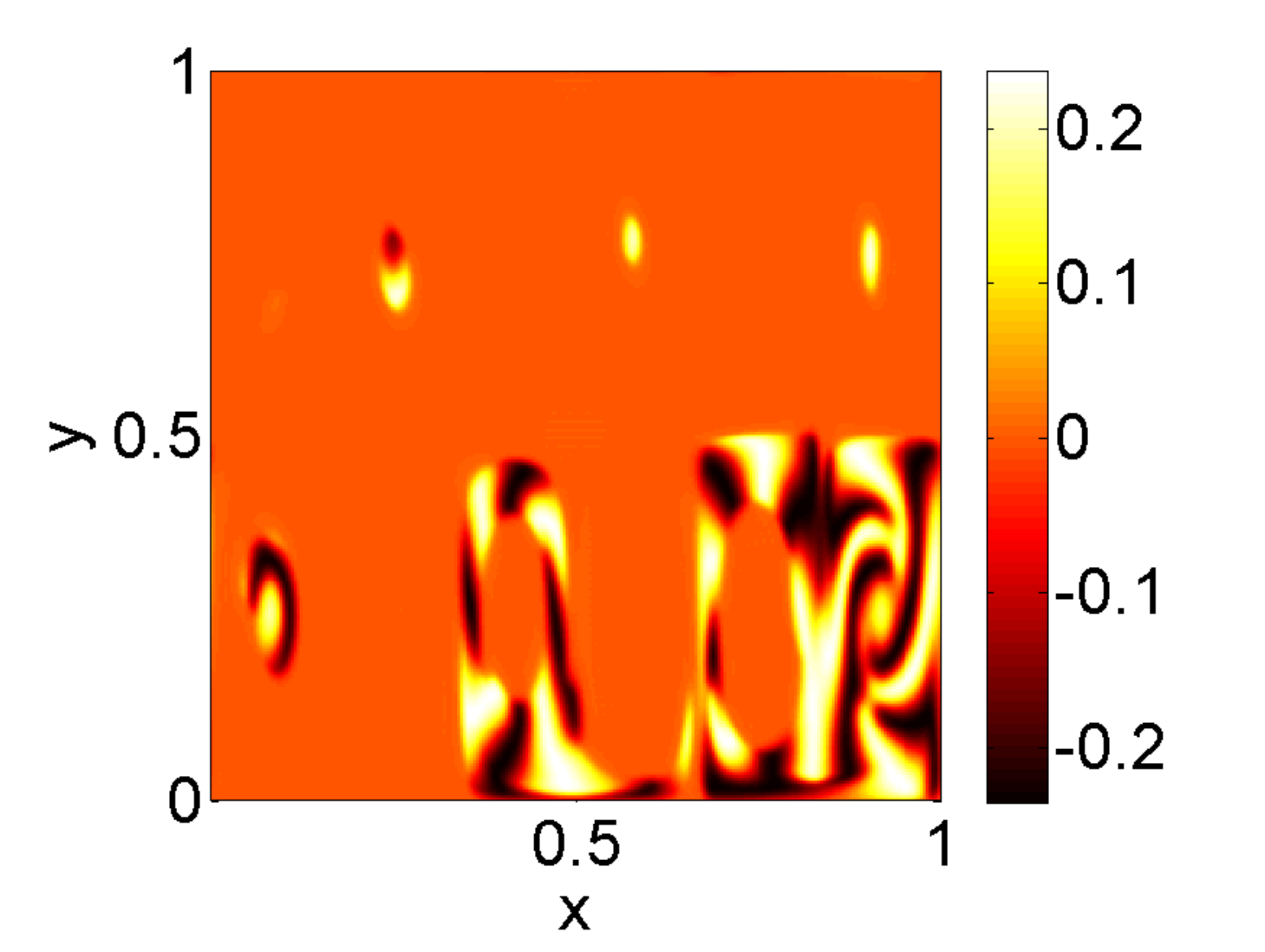}}\\
		\caption{Across the top: snapshots of scalar order parameter for various values of $\omega$ and for $\mytu=0$, for the oscillating cell flow.  
		Across the bottom: corresponding snapshots of $r$.}
	\label{fig:op_omega_all}
\end{figure}

Consideration is also given to an oscillating cell flow which can be realised in the lab setting~\cite{solomon2003lagrangian}, such that $u=\partial\psi/\partial y$ and $v=-\partial\psi/\partial x$, where $\psi(x,y,z)$ is the time-dependent streamfunction
\begin{equation}
\psi(x,y,z,t)=A_0\sin\left(k_x x+A_1\cos(\omega t)\right)\sin(k_y y).
\label{eq:mypsi}
\end{equation}
It is emphasized that Equation~\eqref{eq:mypsi} is applied using periodic boundary conditions in both spatial directions, whereas in the laboratory setting, walls matter.  However, since the focus on the present case is on the $Q$-tensor dynamics in the bulk of the sample (i.e. far from walls and wall effects), the choice of periodic boundary conditions is thereby justified.
As such, for $A_1=0$ the flow is time-dependent and regular, such that Lagrangian trajectories are confined within periodic cells; the presence of the time-dependent term causes particles near cell boundaries to move into neighbouring cells thereby making the flow chaotic.  Accordingly, Figure~\ref{fig:Rt_omega_all} shows the time series of $R(t)$ for various values of the forcing frequency $\omega$ (the other flow parameters are given in the caption).  
The figure indicates that the system relaxes to a uniform biaxial state, the relaxation is slower for larger values of $\omega$ corresponding to shorter flow periods.  There is therefore some consistency between these results and the earlier results for the constant-phase sine flow, wherein model flows with a shorter flow period did not relax entirely to the biaxial fixed point.  Snapshots of the scalar order parameter and the corresponding snapshots of $r(x,y,t)$ (Figure~\ref{fig:op_omega_all}) confirm this description.
The snaphshots further illustrate the slow convergence of the $Q$-tensor dynamics to the uniform biaxial fixed point under the model cell flow: mixed uniaxial/biaxial domains persist until late time, especially for $\omega=1$ and $\omega=10$.  However, the ultimate state in all considered cases is for the system to relax uniformly to the only (stable) fixed point that exists in the $Q$-tensor equations with flow, this being the biaxial fixed point.

Summarizing, the picture concerning the effects of flow on the $Q$-tensor dynamics are mixed:
\begin{itemize}
\item For the constant-phase sine flow, the domain structure obtained in the unstirred case persists when the flow period is very short (specifically, $\tau\ll 1$); namely, the system exhibits an interpenetrating domain structure consisting of coherent biaxial regions but also of further distinct mixed uniaxial/biaxial regions.  In all other cases, the system relaxes ultimately to a stable biaxial fixed point, uniform in space.
\item For the random-phase sine flow the system relaxes rapidly to a uniform biaxial state; the mixed domains are no longer present.
\item For the oscillating cell flow, the system again relaxes to the stable biaxial fixed point; the relaxation is most slowest for the short flow periods.
\end{itemize}
The aim of the next section is to place this picture into a wider theoretical framework.

\section{Discussion and Conclusions}
\label{sec:discussion}

In this section the results concerning the flow in the absence of tumbling are placed in a theoretical context.  
The approach is to consider Equations~\eqref{eq:qtensor_2d} in the absence of diffusion.  This is justified in a first approximation since $\epsilon^2$ is small, meaning that diffusion can be neglected over times that are short compared to the flow timescale $L/A$.  Then, Equation~\eqref{eq:qtensor_2d} can be regarded as as a set of rate equations along Lagrangian trajectories:
\begin{equation}
\frac{\mathd}{\mathd t}\left(\begin{array}{c}q\\r\\s\end{array}\right)=
\left(\begin{array}{c}F_1(q,r,s)\\F_2(q,r,s)\\F_3(q,r,s)\end{array}\right)+\left(\begin{array}{c}2r\Omega_{12}\\\Omega_{12}(s-q)\\-2r\Omega_{12}\end{array}\right),\qquad \mytu=0.
\label{eq:qtensor_2d_rde}
\end{equation}
where $(F_1,F_2,F_3)^T$ encode the $Q$-tensor dynamics, and where $\mathd/\mathd t$ is the Lagrangian derivative along particle trajectories.  It is clear that the stable biaxial fixed point $r=0$ and $s=q$ is preserved by the addition of the co-rotational term in Equation~\eqref{eq:qtensor_2d_rde}.  The other stable (uniaxial) fixed points are not preserved; this can be shown rigorously 
(e.g. Appendix~\eqref{sec:app:fixedpoints}).  Therefore, unless a prescribed flow has highly specific characteristics, such that $\Omega_{12}=0$ identically, or such that $\Omega_{12}$ is effectively zero along trajectories, the $Q$-tensor will relax towards the stable biaxial fixed point along trajectories.  Allowing again for a small amount diffusion, its presence in the relevant $Q$-tensor equations will only accelerate the relaxation to the stable biaxial fixed point, by smoothing out small-scale spatial variations in the $Q$-tensor components.  The focus for the remainder of this section therefore is to investigate the properties of the co-rotation term $\Omega_{12}$ for the different model flows, and to relate this to the evolution of the $Q$-tensor under the relevant flow.  

Some general discussion is possible concerning those flows for which all fixed points (whether uniaxial or biaxial) survive in the presence of the co-rotational term in Equation~\eqref{eq:qtensor_2d_rde}.  For cases where $\Omega_{12}=0$ identically (i.e. irrotational flows), this discussion is trivial.  Therefore, consideration is given to flows with rotation for which $\Omega_{12}$ is \textit{effectively zero} along trajectories, according to the following prescription.
Starting with Equation~\eqref{eq:qtensor_2d_rde},
 it can be seen that $q$, $r$, and $s$ vary on multiple timescales.   First, because $(F_1,F_2,F_3)$ are $O(1)$ quantities, the time derivative $\mathd/\mathd t$ induces temporal variations in $q$, $r$, and $s$ that are themselves $O(1)$ in the dimensionless timescale of the problem.  Equally, the terms proportional to $\Omega_{12}$ induce a second independent temporal variation in $q$, $r$, and $s$ that takes place on the timescale set by the applied flow.  Thus, for specific cases involving periodic or quasi-periodic flows with period $\tau$, the $Q$-tensor will vary on independently on an $O(1)$ timescale but also on an $O(\tau)$ timescale.  For $\tau\ll 1$ in the dimensionless units of the problem, the flow timescale will be rapid.  For these reasons, and motivated by a similar scenario in advection-diffusion~\cite{Lapeyre2002,Lapeyre1999},  a rotating-wave-type approximation is made in the limit as $\tau\rightarrow 0$: $\Omega_{12}$ is thereby replaced by its average value over a Lagrangian trajectory, 
\[
\langle\Omega_{12}\rangle(\vecx_0)=\lim_{T\rightarrow\infty}\frac{1}{T}\int_0^T \Omega_{12}(\vecx(t),t)\mathd t,\qquad \frac{\mathd\vecx}{\mathd t}=\vecv(\vecx(t),t),\qquad \vecx(t=0)=\vecx_0.
\]
For those flows where $\langle\Omega_{12}\rangle(\vecx_0)=0$ in parts of the flow domain, 
and for $\tau\rightarrow 0$, the co-rotational term on the right-hand side of Equation~\eqref{eq:qtensor_2d_rde} becomes negligible, meaning that the fixed points of Equation~\eqref{eq:qtensor_2d_rde} are the same as the fixed points of the corresponding system without flow, i.e. the uniaxial and biaxial families of fixed points previously introduced in Section~\ref{sec:fixed}.   

Because diffusion does not alter the fixed points of the unadvected system of equations, these conclusions carry over to scenarios where the diffusion is small but finite, with one caveat: if 
$\langle\Omega_{12}\rangle(\vecx_0)=0$ only in small isolated parts of the flow domain, then the mixed uniaxial/biaxial domain structures will only persist in such small regions, which may eventually by smoothened out by the effects of diffusion, such that a homogeneous final state is favoured.  Since only the biaxial fixed point corresponds to a homogeneous state (as the multiplicity of stable uniaxial fixed points give rise to an inhomogeneous domain structure), it is thereby expected that the twin effects of diffusion and co-rotation will favour the biaxial fixed point.



\begin{figure}
	\centering
		\subfigure[$\,\,A\tau=0.04$]{\includegraphics[width=0.32\textwidth]{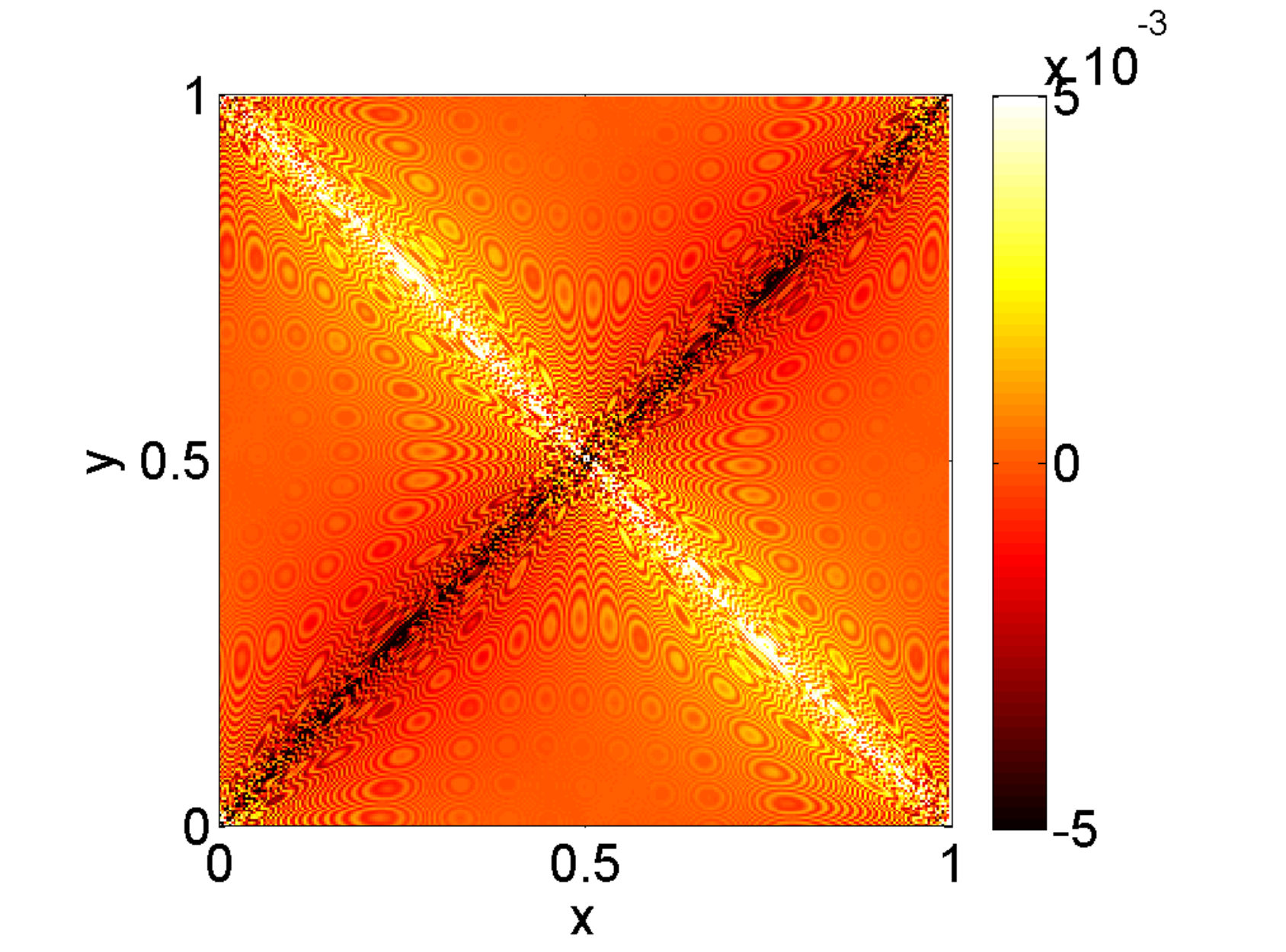}}
		\subfigure[$\,\,A\tau=0.6$]{\includegraphics[width=0.32\textwidth]{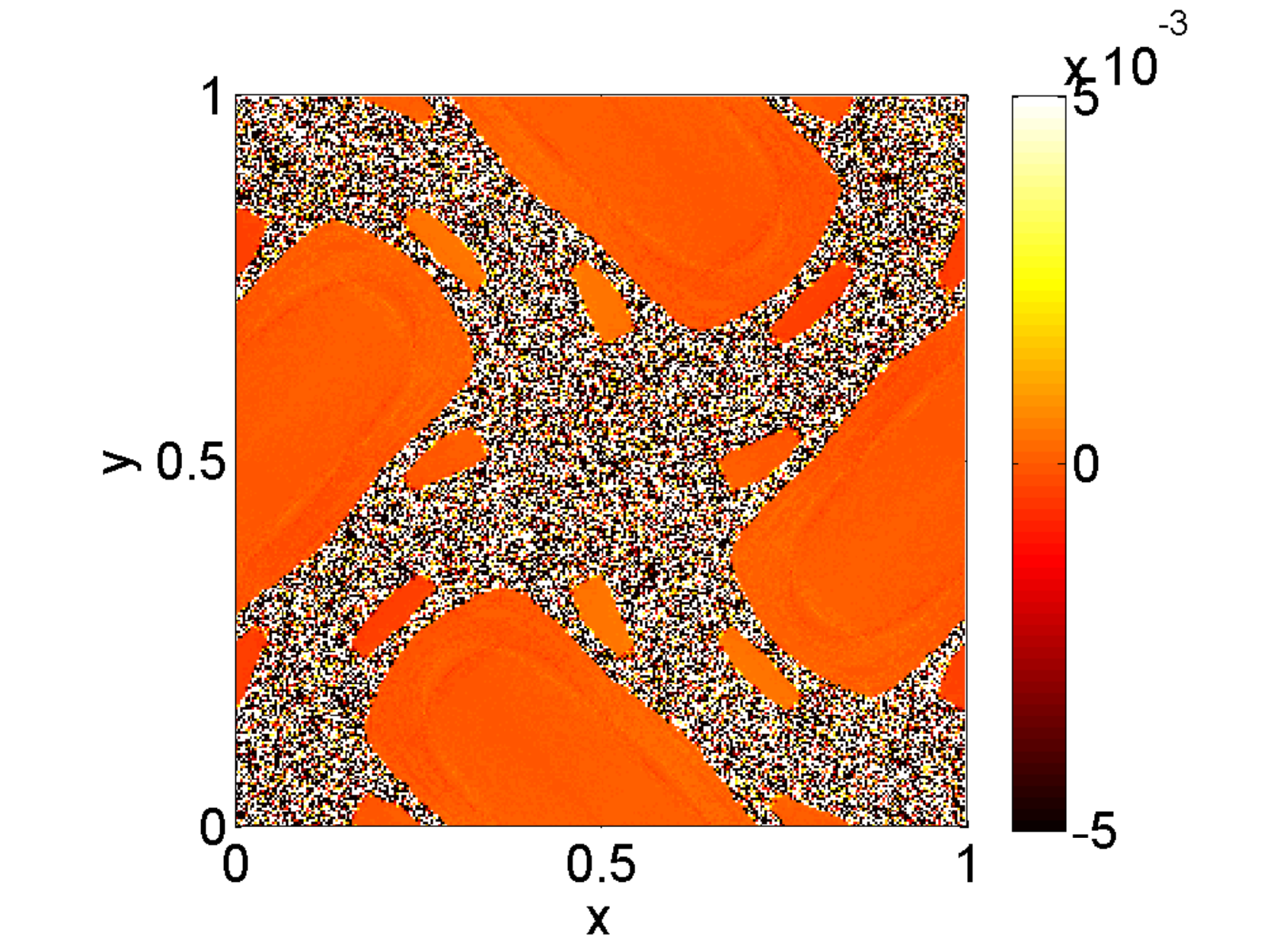}}
		\subfigure[$\,\,A\tau=1.6$]{\includegraphics[width=0.32\textwidth]{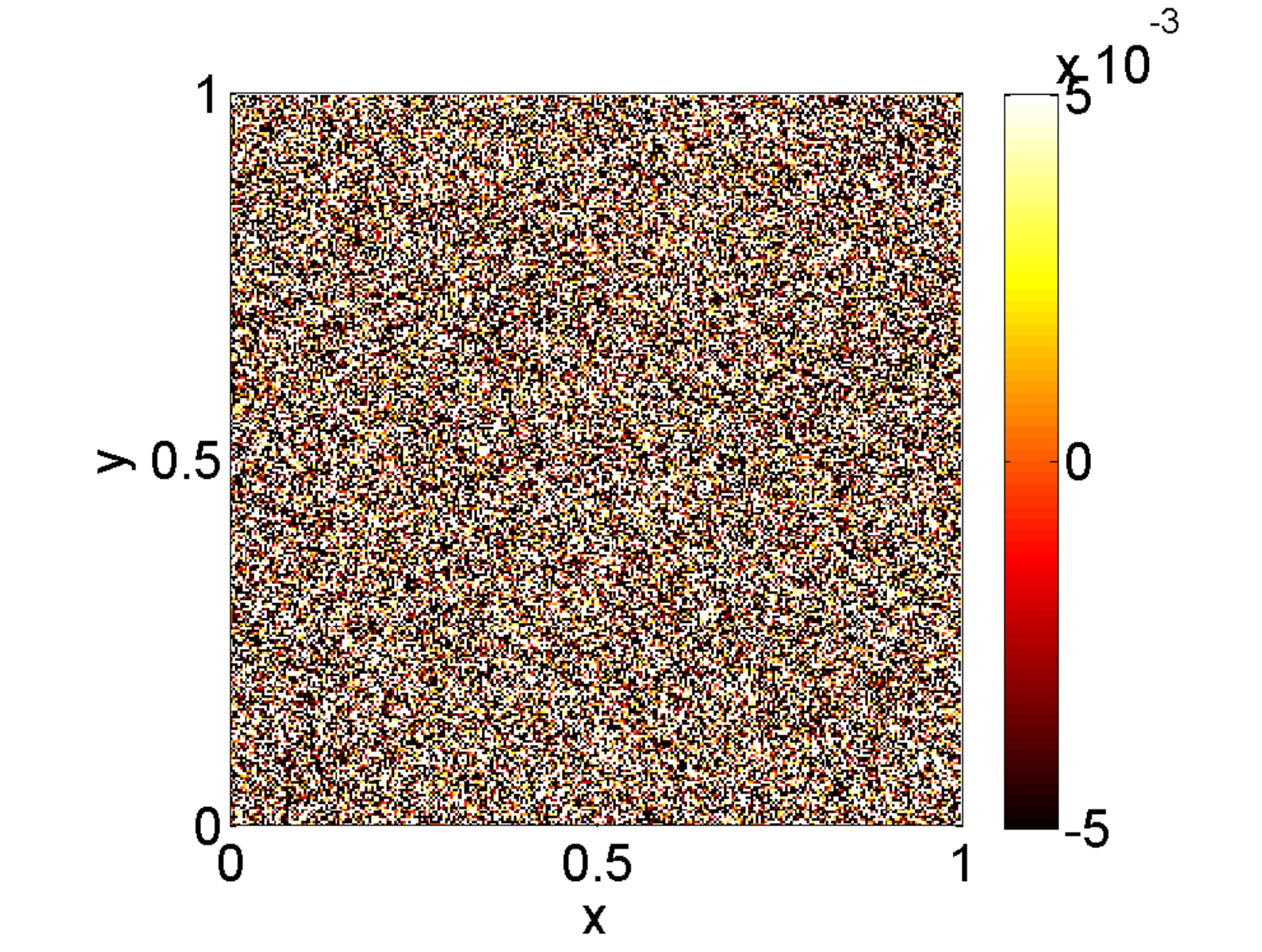}}
		\caption{The Lagrangian time average $\langle \Omega_{12}\rangle(\vecx_0)$ for the constant-phase sine flow, for various values of $A\tau$.  The time average is computed over a sufficiently long time interval such that the pictured result is independent of the time interval.}
	\label{fig:omega12_all}
\end{figure}
These ideas are now used to explain the previously observed features of the $Q$-tensor dynamics for the specific model chaotic flows.
Accordingly, the Lagrangian average $\langle \Omega_{12}\rangle(\vecx_0)$ is plotted for the constant-phase sine flow, for different values of $A\tau$ in Figure~\ref{fig:omega12_all}.    For $A\tau\leq 0.6$ the plots indicate extended regions where $\langle \Omega_{12}\rangle(\vecx_0)\approx 0$.  These regions correspond clearly to the mixed uniaxial/biaxial domain structures observed in Figure~\ref{fig:flow_no_tumble1}.
In contrast, for $A\tau=1.6$,  $\langle \Omega_{12}\rangle(\vecx_0)$ varies on (arbitrarily) small spatial scales, which is a consequence of the fully chaotic nature of the flow in this regime, whereby initially neighbouring trajectories diverge exponentially in time, meaning that the fluid properties along initially neighbouring trajectories decorrelate.  In this case, it can be expected that the co-rotational term will induce corresponding small-scale variations in $(q,r,s)$, which will enhance the effect of diffusion, thereby promoting a homogeneous final state.  This is consistent with what is observed in Figure~\ref{fig:flow_no_tumble1}(c,f), where only the biaxial fixed point survives at late times.  
The random-phase sine flow is investigated along similar lines.  In this case the, Lagrangian average $\langle \Omega_{12}\rangle(\vecx_0)$ resembles Figure~\ref{fig:omega12_all}(c) for all considered values of $A\tau$, which makes sense, since the random-phase version of the sine flow is chaotic for all values of $A$.  Correspondingly, only the biaxial fixed point survives at late time, for all considered values of $A\tau$ , as observed previously in Figures~\ref{fig:Rt_random}--\ref{fig:randomphase_r_op}.

\begin{figure}
	\centering
		\subfigure[$\,\,\omega=0$]{\includegraphics[width=0.24\textwidth]{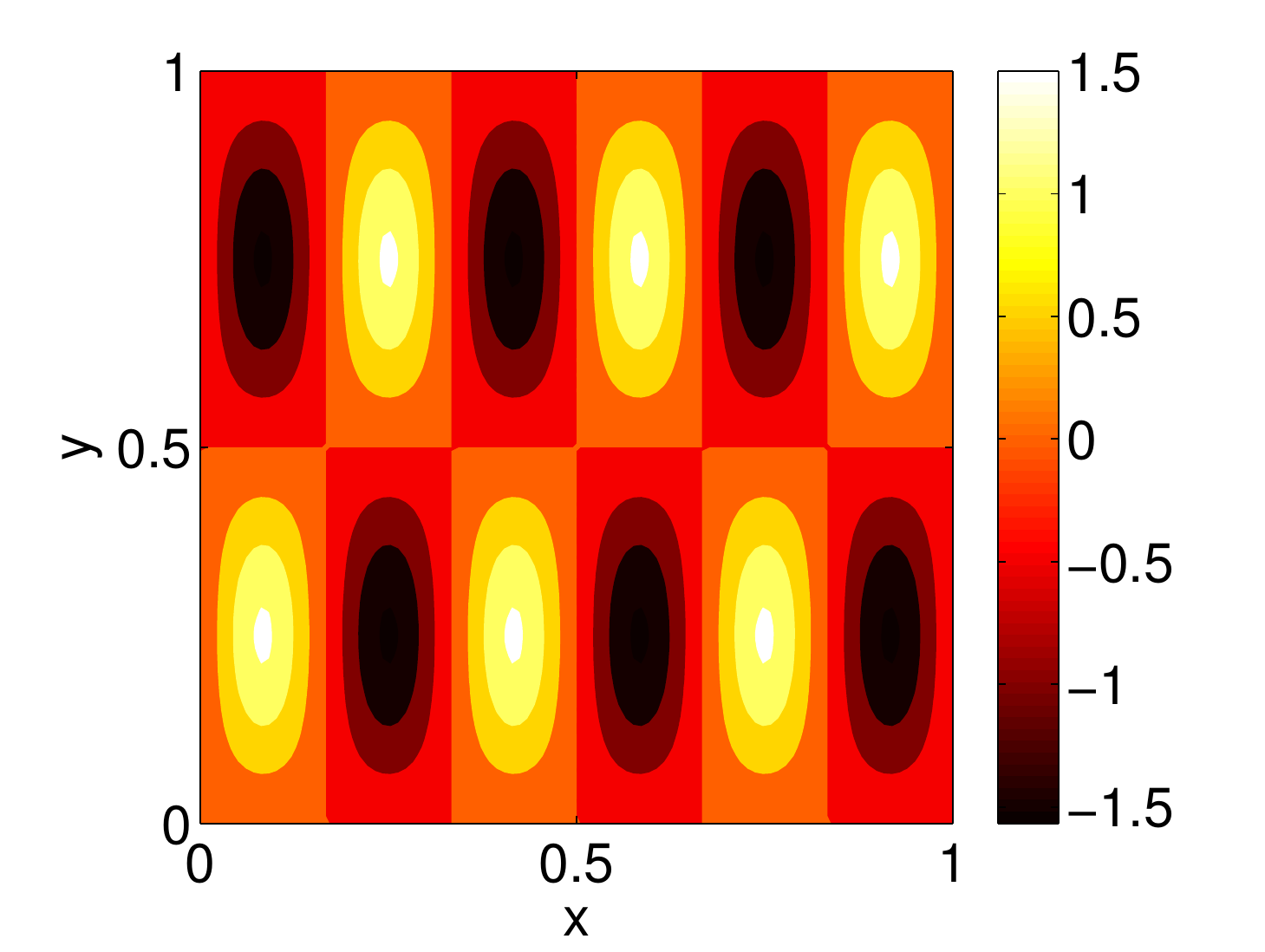}}
		\subfigure[$\,\,\omega=0.1$]{\includegraphics[width=0.24\textwidth]{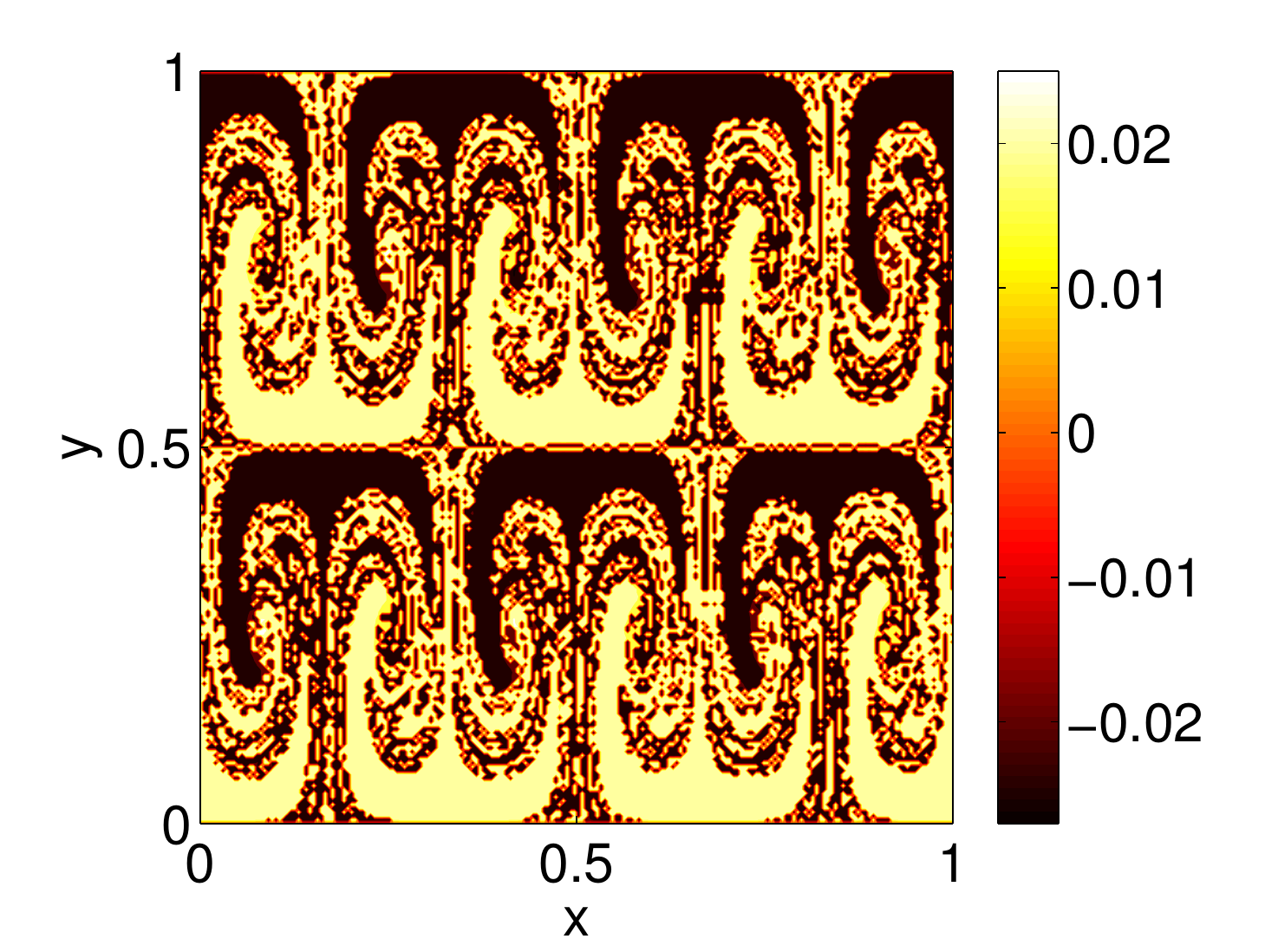}}
		\subfigure[$\,\,\omega=1$]{\includegraphics[width=0.24\textwidth]{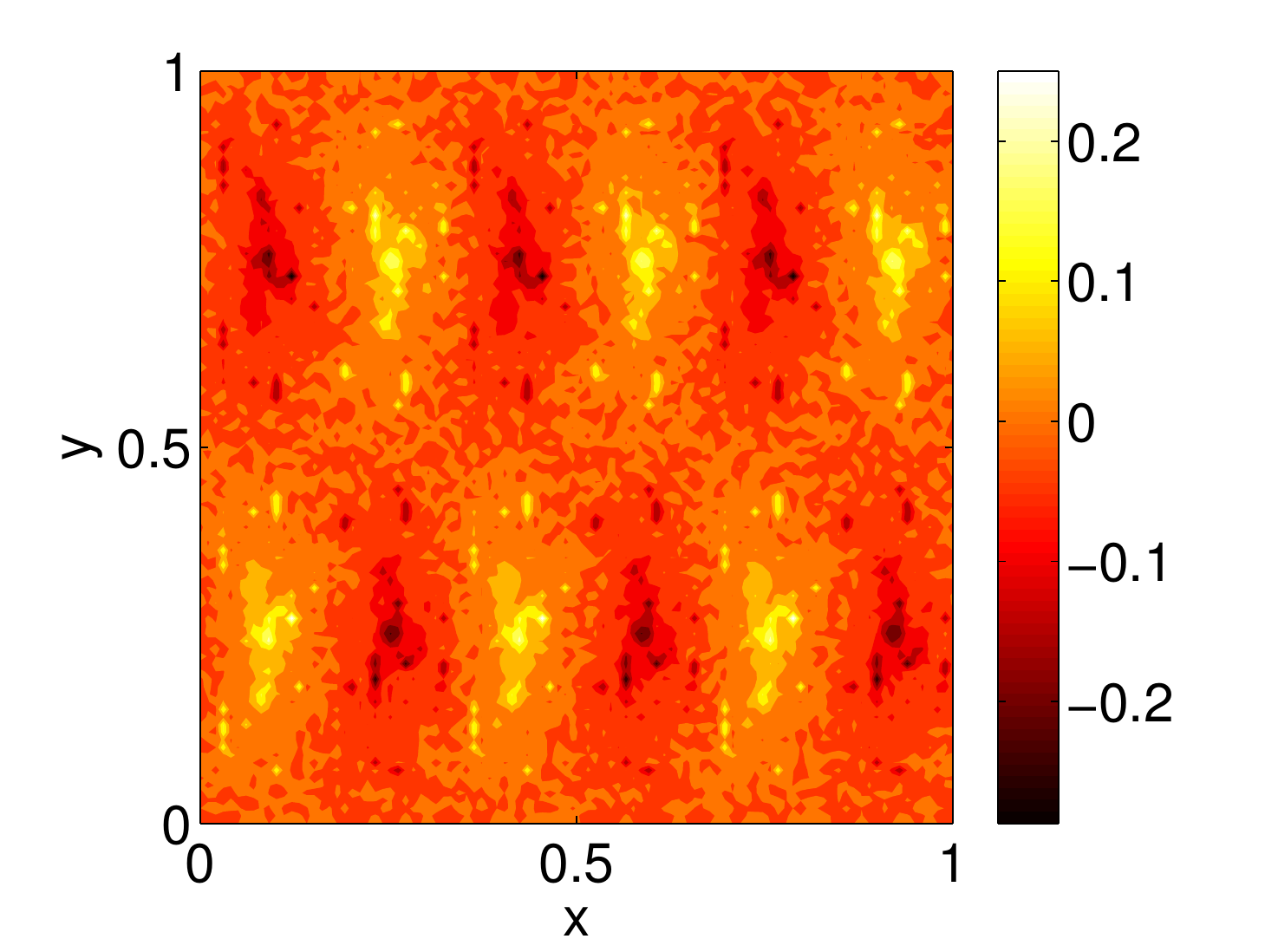}}
		\subfigure[$\,\,\omega=10$]{\includegraphics[width=0.24\textwidth]{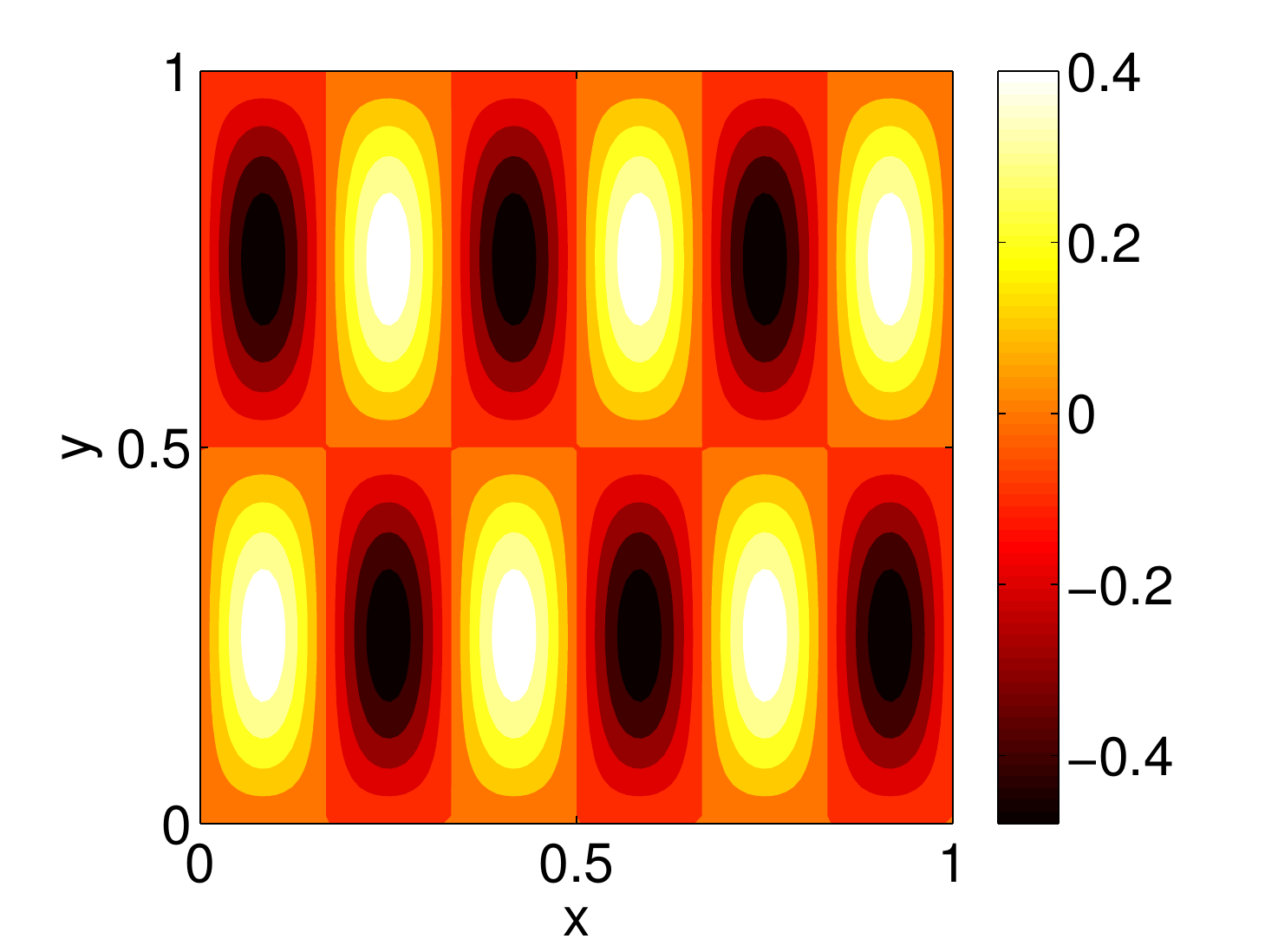}}
		\caption{The Lagrangian time average $\langle \Omega_{12}\rangle(\vecx_0)$ for the cell flow, for various values of $\omega$.  The time average is computed over a sufficiently long time interval such that the pictured result is independent of the time interval.}
	\label{fig:omega12_cell}
\end{figure}

The Lagrangian average $\langle\Omega_{12}\rangle(\vecx_0)$ is shown for the case of the cell flow in Figure~\ref{fig:omega12_cell}.  Evidence of a chaotic flow is visible in panels (b)--(c) corresponding to the intermediate values of the oscillating frequency $\omega$.  The flow is not chaotic at the large values of $\omega$ (panel (d)), as the rapid oscillation in the boundaries between the different cells in the flow effectively averages to zero over a Lagrangian trajectory, and the result for $\omega=10$ is qualitative very similar to the other extreme case with $\omega=0$.  In no case is $\langle \Omega_{12}\rangle(\vecx_0)$ zero in extended regions of the domain, and hence, only the biaxial fixed point survives as a steady solution of the $Q$-tensor equations.  It is noteworthy however that the magnitude of $\langle\Omega_{12}\rangle(\vecx)_0$ is much smaller for the $\omega=10$ case than for the $\omega=0$ case, while these quantities otherwise have the same spatial structure.  This explains why the uniaxial fixed points persist in the $\omega=10$ simulations for late times (e.g. Figure~\ref{fig:omega12_cell} (d,h)), before eventually being overwhelmed by the effect of the co-rotation term.


Summarizing, a planar form of the Landau--de Gennes equations coupled to hydrodynamics has been introduced.  
A limiting scenario has been identified whereby the hydrodynamics decouples from the $Q$-tensor dynamics, meaning that the $Q$-tensor is driven by an externally-prescribed flow field.  This is an particularly appropriate regime in which to study the chaotic advection of the $Q$-tensor.
The main tool for analyzing the resulting equations (both with and without flow) has been the identification of the fixed points of the dynamical equations without flow, which are relevant (to varying degrees) when flow is added to the model.  The fixed points are classified as stable/unstable and are further classified as either uniaxial or biaxial.

Accordingly, various model chaotic flows have been investigated, with the focus being primarily on cases where tumbling is not important (although tumbling -- effectively an inhomogeneous contribution to the advected $Q$-tensor equations -- has been investigated briefly also).  It is found that only the biaxial fixed point survives as a solution of the $Q$-tensor dynamics under the imposition of a general flow field.  This is due to the co-rotation term, whose presence means the uniaxial fixed points no longer solve the (advected) $Q$-tensor dynamics.  For certain highly specific flows, for which the co-rotation term $\Omega_{12}$ is zero or effectively zero on trajectories, both families of fixed points survive.  It is anticipated that these insights can be used as a means of controlling liquid-crystal morphology in planar geometries.

\appendix

\section{Derivation of the coupled equations for the hydrodynamics and the Q-tensor}
\label{app:derivation}

In this section  the general framework for coupling hydrodynamics to the $Q$-tensor dynamics is described, with a view to deriving Equation~\eqref{eq:dynam2} in the main text from first principles.  The starting-point is a variational principle wherein the Lagrangian for the system is given as follows~\cite{lowengrub1998}:
\begin{equation}
L=\int_\Omega \mathd^3a \bigg\{\tfrac{1}{2} \left(\frac{\partial\vecx}{\partial\tau}\right)^2+\left(1-\frac{\rho}{\rho_0}\right)\frac{p}{\rho_0}-
\frac{1}{\rho_0} \left[\chi(\mxQ)+W(\mxQ)-\lambda \mxQ\cdot\mathbb{I}\right]\bigg\},
\end{equation}
where $\mathd^3 a=\mathd m=\rho \mathd^n x$, and $\bm{a}$ is a Lagrangian label.  Thus, $\partial/\partial \tau$ denotes the time derivative at a fixed particle label.  Also, the pressure $p$ appears as a Lagrange multiplier that enforces the incompressibility of the fluid (thereby assigning to the fluid a constant density $\rho_0$); $\lambda$ is a further Lagrange multiplier that enforces the tracelessness of $\mxQ$, with $\mxQ\cdot\mathbb{I}=Q_{ij}\delta_{ij}=Q_{ii}$.   The generalized velocities are identified as $\vecv=\partial\vecx/\partial \tau$, and $\dot \mxQ=\partial \mxQ/\partial \tau$, and the relation
\[
\dot \mxQ\equiv\frac{\partial \mxQ}{\partial\tau}=\frac{\partial \mxQ}{\partial t}+\vecv\cdot\nabla \mxQ
\]
connects the Lagrangian time derivative $\dot\mxQ$  with its Eulerian counterpart.  Thus, the generalized forces are identified as
\begin{equation}
\bm{F}_{\vecv}=\frac{\partial}{\partial \tau}\frac{\delta L}{\delta (\partial\vecx/\partial\tau)}-\frac{\delta L}{\delta \vecx},\qquad
\bm{F}_{\mxQ}=\frac{\partial}{\partial \tau}\frac{\delta L}{\delta(\partial \mxQ/\partial \tau)}-\frac{\delta L}{\delta \mxQ},
\label{eq:app:gen_force}
\end{equation}
where $\delta L/\delta (\partial\vecx/\partial \tau)$ etc. denote functional derivatives, taken in the appropriate sense (e.g. References~\cite{lowengrub1998,phdlennon}).  Following References~\cite{sonnet2001dynamics,sonnet2004continuum},  the theory is made into a dynamical one by assuming that the generalized (conservative) forces in Equation~\eqref{eq:app:gen_force} are related to the dissipative forces via the equations
\begin{equation}
\bm{F}_{\vecv}=-\frac{\delta \mathcal{R}}{\delta \vecv},\qquad
\bm{F}_{\mxQ}=-\frac{\delta \mathcal{R} }{\delta \dot \mxQ},
\label{eq:app:gen_force_bal}
\end{equation}
where $\mathcal{R}$ is the dissipation function, with 
\begin{equation}
\mathcal{R}=\frac{1}{\rho_0}\int_\Omega \mathd^3 a R.
\label{eq:app:rara}
\end{equation}
This amounts to a balance between power production via the conservative forces and power dissipation.

In order to make the theory materially frame-indifferent, it is necessary that the dissipation function $\mathcal{R}$ depend on a frame-indifferent time derivative of the $Q$-tensor; as in Reference~\cite{qian1998},  this is chosen to be the co-rotational derivative,
\[
\stackrel{\circ}{\mxQ}\equiv\frac{\partial \mxQ}{\partial t}+\vecv\cdot\nabla \mxQ-\mxOm \mxQ+\mxQ\mxOm,\qquad
\Omega_{ij}=\tfrac{1}{2}\left(\frac{\partial u_j}{\partial x_i}-\frac{\partial u_i}{\partial x_j}\right).
\]
Also, in order for the theory to reduce to the Navier--Stokes equations in the limit $\mxQ\rightarrow 0$, it is necessary for $\mathcal{R}$ to depend on the velocity $\vecv$ through the symmetric rate-of-strain tensor $\mxD$, where $D_{ij}=(1/2)(\partial_i u_j+\partial_j u_i)$.  Finally, in order for the dissipative forces to be linear in the generalized velocities, the dissipation function must be  bi-linear in $\stackrel{\circ}{\mxQ}$ and $\mxD$~\cite{sonnet2004continuum}.  These assumptions can be used to compute the following variation in $\mathcal{R}$ with respect to the generalized velocities:
\begin{equation}
\delta \mathcal{R}=\frac{1}{\rho_0}\int_{\Omega} \left[\frac{\partial R}{\partial\stackrel{\circ} \mxQ}\cdot\delta\dot \mxQ
-\mathrm{div}\left(\frac{\partial R}{\partial \mxD}+
\mxQ\frac{\partial R}{\partial\stackrel{\circ}{\mxQ}}-\frac{\partial R}{\partial\stackrel{\circ}{\mxQ}}\mxQ\right)\cdot\delta\vecv\right]d^3a.
\label{eq:app:dr}
\end{equation}
Combining Equations~\eqref{eq:app:gen_force_bal} and~\eqref{eq:app:dr} gives the following set of equations coupling the hydrodynamics to the orientational dynamics of the liquid crystal:
\begin{subequations}%
\begin{eqnarray}%
\rho_0 \dot\vecv&=&\nabla\cdot\mxT,\\
-\frac{\partial R}{\partial \stackrel{\circ}{\mxQ}}&=&
\frac{\partial \chi}{\partial\mxQ}-\mathrm{div}\left(\frac{\partial W}{\partial\nabla\mxQ}\right)-\lambda\mathbb{I},
\end{eqnarray}%
where
\begin{equation}
\mxT=-p\mathbb{I}-\nabla \mxQ\odot\frac{\partial W}{\partial\nabla \mxQ}
+\frac{\partial R}{\partial \mxD}+
 \mxQ\frac{\partial R}{\partial\stackrel{\circ}{ \mxQ}}-\frac{\partial R}{\partial\stackrel{\circ}{ \mxQ}} \mxQ,\qquad
\left(\nabla\mxQ\odot\frac{\partial W}{\partial\nabla\mxQ}\right)_{ij}=
\frac{\partial Q_{i'j'}}{\partial x_i}\frac{\partial W}{\partial (\partial Q_{i'j'}/{\partial x_j})}
\label{eq:app:myTdef}
\end{equation}%
is the stress tensor for the fluid, and where a sum over repeated indices is implied in Equation~\eqref{eq:app:myTdef}.  
\label{eq:app:dynam1}%
\end{subequations}%

The constitutive model for $R$ is outlined briefly.   Following Reference~\cite{ramage2015}, one takes
\[
R=\tfrac{1}{2}\zeta_1 \stackrel{\circ}{ \mxQ}\cdot \stackrel{\circ}{ \mxQ}
+\zeta_2  \mxD\cdot\stackrel{\circ}{ \mxQ}
+\tfrac{1}{2}\zeta_3  \mxD\cdot \mxD
+\tfrac{1}{2}\zeta_{31}  \mxD\cdot( \mxD \mxQ)
+\tfrac{1}{2}\zeta_{32} ( \mxD\cdot \mxQ)^2,\qquad \mxD\cdot\mxQ=D_{ij}Q_{ij}\text{ etc.}
\]
where $\zeta_1$, $\zeta_2$, $\zeta_3$, $\zeta_{31}$ and $\zeta_{32}$  are phenomenological viscosity coefficients, which gives the following a closed-form set of evolution equations based on Equations~\eqref{eq:app:dynam1}.  In particular, one obtains the following balance for the $Q$-tensor:
\begin{subequations}
\begin{multline}
%
%
\zeta_1\left(\frac{\partial\mxQ}{\partial t}+\vecv\cdot\nabla\mxQ-\mxOm\mxQ-\mxQ\mxOm\right)+\zeta_2\mxD
\\
=k\nabla^2\mxQ-\left(\alpha_F\mxQ-3\beta_F\mxQ^2+4\gamma_F\mathrm{tr}(\mxQ^2)\mxQ\right)+\tfrac{1}{3}\mathbb{I}
\left[\zeta_2 \mathrm{tr}(\mxD)-3\beta_F \mathrm{tr}(\mxQ^2)\right].
\label{eq:app:q_tensor}
\end{multline}
For the hydrodynamics,  $\rho_0 \dot\vecv=\nabla\cdot\mxT$ (as before), with now an explicit form for the stress tensor $\mxT$:
\begin{equation}
\mxT=-p\mathbb{I}-k\nabla\mxQ\odot\nabla\mxQ+\zeta_2\stackrel{\circ}{\mxQ}+\zeta_3\mxD+\zeta_{31}(\mxD\mxQ+\mxQ\mxD)+\zeta_{32}(\mxD\cdot\mxQ)\mxQ.
\label{eq:app:ttensor}
\end{equation}
\label{eq:app:dynam2}%
\end{subequations}%
In this way, the presented model agrees with the model derived elsewhere but by equivalent means by Qian~\cite{qian1998}.

\section{Fixed points of the $Q$-tensor dynamics for an arbitrary flow}
\label{sec:app:fixedpoints}

In this section, consideration is given to the properties of the following modified system of equations for the $Q$-tensor components in the planar geometry:
\begin{equation}
\frac{\mathd}{\mathd t}\left(\begin{array}{c}q\\r\\s\end{array}\right)=
\left(\begin{array}{c}F_1(q,r,s)\\F_2(q,r,s)\\F_3(q,r,s)\end{array}\right)+\left(\begin{array}{c}2r\omega_0\\\omega_0(s-q)\\-2r\omega_0\end{array}\right),
\label{eq:app:qtensor_2d_rde}
\end{equation}
where $(F_1,F_2,F_3)^T$ encode the $Q$-tensor dynamics (\textit{cf.} Equation~\eqref{eq:qtensor_2d_rde}) and $\omega_0$ is a constant.  The aim of this section is to show that only the biaxial fixed points exist under the deformation~\eqref{eq:app:qtensor_2d_rde} of the basic equations for $(q,r,s)$.  Referring back to Section~\ref{sec:discussion}, this further establishes that
\begin{itemize}
\item Only the biaxial fixed point with $r=0$ and $s=q$ exists under a general flow for which $\Omega_{12}$ is a function of time;
\item Again, only the biaxial fixed point exists under a specific flow for which $\Omega_{12}$ is constant along Lagrangian trajectories.
\end{itemize}
This thereby places the discussion in Section~\ref{sec:discussion} on a firm theoretical footing.

\begin{figure}
	\centering
		\includegraphics[width=0.6\textwidth]{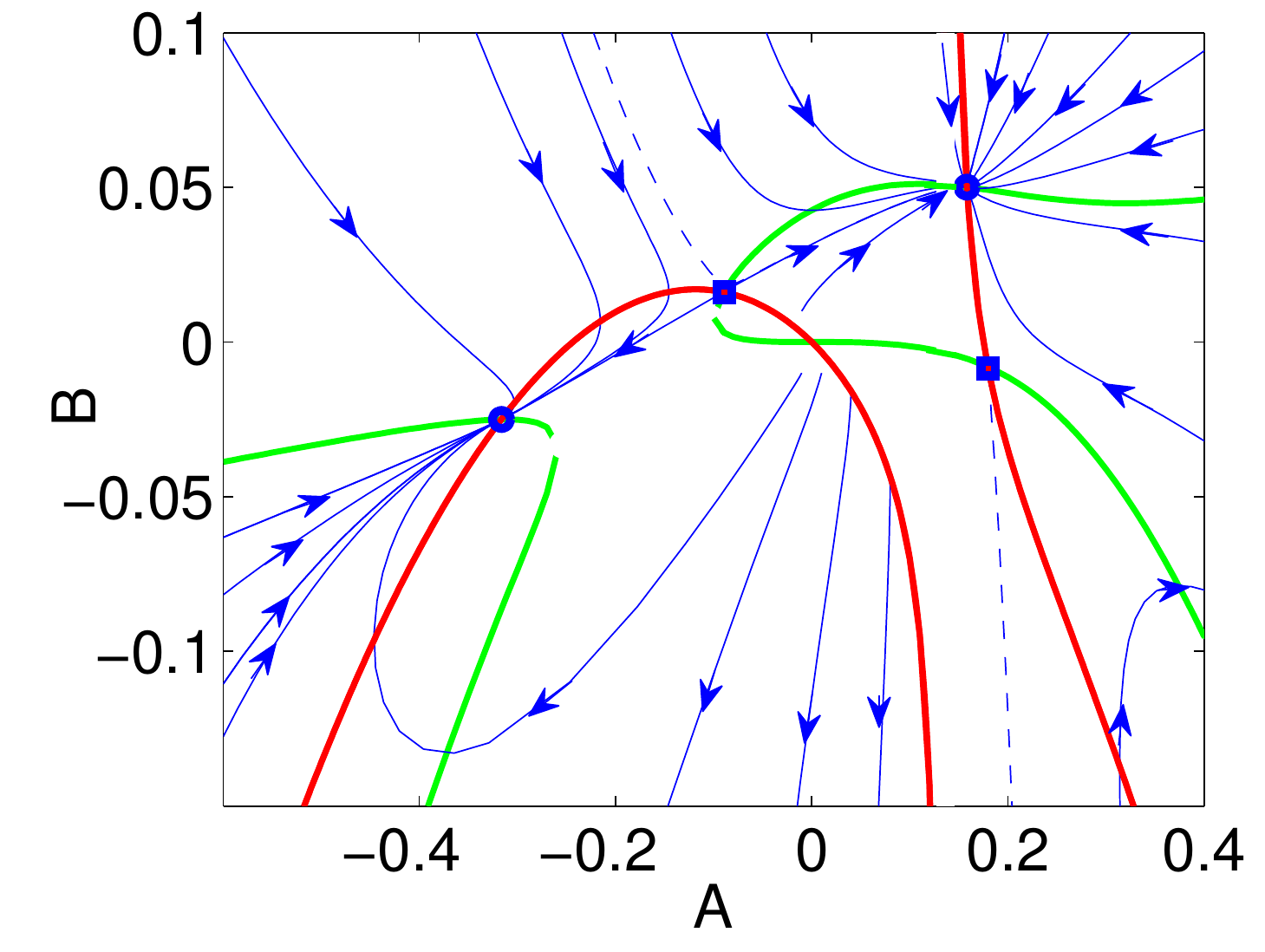}
		\caption{Phase portrait for the dynamics of Equations~\eqref{eq:app:qtensor_2d_AB}.  The thick solid lines denote the nullclines, the points of intersection of which are the fixed points.  Thus, the circles show the stable fixed points and the squares show the unstable ones.  The broken lines show sepratrices between different dynamical regimes (e.g. the separatrix in the $B>0$ half-plane marks out the boundaries between the basins of attraction of the stable uniaxial and biaxial fixed points).}
	\label{fig:phase_portrait_zoom}
\end{figure}

To make analytical progress,   new variables $A=q+s$ and $B=r^2-qs$, and $C=q-s$ are introduced, such that Equations~\eqref{eq:app:qtensor_2d_rde} simplify:
\begin{subequations}
\begin{align}
\frac{\mathd A}{\mathd t}&=g_1(1-\theta)A -g_2\left(A^2-2B\right)-A(A^2+B) ,\\
\frac{\mathd B}{\mathd t}&=2g_1(1-\theta)B+3g_2AB-2B(A^2+B)+2g_2A (A^2+B),\\
\frac{\mathd C}{\mathd t}&=g_1(1-\theta)C+3g_2A-(A^2+B)C+4\omega_0r.
\end{align}
\label{eq:app:qtensor_2d_AB}%
\end{subequations}%
Thus, $(A,B)$ form a two-dimensional dynamical system, the behaviour of which can be characterized by a corresponding phase portrait, the fixed points of which are obtained by setting
\begin{subequations}
\begin{align}
g_1(1-\theta)A -g_2\left(A^2-2B\right)-A(A^2+B)&=0,\\
2g_1(1-\theta)B+3g_2AB-2B(A^2+B)+2g_2A (A^2+B)&=0.
\end{align}
\end{subequations}%
The fixed points for $A$ and $B$ are thus independent of $C$ and $\omega_0$. By inspection of the pertinent phase portrait (Figure~\ref{fig:phase_portrait_zoom}), all trajectories are either unbounded, or flow into one or other of the two fixed points.  
These fixed points are further identified with the biaxial or type-3 uniaxial fixed points.
However, to reconstruct the full solution in terms of the primitive variables $(q,r,s)$ (and thus pick out precisely which of the fixed points of the full system of equations is selected by the dynamics in Figure~\ref{fig:phase_portrait_zoom}), a third equation is necessary, which is obtained by setting $\mathd C/\mathd t=0$ in Equation~\eqref{eq:app:qtensor_2d_AB}(c):
\begin{equation}
g_1(1-\theta)C+3g_2AC-(A^2+B)C+4\omega_0r=0.
\end{equation}
Using $r^2=B+(1/4)A^2$, this becomes
\begin{equation}
\bigg\{\left[g_1(1-\theta)+3g_2A-(A^2+B)\right]^2+16\omega_0^2\bigg\}^2C^2=16\omega_0^2\left(B+\tfrac{1}{4}A^2\right).
\end{equation}
For $\omega_0^2\neq 0$, the only fixed point that survives is therefore $C=B+(1/4)A^2=0$, which is precisely the biaxial fixed point.  In contrast, if $\omega_0^2=0$, then either $C=0$ or $g_1(1-\theta)+3g_2A-(A^2+B)=0$, which corresponds to a uniaxial fixed point.  Thus, in the presence of an arbitrary flow, it can be expected that the system will tend to a uniform state consisting only of the biaxial fixed point.


\end{document}